\documentclass[aip,jcp,reprint,floatfix]{revtex4-2}

\usepackage{amssymb}
\usepackage{amsmath}
\usepackage{amsfonts}
\usepackage{graphicx}

\usepackage{cancel}

\usepackage{newfloat}
\DeclareFloatingEnvironment[
  listname = {Case studies} ,
  name = {CASE STUDY}https://www.overleaf.com/project/639065ef54e709d395b47d76
]{casestudy}

\newcommand{\FE}{\kappa}

\newcommand{\wv}{\vec{w}}

\renewcommand{\vec}[1]{\boldsymbol{#1}}

\newcommand{\mat}[1]{\mathbb{#1}}
\newcommand{\Dp}[1]{\partial_{#1}}

\renewcommand{\Im}{\text{Im}}
\renewcommand{\Re}{\text{Re}}

\newcommand{\beq}{\begin{eqnarray}}
\newcommand{\eeq}{\end{eqnarray}}

\newcommand{\tr}{\text{Tr}}
\newcommand{\Tr}{{\tr}}
\newcommand{\sgn}{\text{sgn}}

\newcommand{\half}{\tfrac{1}{2}}

\newcommand{\ansatzeq}{:=}

\newcommand{\rcite}[1]{Ref.~\onlinecite{#1}}
\newcommand{\Rcite}[1]{Ref.~\onlinecite{#1}}

\newcommand{\rcites}[1]{Refs~\onlinecite{#1}}

\newcommand{\HH}{\text{H}}
\newcommand{\Hx}{\text{Hx}}
\newcommand{\xx}{\text{x}}

\newcommand{\xrm}{\text{x}}
\newcommand{\crm}{\text{c}}
\newcommand{\Hrm}{\text{H}}
\newcommand{\xc}{\text{xc}}
\newcommand{\Hxc}{\text{Hxc}}

\newcommand{\W}{{\cal W}}

\newcommand{\Ts}{{\cal T}_{s}}
\newcommand{\T}{{\cal T}}
\newcommand{\E}{{\cal E}}

\newcommand{\F}{{\cal F}}

\newcommand{\EHxc}{\E_{\Hxc}}
\newcommand{\EHx}{\E_{\Hx}}
\newcommand{\Ec}{\E_{\crm}}
\newcommand{\EH}{\E_{\Hrm}}
\newcommand{\Ex}{\E_{\xrm}}

\newcommand{\HInt}{U}

\newcommand{\CSF}{\text{CSF}}

\newcommand{\DFA}{{\text{DFA}}}
\newcommand{\EDFA}{\text{EDFA}}

\newcommand{\level}{\text{level}}

\newcommand{\SD}{{\text{SD}}}

\renewcommand{\sd}{{\text{sd}}}
\newcommand{\dd}{{\text{dd}}}

\newcommand{\RES}{\text{RES}}
\newcommand{\ENS}{\text{DENS}}

\newcommand{\gs}{\text{gs}}
\newcommand{\ts}{\text{ts}}

\newcommand{\pr}{^{\prime}}

\renewcommand{\vr}{\vec{r}}

\newcommand{\vrp}{\vec{r}\pr}

\newcommand{\tint}{{\begingroup\textstyle\int\endgroup}}

\newcommand{\iket}[1]{|#1\rangle}
\newcommand{\ibraket}[2]{\langle#1|#2\rangle}
\newcommand{\ibraketop}[3]{\langle#1|#2|#3\rangle}
\newcommand{\ibkouter}[1]{|#1\rangle\langle#1|}

\newcommand{\iout}{\ibkouter}

\newcommand{\FDT}{{\text{FDT}}}

\newcommand{\LDA}{\text{LDA}}

\newcommand{\up}{\mathord{\uparrow}}
\newcommand{\down}{\mathord{\downarrow}}

\newcommand{\nh}{\hat{n}}
\newcommand{\Th}{\hat{T}}
\renewcommand{\th}{\hat{t}}
\newcommand{\Wh}{\hat{W}}
\newcommand{\vh}{\hat{v}}

\newcommand{\Hh}{\hat{H}}

\newcommand{\Gammah}{\hat{\Gamma}}

\newcommand{\zp}{0^{+}}
\newcommand{\zm}{0^{-}}

\newcommand{\SCE}{\text{SCE}}
\newcommand{\ESCE}{V_{ee}^{\SCE}}

\newcommand{\ZPE}{\text{ZPE}}

\usepackage[normalem]{ulem}
\usepackage{color}
\usepackage{xcolor}

\definecolor{Mygrey}{gray}{0.80}
\definecolor{lteal}{rgb}{0.10,0.60,0.70}
\definecolor{dkred}{rgb}{0.80,0.10,0.00}
\definecolor{Purple}{rgb}{0.40,0.00,0.50}

\newcommand{\comment}[1]{}

\ifdefined\NoComments

 \newcommand\TGD[2]{}
 \newcommand\SPD[2]{}
 \newcommand\LKD[2]{}
\else
 \ifdefined\LightComments

 \newcommand\TGD[2]{{\bf !\textcolor{Purple}{#2}}}
 \newcommand\SPD[2]{{\bf !\textcolor{lteal}{#2}}}
 \newcommand\LKD[2]{{\bf !\textcolor{red}{#2}}}
\else

 \newcommand\TGD[2]{{\bf !\textcolor{Purple}{#2}}}
 \newcommand\SPD[2]{{\bf !\textcolor{lteal}{#2}}}
 \newcommand\LKD[2]{{\bf !\textcolor{orange}{#2}}}
 \fi
\fi

\newcommand{\Response}[2]{\textcolor{blue}{#2}}

\begin{document}

\title{
``Ensemblization'' of density functional theory}

\author{Tim Gould}
	\affiliation{Qld Micro- and Nanotechnology Centre, Griffith University, Nathan, Qld 4111, Australia}
\email{t.gould@griffith.edu.au}
\author{Leeor Kronik}
	\affiliation{Department of Molecular Chemistry and Materials Science, Weizmann Institute of Science, Rehovoth 7610001, Israel}
 \email{leeor.kronik@weizmann.ac.il}
\author{Stefano Pittalis}
	\affiliation{CNR-Istituto Nanoscienze, Via Campi 213A, I-41125 Modena, Italy}
 \email{stefano.pittalis@cnr.it}

\begin{abstract}
Density functional theory (DFT) has transformed our ability to investigate and understand electronic ground states.   
In its original formulation, however, DFT is not suited to addressing (e.g.) degenerate ground states, mixed states with different particle numbers, or excited states.
All these issues can be handled, in principle exactly, via ensemble DFT (EDFT).
This Perspective provides a detailed introduction to and analysis of EDFT, in an in-principle exact framework that is constructed to avoid uncontrolled errors and inconsistencies that may be associated with {\it ad hoc} extensions of conventional DFT.
In particular, it focuses on the ``ensemblization'' of both exact and approximate density functionals, a term we coin to describe a rigorous approach that lends itself to the construction of novel approximations consistent with the general ensemble framework, yet applicable to practical problems where traditional DFT tends to fail or does not apply at all.
Specifically, symmetry considerations and ensemble properties are shown to enable each other in shaping a practical DFT-based methodology that extends beyond the ground state and, in doing so, highlights the need to look outside the standard ground state Kohn-Sham treatment.
\end{abstract}

\maketitle
\tableofcontents

\section{Introduction}
\label{sec:Introduction}

Density functional theory (DFT)\cite{HohenbergKohn} is a first-principles approach
to the many-electron problem, in which the electron density,
rather than the many-electron wave function, plays the central
role.~\cite{Parr89,Gross90,Koch01,Tsuneda14} By virtue of offering an excellent balance between accuracy and computational cost, DFT has become the method of choice across a wide range of applications in diverse research fields within (at least) chemistry, condensed matter physics, and materials science.\cite{Martin04,Cramer2006,Sholl09,Giustino2014,Cohen2016} 

The overwhelming majority of practical DFT applications rely on mapping the original interacting-electron problem into an equivalent system of a fictitious electron gas, described by a single Slater determinant, such that the fictitious system retains the same ground-state density as the real one.\cite{KohnSham,Seidl1996}
This mapping is exact in principle, but almost always approximate in practice. Nonetheless, over fifty years of research have resulted in highly sophisticated approximations that offer quantitative accuracy for a wide range of realistic scenarios, thus facilitating the wide reach of DFT.\cite{Perdew2001-Jacob,Fiolhais03,Kummel08,Becke14,Jones15,Teale22}

Normally, the reference fictitious electron gas, into which the original system is mapped, is a pure quantum mechanical state. However, there are important cases where use of a reference {\it ensemble} state, i.e., a statistical mixture of pure quantum states,\cite{Fano1957} is either desirable or outright necessary. Four scenarios where ensemble reference states arise are: (i) systems with degenerate ground states,\cite{Valone1980,Ullrich2001} (ii) systems possessing a fractional number of electrons,\cite{Perdew1982} (iii) systems in an excited stationary state,\cite{Theophilou1979,GOK-1,GOK-2} and (iv) systems at a finite temperature.\cite{Mermin1965}
Collectively, approaches dealing with any of the four scenarios (and related ones) are known as ensemble DFT (EDFT).

In this Perspective, we explore the first three scenarios. We do not review thermal (or free-energy) DFT as: a) In this work we do not aim at addressing  properties requiring  consideration of statistical mechanics; and b) Thermal DFT is a more mature field than the three cases we do consider -- see, e.g., Refs.\ \onlinecite{Pastorczak2013,Pribram-Jones2014bookchap,Karasiev2014,vorberger2025roadmapwarmdensematter}. 
For each of the three types of ensembles reviewed here, we explain why EDFT is needed and the framework within which it is rigorously defined.
We do not aim to provide a comprehensive overview of all ensemble types and approaches, or a complete historical survey of EDFT, and refer the reader to Refs.\ \onlinecite{Nagy2003,Filatov2015-Review,Cernatic2022} for a more comprehensive picture. 

Here, we focus on presenting a unified framework for the extension of standard approximate density functionals into novel ensemble density functionals, a process which we call ``ensemblization'', and on explaining how ensemblization can solve problems with which traditional DFT struggles.
We stress that this provides our perspective to key steps in the ensemblization strategy. Although the point of view taken and solutions presented here are based on first principles throughout, other first principles inspired options exist, some of which are reviewed briefly in Section \ref{sec:Others} below.

\section{Ensembles in density functional theory}
\label{sec:EDFT}

In electronic structure theory, the primary goal of calculations is usually to evaluate the energy of electrons, given an external (nuclear) potential $v_N(\vr)$.
For the ground state, this can be achieved using the variational theorem,
\begin{align}
E_0[v_N]:=\min_{\Psi}\ibraketop{\Psi}{\Th+\vh_N+\Wh}{\Psi}
\label{eqn:Ev_Min}
\end{align}
where $\iket{\Psi}$ is a Fermionic wavefunction, $\Th$ is the kinetic energy operator and $\Wh$ is the electron-electron interaction operator.
We note that all wavefunctions occurring throughout the manuscript are considered to be normalizable and to obey $\ibraket{\Psi}{\Psi}=1$.
We also stress that the theory reviewed in the present manuscript is restricted to finite systems with discrete spectra, and thus needs to be treated with caution if considered in the context of extended solids, which is an issue of recent interest.\cite{Leano2024,Ohad2025}

\subsection{Kohn-Sham theory}
\label{sec:KS}

In the pure-state Kohn-Sham (KS) formulation of DFT,\cite{KohnSham} the interacting many-electron variational problem is circumvented. This is accomplished by  mapping a {\em non-degenerate} ground state of the physical interacting-electron
system -- or a  compatible selected state in a degenerate ground state -- into the ground state of auxiliary non-interacting electrons in the form of a single Slater determinant of orbitals, \{$\varphi_k(\vr)$\}. The orbitals are subject
to a common external auxiliary potential, $v_s$, i.e., 
\begin{equation}
\left( -\frac{\nabla^2}{2}+ v_s(\vr) \right) \varphi_k(\vr)=\epsilon_k\varphi_k(\vr),
\label{eqn:KS}
\end{equation}
where $\epsilon_k$ and $\varphi_k(\vr)$ are  energy levels and orbitals, respectively, of the non-interacting electron system. 
$v_s(\vr)$ is constructed such that the density obtained from the occupied KS one-electron orbitals, $n(\vr) = \sum_{i, {\rm occ.}}|\varphi_i(\vr)|^2$, is the same as that real system.\footnote{Throughout this work, we assume that non-interacting $v$-representability of interacting densities holds true either in the pure state or, more generally (and safely), in the ensemble generalization formulation. We further assume that the order of states is the same in the non-interacting and interacting systems.}
Note that we use Hartree units here and throughout.

In the KS framework, the ground-state energy of the original system is expressed as a functional of the density in the form
\begin{equation}
E_0[n]=T_s[n] + \int n (\vr) v_N (\vr) d\vr + E_{\Hxc}[n]\;.
\label{eqn:E0DFT}
\end{equation}
Then, we can evaluate Eq.~\eqref{eqn:Ev_Min} via,
\begin{align}
    E_0[v_N]:=\min_n E_0[n]
    \label{eqn:E0_v_Min}
\end{align}
Here, $T_s[n] = -\frac{1}{2} \sum_{i, {\rm occ.}} \ibraketop{\varphi_i}{\nabla^2}{\varphi_i}$ is the kinetic energy of the KS electrons,
$v_N(\vr)$ is the potential attracting electrons to the nuclei; $E_{\Hxc}[n]$ encompasses three energy terms (discussed individually in Section \ref{sec:Ensemblisation}): i) $E_{\Hrm}[n]$ -- the Hartree energy -- is the semi-classical electron repulsion energy; ii) $E_{\xrm}[n]$ -- the exchange energy -- arises from the need to anti-symmetrize the wave function; and iii) $E_{\crm}[n]$ -- the correlation energy -- arises from all remaining quantum effects not included in the previous terms, including the difference between the true and KS kinetic energies.

One can show that the KS potential has to be given by,
\begin{equation}
\label{Kohn_Sham_potential}
v_s[n](\vr) = v_N(\vr) + v_{\Hxc}[n](\vr) 
\end{equation}
where $v_{\Hxc}[n](\vr)$ is the functional derivative with respect to the density of $E_{\Hxc}[n]$.
It is the ability to effectively approximate $E_{\xc}$ (and therefore $v_{\xc}[n](\vr)$) that has led to the incredible success of KS theory. 

In anticipation of further considerations presented below, we provide some additional aspects of KS theory.\cite{Perdew2003} The KS kinetic energy, defined just below \Response{}{Eq.\ \eqref{eqn:E0_v_Min}}, can equivalently  be expressed as
\begin{align}
    T_s[n]:=&\min_{\Phi\to n}\ibraketop{\Phi}{\Th}{\Phi}
    := \ibraketop{\Phi_s[n]}{\Th}{\Phi_s[n]}\;,
    \label{eqn:Ts_gs}
\end{align}
where the minimization is subject to the constraint ($\Phi\to n$) that the minimization is taken over wave functions that yields a given target density, $n$.
The minimizing argument is the KS wave function, $\Phi_s[n]$ (a Slater determinant formed from the orbitals  \{$\varphi_k$\}).

The complementary Hartree-exchange-correlation (Hxc) energy is then defined as 
\begin{align}
    E_{\Hxc}[n]=&\min_{\Psi\to n}\ibraketop{\Psi}{\Th+\Wh}{\Psi} - \min_{\Phi\to n}\ibraketop{\Phi}{\Th}{\Phi}\;,
    \label{eqn:Exc_gs}
\end{align} 
where $\Psi$ is an interacting-electron wave function yielding the density $n$. 
Eqs.\ \eqref{eqn:Ts_gs} and \eqref{eqn:Exc_gs} can both be obtained from a universal functional, 
\begin{align}
F^{\lambda}[n]=&\min_{\Psi\to n}
\ibraketop{\Psi}{\Th+\lambda\Wh}{\Psi}
\;,
\label{eqn:AC_KS}
\end{align}
where $\lambda$ continuously and adiabatically connects~\footnote{The connection is ``adiabatic'' if the target density is the density of a ground state for any $\lambda$.} the KS system ($\lambda=0$) to the original interacting-electron systems ($\lambda=1$).\cite{Harris1974,Langreth1975}
By inspection, $T_s[n] \equiv F^0[n]$ and $E_{\Hxc}[n]=F^1[n] - F^0[n]$.

For completeness, we also mention briefly that the original KS theory has been generalized to the case of mapping to a {\it partially} interacting electron gas (namely, one that retains some but not all of the Coulomb repulsion between electrons) that can still be described by a single Slater determinant.\cite{Seidl1996} This allows the use of non-multiplicative potentials in the fictitious system (as opposed to the strictly multiplicative KS potential $v_s[n](\vr)$). In particular, this allows us to introduce Fock and Fock-like non-multiplicative exchange potentials at no loss of rigor and provides an exact framework for the use of hybrid functionals.\cite{Goerling1997, Garrick2020,Garrick2022} For simplicity, we mostly focus below on KS theory, but discuss pertinent aspects of generalized KS (GKS) theory where appropriate.

\subsection{From pure to ensemble states}

In quantum mechanics, a pure state can be described by a wave function, $\Psi$, or equivalently by a ``ket'' (vector in Hilbert space) $\iket{\Psi}$.
A Hermitian operator (observable), $\hat{O}$, is associated with any physical quantity, and the expectation value of that quantity is given by $\bar{O}=\ibraketop{\Psi}{\hat{O}}{\Psi}$.
As mentioned above, an ensemble state consists of a statistical average of multiple pure quantum states.
In the ensemble, the statistical averaging is performed by assigning a fixed probability, or weight, $w_{\FE}$ to each pure quantum state $\iket{\Psi_{\FE}}$ in the ensemble, such that $0<w_{\FE}<1$ and $\sum_{\FE} w_{\FE}=1$.
The expectation value of any physical quantity in the ensemble state is then given by the statistical average -- also known as a weighted average -- of the different quantum expectation values of the corresponding observable, i.e., 
\begin{equation}
\label{expectation_ensemble_explicit}
\bar{O}^{\wv}=
\sum_{\FE} w_{\FE}\ibraketop{\Psi_{\FE}}{\hat{O}}{\Psi_{\FE}},
\end{equation}
with the superscript $\wv$ indicating the set of weights used.
Mathematically, it is convenient to represent the ensemble by the operator
\begin{equation}
\label{Gamma_ensemble}
\Gammah^{\wv}=\sum_{\FE} w_{\FE}\iout{\Psi_{\FE}},
\end{equation}
which is a weighted sum over the density matrix operators formed from each of the pure states. To understand why, consider  
$\Tr[\Gammah^{\wv} \hat{O}]$, i.e., the trace of ensemble operator multiplied by the observable. 
We find that
$\Tr[\Gammah^{\wv} \hat{O}] =\Tr[\sum_{\FE} w_{\FE}\iout{\Psi_{\FE}}\hat{O}]
=\sum_{\FE} w_{\FE}\Tr[\iout{\Psi_{\FE}}\hat{O}]=\sum_{\FE} w_{\FE}\Tr[\ibraketop{\Psi_{\FE}}{\hat{O}}{\Psi_{\FE}}]
=\sum_{\FE} w_{\FE}\ibraketop{\Psi_{\FE}}{\hat{O}}{\Psi_{\FE}}=\bar{O}^{\wv}$,
where the first equality follows from the definition \eqref{Gamma_ensemble}, the second one from the linearity of the trace operator, the third one from the invariance of the trace under a cyclic permutation of matrices, the fourth one from the fact that the trace of a scalar is simply that scalar, and the last one from relation \eqref{expectation_ensemble_explicit}. 
We therefore have verified that the natural extension of 
the well-known pure-state expression $\bar{O}=\ibraketop{\Psi}{\hat{O}}{\Psi}$
to an ensemble state is 
\begin{equation}
\label{expectation_ensemble_Gamma}
\bar{O}^{\wv} = \Tr[\Gammah^{\wv} \hat{O}]\;,
\end{equation}
from which the pure state result is obtained as a special case by setting $\Gammah=\iout{\Psi}$.

Before proceeding, we mention a special case of Eq.\ \eqref{expectation_ensemble_Gamma} that is of particular interest in DFT: If $\hat{O}$ is chosen to be the density operator, $\nh(\vr)$, then the electron density of any ensemble is given by
\begin{align}
    n^{\wv}(\vr)=&\Tr[\Gammah^{\wv}\nh(\vr)]
    =\sum_{\FE}w_{\FE}n_{\FE}(\vr)
\end{align}
where $n_{\FE}(\vr)$ are the electron densities associated with individual pure states, $\iket{\Psi_{\FE}}$, within the ensemble.

\subsection{Ensemble Kohn-Sham theory}\label{eKSDFT}

How can one rigorously extend the KS framework to the case of ensemble states?
As discussed above, ensembles extend the usual quantum mechanical description to incorporate statistical averaging, a procedure that applies to any set of pure states, including those describing non-interacting electrons.
Thus, if $\iket{\Phi_s}$ is a wave function in the space of solutions of some non-interacting Hamiltonian, $\Hh_s$, it follows that $\bar{O}_s:=\ibraketop{\Phi_s}{\hat{O}}{\Phi_s}$ is the expectation value of an operator $\hat{O}$ therein. Then, using Eq.\ \eqref{Gamma_ensemble} we can construct an operator that is given by
\begin{align}
\label{Gamma_ensemble_KS}
\Gammah_s^{\wv}=&\sum_{\FE} w_{\FE}\iout{\Phi_{s,\FE}}\;,
\end{align}
and following Eq.\ \eqref{expectation_ensemble_Gamma} its expectation value is
\begin{align}
\bar{O}_s^{\wv}
=&\sum_{\FE} w_{\FE}\ibraketop{\Phi_{s,\FE}}{\hat{O}}{\Phi_{s,\FE}}
=\tr[\Gammah_s^{\wv}\hat{O}]\;.
\label{KS_expectation}
\end{align}

A KS ensemble is a special case of an ensemble of non-interacting pure states, the density of which, $n_s^{\wv}=\tr[\Gammah_s^{\wv}\nh]$, is equal to the  density, $n^{\wv}$, of the target interacting ensemble.
KS approaches have been motivated and derived for a variety of different ensembles and purposes.\cite{Valone1980,Theophilou1979,GOK-1,GOK-2,Perdew1982,Nagy1995,Senjean2020}
In all cases, it is the ensemble density, $n^{\wv}(\vr)$, that serves as the basic variable in EDFT.

In analogy to Eq.\ \eqref{eqn:AC_KS}, we can now define a universal ensemble functional,\cite{GOK-2,Nagy1995}
\begin{align}
    \F^{\wv,\lambda}[n]\equiv \bar{F}^{\wv,\lambda}[n]:=&\min_{\Gammah^{\wv}\to n}\tr\big[\Gammah^{\wv}(\Th+\lambda\Wh)\big]
    \nonumber
    \\
    :=& \tr\big[\Gammah^{\wv,\lambda}[n](\Th+\lambda\Wh)\big]\;.
    \label{eqn:Gammalambda}
\end{align}
which serves a role similar to $F^{\lambda}[n]$ in KS theory.\footnote{
Following the analogy with the ground state formulation, the adiabatic hypothesis is upgraded by requiring that the  structure of the relevant excitations does not change with $\lambda$. 
However, as elaborated in Section~\ref{sec:Hx},  we must allow  for extra degeneracy in the KS systems at $\lambda =0$.}
Note that in Eq.\ (\ref{eqn:Gammalambda}) above, and henceforth, we denote averages of observables (besides densities) with calligraphic capital letters rather than with bars on capital letters, e.g., $\bar{F} \rightarrow {\cal F}$.
Here, $\Gammah^{\wv,\lambda}[n]$ is the minimizing argument (which we assume to exist), where 
$\Gammah^{\wv}\to n$ indicates that we restrict ourselves to valid ensembles obeying $\tr[\Gammah^{\wv}\nh]=n$.

Pursuing the analogy with KS theory further, we may extend Eq.\ \eqref{eqn:E0DFT} to a KS ensemble as~\footnote{
It is not accidental that we have grouped H, x, and c together when writing the expression for ensembles; i.e., at this point we deliberately refrain from stating any forms for H, x, or c separately. This important point is discussed in much more detail in Section \ref{sec:Ensemblisation} below.
}
\begin{equation}
\E_0^{\wv}[n]=\T_s^{\wv}[n] + \int n(\vr) v_N(\vr) d\vr + \EHxc^{\wv}[n]\;,
\label{eqn:EE0DFT}
\end{equation}
where 
\begin{align}
\Ts^{\wv}[n]=&\F^{\wv,0}[n]\;,
&
\EHxc^{\wv}[n]=&\F^{\wv,1}[n] - \F^{\wv,0}[n]
\label{eqn:TsEHxc}
\end{align}
are the \emph{ensemble} KS kinetic energy and Hxc energy, i.e. the counterparts to $T_s$ and $E_{\Hxc}$ of Eqs.~\eqref{eqn:Ts_gs} and \eqref{eqn:Exc_gs}.
It then follows, using the relevant KS states (an aspect elaborated below), that
\begin{align}
    n=&\sum_i f_i^{\wv}|\varphi_i^{\wv}|^2\;,
    &
    \Ts^{\wv}=&-\half\sum_i f_i^{\wv}\int \varphi_i^{\wv*}\nabla^2\varphi_i^{\wv}d\vr\;,
    \label{eqn:Ts_ens}
\end{align}
where $f_i^{\wv}$ are the occupation factors of  $\varphi_i^{\wv}$ as relative to the KS ensemble.
Note that $f_i^{\wv}$ can be either an integer or a non-integer number; its value follows from the weighted average of the occupations relative to the individual KS states in the auxiliary ensemble, $\Gammah^{\wv}_s[n]$, keeping in mind the need to map KS states to pure many-body states, as per the state tracking argument in comment 2 above.

The orbitals, $\varphi_k^{\wv}$, obey an effective potential of the form of Eq.~\eqref{eqn:KS}, but with $v_s$ replaced by $v_s^{\wv}$:
\begin{equation}
\left( -\frac{\nabla^2}{2}+ v^{\wv}_s(\vr) \right) \varphi_i^{{\wv}}(\vr)=\epsilon_i^{{\wv}} \varphi_i^{{\wv}}(\vr)\;.
\label{eqn:KS_ens}
\end{equation}
The above notation emphasizes that the KS potential, and therefore the KS orbitals in EDFT, can vary with the weights chosen.~\footnote{This means that the KS states, $\iket{\Phi_{s,\FE}}$, are also  weight-dependent and formally should be written as  $\iket{\Phi^{(\wv)}_{s,\FE}}$.}
For notational simplicity, we shall allow ourselves {\em not} to stress all the dependence on the weights explicitly (via a superscript $^{\wv}$), unless it is essential.

Before proceeding, four important comments are in order, each of which is further explained and elaborated below:
\begin{enumerate}
\item Regular ground-state DFT is reproduced for the special case of an integer electron number, where the ensemble includes only a non-degenerate ground state.
\item In all cases considered here, we choose the weights, $\wv$, used in the ensemble of the original interacting system and in the KS system, to be the same. 
This involves an assumption that the relevant auxiliary states and interacting states can be `matched' to one another, e.g., by tracking them along the {\it ensemble} adiabatic connection path (namely, an adiabatic connection that preserves the overall ensemble density) and/or by exploiting pure state properties, e.g., electron number and/or spatial symmetry and/or spin symmetry. This may have specific ramifications for excited-state ensembles, a point we return to at the end of Section \ref{sec:Excited}.
\item Despite the above, the  density obtained from each auxiliary state in the ensemble, $\iket{\Phi_{s,\FE}}$, generally differs from its interacting counterpart, $\iket{\Psi_{\FE}}$, i.e., generally $n_{s,\FE}(\vr)=\ibraketop{\Phi_{s,\FE}}{\nh}{\Phi_{s,\FE}} \neq n_{\FE}(\vr)=\ibraketop{\Psi_{\FE}}{\nh}{\Psi_{\FE}}$.
It is only the \emph{ensemble-averaged} density,
\begin{align}
~~~~~~n(\vr)=\tr[\Gammah^{\wv}\nh(\vr)]
=\tr[\Gammah_s^{\wv}\nh(\vr)]
=n_s(\vr)\;,
\label{ensemble_density}
\end{align}
that must be the same, by construction, 
in the interacting ($\Gammah^{\wv}$) and the KS
($\Gammah_s^{\wv}$) systems, i.e., $\sum_{\FE} w_{\FE}n_{s,\FE}(\vr)=\sum_{\FE} w_{\FE}n_{\FE}(\vr)$.
In other words, we stress that the task of the ensemble KS system is to reproduce the density of the interacting ensemble, and not to reproduce the  density of each individual interacting state in the ensemble.
This task is achieved via a common set of single-particle KS orbitals,  derived from a single effective potential.
\item Because the states used to form Eq.~\eqref{Gamma_ensemble_KS} are by construction  expressed  in terms of single-particle orbitals,
the properties derived therefrom using Eq.~\eqref{KS_expectation} can also be expressed in terms of single-particle orbitals.
However, for reasons that discussed in detail below, the auxiliary states $\iket{\Phi_s}$  are not necessarily of single Slater determinant form.
In fact, it is allowed, often convenient, and sometimes even required to represent the ensemble auxiliary systems via individual states, $\iket{\Phi_{s,\FE}}$, that go beyond the standard KS Slater determinant, by using superpositions of \emph{several} degenerate Slater determinants, known as configuration state functions (CSFs).
This is addressed in Section~\ref{sec:HxResolution} below.
\end{enumerate}

\subsection{Systems in a degenerate ground state}
\label{sec:Symmetries}

We start our discussion by considering how to systematically deal with degenerate ground states of isolated atoms. These are highly symmetric quantum systems that form the building blocks of matter and therefore serve as elementary realistic systems that illustrate the issues brought about by degeneracy in DFT.
Symmetries are sometimes considered to be an esoteric issue in DFT, but in fact they raise serious issues of both formal and practical nature that are often overlooked. Importantly, many issues for atoms with degenerate ground states apply to general open shell ground and excited state problems. 

We first stress that standard pure-state DFT~\cite{HohenbergKohn,KohnSham} can be and has been extended to the case of a degenerate ground state.\cite{Gross90,Parr89,Kohn1985,Capelle2007}
Assuming the existence of valid solutions allows the KS approach to be applied to individual degenerate ground states.\cite{Erhard2025} However, pure-state DFT solutions are not without issues.
In some cases, \emph{aufbau} solutions do not exist~\cite{Trushin2024} and even when solutions do exist the DFT solution can fail to reproduce the symmetries of the interacting system.~\cite{Fertig2000}
These issues are successfully addressed by EDFT and therefore allow for a natural introduction to the advantages offered by ensembles in DFT.

To motivate the usefulness of EDFT for preserving physical symmetries, consider the ground state of the boron atom. 
This atom has five electrons that are arranged around a spherical
$-5/r$ nuclear potential. 
Neglecting relativistic interactions, it has a six-fold degenerate ground state.
We can classify the six ground states in terms of their leading Slater determinants, each having one electron in an open 
$p$-shell. An initial guess using symmetry-adapted single particle orbitals assigns the ``last'' electron to one of three real-valued $2p$ orbitals in the $x$, $y$ or $z$ directions  ($2p_x$, $2p_y$, or $2p_z$) and assigns the atom a net spin  ($\sigma$ = $\up$ or $\down$). 

Importantly, despite the nuclear potential being spherically symmetric, each of the six above-mentioned wave functions has a lower spatial symmetry; namely, \emph{cylindrical} symmetry along a specific axis. This cylindrical symmetry must be replicated by the Slater determinants of KS orbitals.
Hence, the corresponding {\em self-consistent} KS potential inherits the cylindrical symmetry of the selected state.~\cite{Fertig2000}
In a narrow sense, such a {\em state-specific} DFT solution behaves exactly as it should, i.e., the  corresponding exact KS potential does indeed produce the energy and density of the {\em selected} state.
But in a broader sense, the corresponding KS Hamiltonian supports neither a unique nor a spherically symmetric model of the boron atom.
In particular, the self-consistent KS orbitals will not be  $2p_x$, $2p_y$, or $2p_z$ orbitals,\cite{Fertig2000} nor will they behave as $p$ orbitals under rotation.

One could be tempted into naively thinking that the above issue can be circumvented by using a superposition of the degenerate initial $p$ orbitals.
Unfortunately, this is not so, because any normalized linear combination of basis functions in the degenerate sub-space simply rotates, rather than symmetrizes, the wavefunction. For example, taking $\tfrac{1}{\sqrt{2}}$ of $2p_x$ (``East'') plus $\tfrac{1}{\sqrt{2}}$ of $2p_y$ (``North'') yields a new $2p$ orbital that points to the ``Northeast''. 

In the absence of an ``easy fix'', can we work with the lower symmetry KS-potential nonetheless?
Unfortunately, this is fraught with practical difficulties, not just formal ones. Recall that for $d$ and $f$ orbitals, as well as for complex-valued $p$ orbitals, the densities associated with the orbitals can differ by more than a rigid transformation in space. For example, the density corresponding to the complex orbital $2p_{-1}$ is doughnut shaped and parallel to the $x-y$ plane. It cannot be rotated onto the $2p_0$ orbital, which is oriented along the $z$ axes and has a cylindrical, rather than doughnut-like, shape. As a consequence, approximate density functionals, designed to handle non-degenerate ground states, will almost unavoidably yield different (i.e., non-degenerate) energies, depending on which of the degenerate states is being considered, even if the exact density is used! 

This problem would immediately carry over into, e.g., calculations of the atomization energy of molecules, which involve breaking the molecule into its corresponding (usually open-shell) atoms. The unacceptable result would then be that the computed atomization energy depends on the specific state and type of atomic orbitals used in the calculation. While this situation may be ameliorated by accounting for the paramagnetic currents of the complex-valued orbitals,\cite{Baerends1997, Becke2002, Johnson2007, Maximoff2004, Pittalis2006b, Pittalis2007} the issue is not solved exactly.
Ensembles, instead, can solve the aforementioned problem exactly and straightforwardly.

For boron, a useful ensemble involves an equal mix of the three spatially degenerate ground states in both the interacting and KS systems. Let us then consider%
\footnote{Note that an unpaired occupation of the spin configurations does not necessarily imply a spin-depended effective potential and spin-dependent orbitals as long as spin-restricted calculations (i.e., DFT) rather that spin-unrestricted calculations (i.e., SDFT) are being carried out; See, for example,  S.\ Pittalis, S.\ Kurth, and E.\ K.\ U.\ Gross, ``On the degeneracy of
atomic states within exact-exchange (spin-) density functional theory'', {\it J.\ Chem.\ Phys.} {\bf 125}, 084105 (2006).}
\begin{equation}
\Gammah_s = \tfrac13 \sum_{m\in x,y,z}\iout{1s^22s^22p_m^{\up}}\;.
\end{equation}
The density of this statistical mixture, 
\begin{align}
n(\vr) & =   |\phi_{1s\up}(\vr)|^2 + |\phi_{1s\down}(\vr)|^2
  + |\phi_{2s\up}(\vr)|^2 + |\phi_{2s\down}(\vr)|^2
\nonumber \\
& +\tfrac13[|\phi_{2p_x\up}(\vr)|^2+|\phi_{2p_y\up}(\vr)|^2+|\phi_{2p_z\up}(\vr)|^2]\; , 
\end{align}
is manifestly spherically symmetric.
This symmetry can also be thought of as arising from treating the part-filled shell in a manner consistent with Uns\"old's theorem.~\cite{Unsold1927}
Thus, constructing this symmetrized KS density in practice simply boils down to fully occupying the $1s$ and $2s$ orbitals, and adding $\tfrac13$ of the density corresponding to each of the three $2p$ orbitals.

\begin{figure}
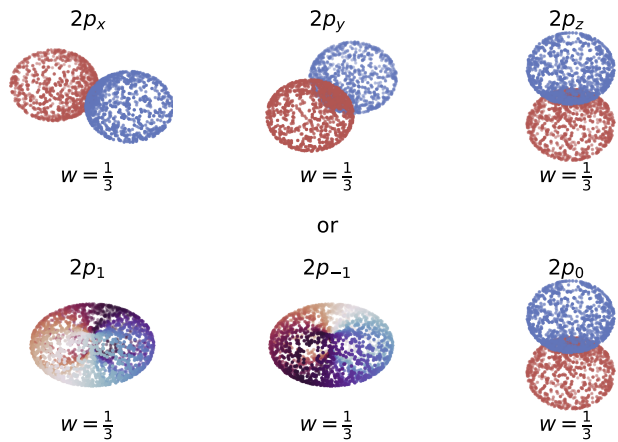

\includegraphics[width=\linewidth]{{{Fig2p}}}
\caption{Real-valued ($2p_x$, $2p_y$, $2p_z$) and complex-valued ($2p_{1}$, $2p_{-1}$, $2p_{0}$) orbitals.
Surface levels correspond to points in space with constant absolute values, whose color represents the phase of the orbitals.
Each $p$ orbital has cylindrical symmetry, whereas combining them as denoted with equal weights, yields an ensemble that retains spherical symmetry.
\label{fig:BoronOrbitals}}
\end{figure}

As illustrated in Fig.~\ref{fig:BoronOrbitals}, the self-consistent orbitals associated with the ensemble KS potential have well-defined atomic quantum numbers. The KS ensemble may be expressed as an average of three Slater determinant states -- each consisting of two electrons in the $1s$ and $2s$ shells, along with one $\up$ electron in the $2p_x$, $2p_y$, or $2p_z$ orbital. Importantly, we obtain the same ensemble density if we choose complex $2p_{\pm 1}$ orbitals instead of $2p_{x,y}$, so that the result \Response{}{becomes} independent of whether real or complex orbitals are employed.
We note that fractional equi-occupation of a degenerate sub-shell of atomic orbitals is a long-standing practice in KS solutions of atoms (see, e.g., Ref.\ \onlinecite{Lehtola2020}, and references therein), typically in the context of constructing pseudopotentials or atomic basis sets. 
The above motivation for invoking ensembles fundamentally justifies using fractional equi-occupations and validates the numerical convenience it enables.

The above example points out that while a density resulting from a weighted-average of densities may indeed be symmetrized, it is no longer the density of a pure ground state of the real system,\cite{Levy1982,Lieb1982} i.e., it cannot be obtained from any wave function in the degenerate ground-state sub-space! However, such a density is still physical, because it can be generated via an ensemble of states. 
A mathematical but convenient description of such a density is that it is not pure-state $v$-representable, but is ensemble $v$-representable.\cite{Gould2023-JCP,Erhard2025}

Importantly, the necessity of invoking ensembles goes beyond mere symmetry requirements. In fact, examples of {\em interacting pure-state} $v$-representable densities  that are {\em non-interacting  ensemble} $v$-representable densities but not {\em non-interacting pure-state} $v$-representable densities have  long been known.
\cite{Wang1996,Schipper1998,Schipper1999,Morrison2002} 
Therefore, it is reassuring to know that densities defined on a discrete lattice  are always non-interacting ensemble $v$-representable.\cite{Chayes1985} Moreover, in the continuum limit, 
it can be shown that  non-interacting and interacting ensemble $v$-representable densities are dense 
with respect to each other.~\cite{Englisch-I,Englisch-II}
This means that essentially any discrete representation of an interacting density may be approximated via an ensemble with a non-interacting $v$-representable density, which basically solves all the situations of practical interest.\footnote{A more general but also more abstract solution was pointed out in P.\ W.\ Ayers, ``Axiomatic formulations of the Hohenberg-Kohn
functional,'' Phys.\ Rev.\ A {\bf 73}, 012513 (2006).}

\begin{figure}[b!]
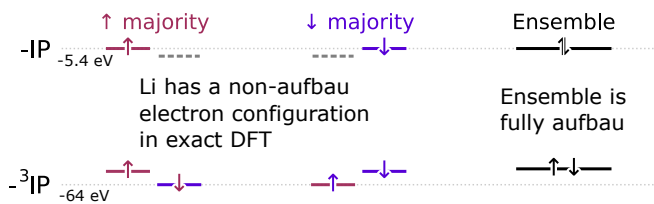

\includegraphics[width=\linewidth]{{{Li_Levels}}}
\caption{KS energy levels of Li in different exact KS approaches. Energy levels of spin-polarized DFT are inconsistent with the aufbau principle, whereas the ensemble is fully consistent with aufbau. IP indicates the first ionisation potential and $^3$IP indicates ionisation to a triplet state, i.e. removal of a $1s$ electron, (with experimental numbers given in the figure).
\label{fig:LiLevels}}
\end{figure}

For the boron atom, we did not pay special attention to spin, other then where necessary to designate individual states. The above considerations, however, are easy to extend to spin degeneracy in a way which is useful even when spatial symmetry is not an issue. Consider the lithium atom, the energy levels of which are illustrated in Fig.~\ref{fig:LiLevels}. The usual way to study this atom is to use the spin-polarized KS formalism, in which the spin-polarized KS potential, $v_s(\vr,\sigma)$, has different values for the two spins. However, it turns out that the exact spin-KS solution of lithium is inconsistent with the usual aufbau principle,\cite{Gritsenko2004}
as (for an $\up$-majority density) the filled $2s_{\up}$ energy level is higher than its empty $2s_{\down}$ counterpart in the minority channel.\footnote{This situation is sometimes referred to as obeying the aufbau principle in the broad sense [E.\ Kraisler, G.\ Makov, N.\ Argaman, and I.\ Kelson, Phys.\ Rev.\ A {\bf 80}, 032115 (2009)], as the aufbau principle is still obeyed for each spin channel separately} Aufbau dictates we should instead fill the minority $\down$-channel, but if we do so this and iterate to self consistency, we simply swap the up and down spin labels and return to the same problem. By contrast, taking an ensemble average, i.e., weighting the $\up$- and $\down$-majority states equally in Eq.~\eqref{Gamma_ensemble_KS}, completely avoids the problem. The resulting density is spin-unpolarized, meaning that the KS potential is also spin-unpolarized, i.e., both spin channels possess the same spatial part, and as a consequence spin up and down KS eigenvalues and associated spatial orbitals are identical. Filling is thus completely consistent with the aufbau principle, namely, the $1s$ orbital is filled with $\up$ and $\down$ electrons and the $2s$ orbital contains half of each.

Spatial averaging is by no means a special feature of boron, nor is  spin averaging a special feature of lithium. Both apply separately and together to any state with symmetries. In general, equally weighted ensembles (equi-ensembles) preserve all fundamental symmetries in densities and potentials\cite{NagyLevy2001,Gould2017-Limits,Gould2020-SP} and lead to separable orbitals with well-defined quantum numbers (e.g. $2p_z^{\up}$ in atoms or $3\pi_y^{\down}$ in linear molecules).
In boron this would involve averaging over all six degenerate states (i.e., including $\up$ and $\down$ in the ensemble) to obtain a KS boron with the same six-fold degenerate ground state as the real system.
As an additional example, equally weighting the nine degenerate triplet ground states of carbon (3 orbital angular momenta $\times$ 3 spin angular momenta) in an ensemble produces a nine-fold degenerate set of triplet KS ground states.%
\footnote{The KS states with two $2p$ electrons (combinations $xy$, $xz$ and $yz$ with $\up\up$, $\up\down$ and $\down\down$) all have the same non-interacting energy, because the $2p$ orbitals are degenerate with respect to the KS Hamiltonian, $-\half\nabla^2 + v_s$. 
The astute reader may realise that this makes the KS ground state 15-fold degenerate (reflecting the number of options of placing two electrons within six orbitals). Only nine of these states are compatible with the triplet ground state of carbon. This issue is resolved by careful ensemblization, as discussed in Section~\ref{sec:Ensemblisation}.}

Generally speaking, degenerate doublet and triplet (spin) states are ubiquitous in chemistry, while quadruplets and quintuplets have important technological uses for, e.g., magnetism.
Eigenstates of spin operators, spin manifolds, and strongly-correlated singlets have long been known to pose a ``symmetry dilemma''~\cite{Lowdin1963} for Hartree-Fock calculations,~\cite{Davidson1983} which also presents itself in KS calculations.~\cite{Perdew2021}, i.e., that in order to improve the treatment of energetics one must sacrifice some symmetry.
Furthermore, even sophisticated methodologies based on pure-state wave functions of definite spin can spuriously break spatial symmetries in the absence of careful spatial-symmetry adaptation.
In these cases, switching to state-averaged calculations (and thus to ensemble states) can offer a way to avoid such issues.

One may argue that ensemble calculations based on approximate functionals typically yield higher energies than symmetry broken solutions. 
However, this may not be an issue when energy differences, rather than absolute energies, are sought (which is nearly always the case in practical applications).
There are also situations in which ensembles yield lower energies than pure-state solutions.
This may occur when the latter are obtained by constraining the occupations of the single-particle orbitals according to an ordering that does not fulfill the {\em aufbau} principle, yet with the goal of targeting specific degenerate ground states.  

Violations of the {\em aufbau} principle 
\Response{}{call into} question the validity of the conventional non-interacting $v$-representability condition.\cite{Trushin2024} 
Examples are open-shell atoms of the first and second rows of the periodic table.
The ensemble approach (see \rcite{Gould2014-KS} for some examples) suffers no such issue, as the corresponding extended {\em aufbau} principle must be satisfied in an average sense.
It is worth noting that G{\"o}rling has proposed a symmetrized  DFT approach which enforces symmetry adaptation in spin and real-space via {\em pure}-state calculations and does not rely on the non-interacting reference state being a ground state.\cite{Goerling1993,Goerling2000,Trushin2023}
Thus, in principle, the approach can equally be applied not only to ground and lowest lying excited state of each given symmetry\cite{Gunnarsson1976:4274} but also to general high-energy eigenstates.
Ensembles built on symmetry considerations can achieve similar objectives -- see Section \ref{sec:HxResolution} below.

In summary, degenerate ground states and states which are energetically lowest in their symmetry come with a number of challenges that include (but are not limited to) those described above in the context of atoms. Equi-EDFT, in which 
degenerate states are assigned equal weights in the interacting and the KS systems, offers an effective way to bypass most of these issues. Importantly, it generates energies that are invariant to state selection, KS potentials with appropriate symmetry, and orbitals that have well-defined quantum numbers.

\subsection{Systems with a non-integer number of electrons}
\label{sec:Fractional}

A different arena where ensemble considerations play a major role is that of systems with a non-integer number of electrons.
At first glance, it may seem strange that such systems are at all a topic of interest, given that it is clearly non-physical for a chemical system to possess an overall non-integer number of electrons.\footnote{Here, the term `` a non-integer number of electrons'' is used literally.
{\it Apparent} fractional electron behavior, as in the fractional quantum Hall effect, is obviously physical.
Literal fractional electron behavior ensures continuity of density functionals as a function of $N$ and thus as a functional of $n$.}
Motivated by the spurious dissociation of a diatomic molecule into atoms with non-integer charges, obtained from DFT calculations within the local density approximation (LDA), the issue has first been raised by Perdew {\it et al.} \cite{Perdew1982}
But even in an exact treatment, quantum mechanics dictates that electrons can localize or delocalize such that the probability for finding an electron in a given region of space (say, around a certain moiety of a molecule) is fractional. Dealing with a non-integer number of electrons can then be viewed as a continuous, rather than a discrete, addition or removal of charge. As elaborated below, this leads to highly useful insights into exact properties of density functionals.

A system with a total fractional number of electrons may be addressed naturally in ensemble theory. 
Consider a system capable of exchanging an electron with a particle reservoir,  such that the classical probability of finding the system to possess $N$ or $N+1$ electrons is $1-\omega$ or $\omega$, respectively, where $0 \leq \omega \leq 1$. 
Based on Eq.\ (\ref{Gamma_ensemble}), the ensemble operator corresponding to such a system is given by:
\begin{equation}
\label{ens_operator_fractional}
\hat \Gamma^\omega = (1-\omega)\iout{\Psi_{N}}+
\omega\iout{\Psi_{N+1}},
\end{equation}
where $\iket{\Psi_{N}}$ and $\iket{\Psi_{N+1}}$ are pure many-electron {\em ground} states with $N$ and $N+1$ electrons,%
~\footnote{we stress that states with different  number of electrons belong to different sectors of the underlying overall Fock space and therefore are orthogonal.}
respectively (for simplicity we  assume these ground states to be non-degenerate).
Clearly, the ensemble becomes a pure state for $\omega=0$ or 1 and, in the absence of degeneracy, becomes amenable to conventional DFT treatment.

\begin{figure}
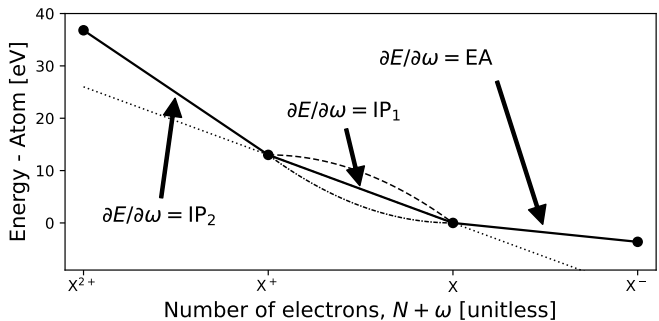

\includegraphics[width=\linewidth]{{{FigPWLinear}}}
\caption{
Solid line: Illustration of piecewise linearity (based on atomic Cl), with energies at arbitrary $N$ shown relative to that of the neutral atom and points indicating integer particle numbers. Gradients are related to electron removal (IP) or addition (EA) energies.
The dotted lines illustrate the concept of piecewise convexity, showing the increasing slope of each linear segment. The dashed curves illustrate (convex or concave)  deviation from linear behaviour that is typical to naive (non-ensemble) use of approximate density functionals.
\label{fig:PWL}}
\end{figure}

Using Eq.\ (\ref{ensemble_density}), the electron density of the ensemble defined by Eq.\ (\ref{ens_operator_fractional}) is readily expressed in terms of $n_N(\vr)$ and $n_{N+1}(\vr)$ , the $N$- and $N+1$ electron densities, as 
\begin{equation}
\label{density_fractional}
n(\vr) =  (1-\omega) n_N(\vr) + \omega n_{N+1}(\vr).
\end{equation}
Integrating this result, the total number of electrons is then $N+\omega$. 
Because minimizing the energy of the ensemble, per a given weight $\omega$, simply entails the minimization of the energy of the two pure states, we immediately obtain the ground-state energy of the ensemble system as:\cite{Perdew1982}
\begin{equation}
\label{E_fractional}
    E(N+\omega) = (1-\omega)E(N) + \omega E(N+1).
\end{equation}
As illustrated in Fig.\ \ref{fig:PWL}, this means that the energy of a system, as a function of a continuous number of electrons, is a piecewise-linear interpolation of energy values at integer electron numbers.
Furthermore, its derivative, $\partial E(N+\omega)/\partial\omega =E(N+1)-E(N)$, is therefore a chemically-relevant constant -- negative the ionisation potential (or electron affinity).

Importantly, the operator defined in Eq.\ (\ref{ens_operator_fractional}) is by no means the only way to construct an ensemble leading to a total number of $N+\omega$ electrons. 
One could achieve the same by creating an ensemble that mixes in additional pure states (e.g., with $N-1$ electrons, $N+2$ electrons, etc.). 
However, such ensembles would not reduce the total energy.
This is because systems based on Coulomb attraction and repulsion (namely, all chemical systems) are nearly always piecewise-convex, i.e., the energy gain upon addition of an electron cannot increase with increasing $N$,\cite{Gross90,Gonis11,Ayers24}
a result recently shown to follow from some common assumptions about the universal functional.\cite{Burgess2023}$^{\rm,}$
\footnote{Note that an example with non-chemical nuclei was recently discovered to break piecewise convexity,{\cite{DiMarino2024}} an example which falls outside the scope of the aforementioned proof as it does not satisfy the assumption of size-consistency.}

Convexity is reflected in Fig.\ \ref{fig:PWL} in the fact that the slope of each linear segment becomes less negative with increasing $N$.
Fig.\ \ref{fig:PWL} also shows the continuation of the straight line between $N$ and $N+1$ and demonstrates that due to the piecewise convexity the energy of the system with $N-1$ or $N+2$ electrons lies above it.
Therefore, energy lost from weighing out some of the contribution of the $N$ or $N+1$ states is larger than that gained by weighing in some new pure states.
Therefore, the two-member ensemble of Eq.\ (\ref{ens_operator_fractional}) is the only one required for attaining the ground-state energy at a given fractional number of electrons (although other useful ensembles with different particle numbers may be constructed\cite{Senjean2018,Senjean2020}).

Having established the ensemble that describes  a mixed state of the real system, creating the commensurate KS ensemble is relatively straightforward. It has to have the same density as the original system and therefore the same fractional number of electrons. Its ground state must therefore also be an ensemble state, given by 
\begin{equation}
\label{fractional_KS_ensemble}
\hat \Gamma^\omega_{s} = (1-\omega)\iout{\Phi_{N}^{(\omega)}} + \omega\iout{\Phi_{N+1}^{(\omega)}},
\end{equation}
where $\iket{\Phi_{N}^{(\omega)}}$ and $\iket{\Phi_{N+1}^{(\omega)}}$ are pure KS ground states with $N$ and $N+1$ electrons, respectively,~\cite{Gross90} arising from a common set of orbitals defined through one KS potential.\cite{Gould2013-LEXX,Kraisler2013,Kraisler2014,Goerling2015,Kraisler2024}
The superscript $(\omega)$ emphasizes that in the Kohn-Sham ensemble the pure states are generally $\omega$-dependent. The KS density is obtained by inserting  Eq.\ (\ref{fractional_KS_ensemble}) into Eq.\ (\ref{ensemble_density}). This leads to an ensemble density given by 
\begin{equation}
\label{fraction_ensemble_density}
n(\vr)= \sum_{i=1}^N |\phi_i^{(\omega)}(\vr)|^2
+ \omega |\phi_{N+1}^{(\omega)}(\vr)|^2,
\end{equation}
where $\phi_i^{(\omega)}(\vr)$ are KS orbitals. In other words, creating the fractional KS ensemble density boils down to partial occupation of the $(N+1)^{\rm th}$ KS orbital, with the appropriate fraction $\omega$.
Using exact EDFT must then also lead to  piecewise-linearity.

\begin{figure}
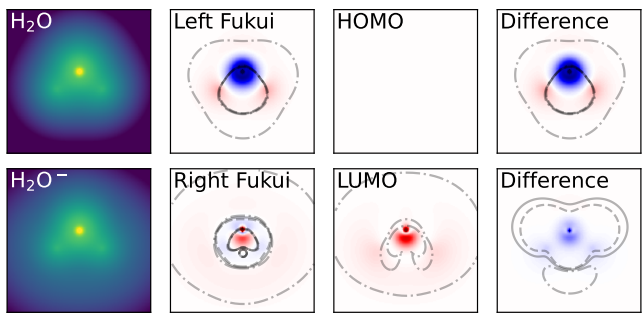

\includegraphics[width=\linewidth]{{{Water-Frac}}}
\caption{Left and right Fukui functions of water (on the bonding plane) compared against HOMO and LUMO densities, respectively. The Fukui functions were computed using CCSD densities and the KS results were obtained from \emph{exact} inversion of these densities.\cite{Gould2023-JCP} Contour lines indicate values of zero (solid) and $\pm 10^{-3}$~Bohr$^{-3}$ (dash-dot/dashes); HOMO is exactly zero on the plane.
In the left-most plots, colors range from navy (essentially no density) through blue to light green (maximal density) and are graded logarithmically.
\label{fig:Water}}
\end{figure}

Importantly, and as mentioned in Section \ref{eKSDFT} above, in general it is only the {\it total} ensemble density of the real system, $n\bf(r)$  of Eq.~\eqref{fraction_ensemble_density}, and not the individual pure-state densities $n_N{\bf(r)} \neq n_{s,N}{\bf(r)}$ and $n_{N+1}{\bf(r)}\neq n_{s,N+1}\bf(r)$, that the KS ensemble needs to reproduce. This is because the KS orbitals themselves, and therefore the individual pure-state densities, depend on $\omega$, per Eqs.~\eqref{fractional_KS_ensemble} and \eqref{fraction_ensemble_density}. 
One way of demonstrating this quirk\cite{Gould2023-JCP} involves the Fukui function, defined as the  derivative of the true electron density with respect to the fractional occupation $\omega$.~\cite{Geerlings2003}
The piecewise linearity of Eq.\ (\ref{density_fractional}) causes the derivative with respect to $\omega$ to be the same as the difference between the integer density. Therefore the exact right Fukui function is given by 
$f_{N_+}(\vr) = n_{N+1}(\vr)-n_{N}(\vr)$ and the exact left Fukui function is given by $f_{N_-}(\vr) = n_{N}(\vr)-n_{N-1}(\vr)$.\cite{Hellgren2012}

For the KS ensemble, Eq.\ (\ref{fraction_ensemble_density}) suggests that a derivative with respect to $\omega$ of the KS density yields the partially occupied orbital. Hence, and again because piecewise linearity equates the derivative with a difference, we find that the right and left Fukui functions are, respectively, \Response{}{$f_{s,N_+}(\vr) =
n^{(\omega)}_{s,{N+1}} (\vr)-
n^{(\omega)}_{s,{N}} (\vr)  =|\phi_l^{(\omega)}|^2$} and $f_{s,N_-}
(\vr) = n^{(\omega)}_{s,{N}} (\vr)-
n^{(\omega)}_{s,{N-1}} (\vr)=  
|\phi_h^{(\omega)}|^2$, where the subscripts $h$ and $l$ denote the highest occupied molecular orbital (HOMO) and lowest unoccupied molecular orbital (LUMO), respectively. Thus, if the KS pure-state densities are the same as the real pure-state densities, then the ensemble frontier orbitals must be equal to the Fukui functions. 
Figure~\ref{fig:Water} compares the exact ensemble Fukui functions and frontier orbitals in the atomic plane of the water molecule. 
Clearly, the right Fukui function bears some resemblance to the LUMO, though they are not identical. But the left Fukui function looks nothing like the HOMO in the plane, as the HOMO has a node in the atomic plane and the Fukui function does not. Clearly, then, the true and KS pure-state densities in the ensemble are not  the same.

Returning to piecewise linearity, Yang et al.\ were able to show that it can also be derived without explicitly invoking EDFT.\cite{Yang2000}
In their approach, a fraction of ``half an electron'' is introduced by creating two far-apart replicas of the same system and adding one electron to the conjoined system.
This extra electron can go to either of the two replicas, creating two possible different systems, each possessing one replica with $N$ electrons and one with $N+1$ electrons in its ground state.
These two systems are obviously degenerate and therefore any linear combination of their wave functions would also possess the same energy. In a balanced linear combination, each subsystem possesses an electron density that corresponds to $N+1/2$ electrons overall and the energy per subsystem is now the average of two energies.
A similar argument would hold for any rational fraction $\omega=M/R$ by adding $M$ electrons to $R$ replicas (and see Ref.\ \onlinecite{Ayers08} for extension to irrational electron numbers), establishing piecewise linearity.
In some sense, this pure-state approach is equivalent to the ensemble one because it avoids explicit use of ensemble states by constructing the ensemble ``manually'' via appropriate replicas.\cite{Kronik2020} 
However, avoiding replicas is clearly preferable for practical computational work and indeed the replica picture has been proposed as a {\it gedanken} calculation.

While piecewise linearity is a requirement for the exact functional, approximate functionals may exhibit a piecewise-convex or piecewise-concave dependence of the energy on the particle number, as illustrated in Fig.\ \ref{fig:PWL}.\cite{MoriSanchez2006,Cohen2008b,Stein2012,Srebro2012} 
To understand why this matters, consider again the above argument of an electron added to two replicas of the system. If the system is convex (concave), the calculation would produce a spurious prediction that it is more (less) energetically favorable to share the electron between the two systems, compared to placing it on one of the two systems, whereas in reality the two scenarios should have the same energy. This phenomenon is usually known as a delocalization (localization) error because, carried over to a general many-electron scenario, it can and often does result in spurious delocalization away from or localization on a specific segment of a system, often resulting in qualitative failures. Understanding and mitigating such errors have therefore attracted much attention as a tool for density functional assessment and/or development (see. e.g.,  \rcites{Cohen2008,Cohen2012,Kronik2012,Autschbach2014,Dabo2014,Kulik2015,Kirpatrick2021,Kronik2020,Linscott2023,Bryenton2023} for perspectives from various points of view) and the issue has even been recently described as ``the greatest outstanding challenge
in DFT''.\cite{Bryenton2023}

Piecewise linearity is important not only for localization/delocalization issues. It was pointed out very early on that capturing the slope discontinuity at integer charge densities is likely to require a derivative discontinuity (DD) in the KS potential, i.e., that it would ``jump'' by a spatial constant as the integer charge point is crossed.\cite{Perdew1982} This DD was then pointed out as the source of the difference between the KS gap and the true fundamental gap (i.e., the difference between ionization potential and electron affinity), even in exact KS theory.\cite{Perdew1983,Sham1983} Originally somewhat controversial, this idea has since been verified numerically by exact KS calculations in a wide range of scenarios,\cite{Godby1986,Chan1999,Allen2002} and minimization of the DD has emerged as a major driving force in GKS schemes that address fundamental gap calculations.\cite{Seidl1996,Onida2002,Kronik2012,Perdew2017,Kronik2020,Wing2021}
Similar issues even arise in many-body perturbation theory.\cite{Dauth2016} Moreover, the DD also manifests itself as a step-like feature in the Kohn-Sham potential across a highly stretched bond, which affects molecular dissociation processes and enforces an integer electron number in each of the dissociating atoms -- see, e.g., Refs.\ \onlinecite{Perdew1990,Teale2008,Karolewski2009,Tempel2009,Makmal2011,Hofmann2012,Hodgson2017,Kraisler2021,Rahat2025}.

Finally, we note that within appropriate circumstances piecewise linearity in DFT can be extended into a ``flat-plane'' condition in plots of the energy as a function of both the spin-up and spin-down electron densities in spin-polarized DFT.\cite{Cohen2008,Cohen2008c,Mori-Sanchez2009,Cohen2009,Capelle2010,Bajaj2017,Su2018,Su2018b,Bajaj2022,Prokopiou2022,Burgess2024,Goshen2024} This has also been pointed out repeatedly as a useful criterion for functional development and assessment, especially in the context of static correlation.

\subsection{Systems in an excited state}
\label{sec:Excited}

The third type of ensemble theory we consider in this Perspective is that concerning higher energy eigenstate (excited state) ensembles at a {\em fixed} electron number. Unlike the previously considered examples (namely, ensembles addressing degenerate ground states or ground states with fractional electron numbers), these type of ensembles  do  not seek  to address a physical state specifically. 
Rather, these ensembles are intended as auxiliary states, the main purpose of which is to access the energy of individual excited states via some extended energy density functionals.
The approach was first formulated by Theophilou\cite{Theophilou1979} (and even earlier used to justify excited states in the context of Slater X$\alpha$ theory~\cite{Trickey1973}, later made rigorous in Ref.~\onlinecite{Theophilou85}), but full generalization along with an enhanced variational flexibility were given and explored in series of  articles by  Gross, Oliveira, and Kohn.\cite{GOK-1,GOK-2,GOK-3} Therefore, the approach is often  denoted as GOK-EDFT, when  stressing the general type of ensembles which can be considered.

Properties of excited states are commonly addressed within density functional theory using time-dependent DFT (TDDFT),\cite{Runge1984-TDDFT,Onida2002,UllrichBook,Marques2012,Burke2012,Maitra2016,Byun2020} which is an in-principle all-purpose extension of DFT to time-dependent phenomena.
When employed in studies of optical excitations at equilibrium, TDDFT is almost invariably used in the linear-response (LR) regime\cite{Casida1995,Gross1996,Casida2012-Review,Tretiak2003} and {\em adiabatic} (A) approximation, which together provide a rather general applicability at substantially lower cost than other perturbation-theory many-body techniques.
LR-ATDDFT suffers from some known limitations, however.\cite{Neepa2004-Double,Maitra2017-CT,Davood2022,Lacombe2023-NA}

GOK-EDFT can avoid key limitations of TDDFT~\footnote{Using the popular short-hand of TDDFT for LR-ATDDFT} because: 1) it deals directly with stationary states without invoking time from the outset, thereby bypassing any issues with adiabatic assumptions; and 2) (as elaborated below) the auxiliary ensemble states can describe the targeted excitation structures unambiguously.
EDFT is ideally suited to situations in which one can focus on specific low-lying excited states, versus the large windows of excitation spectra available through TDDFT.
When used for this purpose it also offers computational advantages over TDDFT.~\cite{Gould2022-HL}
However, developing proper approximate extended density functionals requires additional insights, as elaborated in Section~\ref{sec:Ensemblisation} below.

To illustrate the ensembles used in GOK-DFT, we  begin with the simplest possible ensemble operator involving only the ground state, $\iket{\Psi_0}$,  and the first excited state,  $\iket{\Psi_1}$:
\begin{align}\label{ens_excited_states}
    \Gammah^w = (1-w)\iout{\Psi_0} + w\iout{\Psi_1}\;.
\end{align} 
For now we  assume that $\iket{\Psi_0}$ and $\iket{\Psi_1}$ are both non-degenerate.
The ensemble operator in Eq. \eqref{ens_excited_states} is similar in spirit to that of the fractional case, given in Eq.~\eqref{ens_operator_fractional}.
The states $\iket{\Psi_0}$ and $\iket{\Psi_1}$ are orthogonal,  i.e., $\ibraket{\Psi_0}{\Psi_1}=0$, but off-diagonal matrix elements of an operator $\hat{O}$, $\ibraketop{\Psi_0}{\hat{O}}{\Psi_1}$, can be non-zero.

One may define a trial ensemble operator, $\Gammah^w_{{\rm trial}} = (1-w)\iout{\Psi_{0,{\rm trial}}} +w\iout{\Psi_{1,{\rm trial}}}$ out of any two orthogonal wave functions.
A variational principle~\cite{GOK-1} for the weighted average energy can then be phrased, namely:
\begin{align}
    \E^w:=&\min_{\Gammah^w_{\rm trial}}
    \tr\big[ \Gammah^w_{\rm trial} \Hh \big]
    = (1-w)E_0 + w E_1
    \;,
    \label{eqn:GOK-twostate}
\end{align}
where $E_{\FE} = \ibraketop{\Psi_\FE}{\Hh}{\Psi_\FE}$ is the eigen-energy of eigen-state, $\iket{\Psi_\FE}$.
For the variation in Eq. \eqref{eqn:GOK-twostate} to yield the correct results, the weight, $w$, should not exceed $1/2$.
The bound on $w$ may be understood by writing,
\begin{align}
    \E^w=(1-2w)E_0 + w(E_0+E_1),
    \label{eqn:EwSumEqui}
\end{align}
where both $E_0$ (the lowest energy) and $E_0+E_1$ (the sum of the lowest two energies) must be bounded from below.\cite{Theophilou1979}
It follows that $\E^w$ is bounded from below provided $w$ and $1-2w$ are positive, i.e. $0\leq w\leq \tfrac12$.
\Response{}{By contrast, if $w$ exceeds $\half$ then the ensemble solution with the trial states swapped will yield a lower ensemble energy.}

In principle, minimization under the GOK conditions also yields the ground and excited state wave functions, $\iket{\Psi_{0,1}}$ (the eigen-states of the minimizing trial ensemble operator), and thereby their energies, $E_{0,1}$.
In practice, our intention is to bypass the interacting problem by using EDFT to evaluate the weighted average energy directly.
Importantly, taking the weight derivative of Eq.~\eqref{eqn:GOK-twostate} yields,
\begin{align}\label{edw}
    E_1 - E_0 =& \frac{\partial}{\partial w}\E^w
    = \frac{\partial}{\partial w}
    \min_{\Gammah^w_{\rm trial}}\tr\big[ \Gammah^w_{\rm trial} \Hh \big]\;,
\end{align}
in a similar way to finding the ionisation potential via differentials with respect to particle number (as per Figure~\ref{fig:PWL}).
This idea can be generalized to address specific excited states.\cite{Gould2024-Stationary,Fromager2024-Stationary,Dupuy2025}

The GOK theorems~\cite{GOK-2} ensure that the external potential in $\Hh$ is a functional of the ensemble density, $n^w=(1-w)n_0 + wn_1$, for any given $w \leq 1/2$, again analogously to the fractional case.
Next, one may assume non-interacting ensemble $v$-representability -- a condition that, as mentioned above, has been recognized to be less restrictive than pure-state non-interacting $v$-representabilty (although issues are not fully avoidable especially in lattice DFT, see, e.g., Section 4.3.3 of \rcite{Cernatic2022}).
Thus, one may define a corresponding KS ensemble via the ensemble operator,
\begin{align}
    \Gammah_s^w=&(1-w)\iout{\Phi_0^{(w)}} + w\iout{\Phi_1^{(w)}}\;,
\end{align}
with the same weights, but using eigenstates of a KS Hamiltonian, $\Hh_s=\Th+\vh_s$, involving \Response{}{an auxiliary} external potential, $v_s$, chosen to ensure that $\tr[\Gammah_s^w\nh]=n$. 
Here, $\iket{\Phi_{\FE}^{(w)}}$ are mutually orthogonal \emph{non-interacting} many-body states built from a common set of single-particles states (orbitals), the orthogonality of which derives from the orthogonality of the orbitals.
For the special case of $w=0$ and a non-degenerate state, we recover conventional KS DFT (but note that the case of $w=0^+$ is different \cite{Gould2022-HL}).

We may also consider more than two states: indeed, the variational principle for GOK-ensembles covers any countable set of bound states.\cite{GOK-1}
This leads to an expression for the ensemble energy,
\begin{align}
    \E^{\wv}=&\min_{\{\Psi\}}\tr\big[\Gammah_{\text{trial}}^{\wv}\Hh\big]
    =\tr\big[\Gammah^{\wv}\Hh\big]
    =\sum_{\FE}w_{\FE} E_{\FE}\;,
    \label{eqn:Ew}
\end{align}
and general (trial) ensemble operator,
\begin{align}
\Gamma^{\wv}_{\text{(trial)}}=&\sum_{\FE}w_{\FE}\iout{\Psi_{\text{(trial,)}\FE}}\;,
\end{align}
where $E_{\FE}=\ibraketop{\FE}{\Hh}{\FE}$ is the $\FE$th-lowest eigen-energy of $\Hh$. Probability dictates that the weights obey $w_{\FE}>0$ and $\sum_{\FE}w_{\FE}=1$.
Notationally, it is convenient to order the weights such the largest one is associated with the lowest energy, etc., i.e $w_{\FE}\leq w_{\FE'}$ when $E_{\FE}\geq E_{\FE'}$.
This notational choice reflects the fact that minimization leads to a pairing of the highest weight with the lowest energy, and so forth.
A side effect is that ``holes'' (i.e. zero-weight energy levels in-between positive-weight levels) are forbidden, which means that addressing a target state, $\iket{\bar{\FE}}$, requires forming a GOK-ensemble with \emph{all} energetically lower states.
We note that the above choice of weights is a sufficient but not a necessary condition, especially when using ensembles to motivate stationary solutions.\cite{Goerling1999,Gould2024-Stationary,Fromager2024-Stationary,Dupuy2025}
In Section~\ref{sec:AppOurs} below we discuss how to work around this requirement, especially by exploiting symmetry.

Consider, now, the most common complication -- degeneracies.
As discussed in Section~\ref{sec:Symmetries}, degenerate ground states require a more sophisticated treatment than the basic KS-DFT approach.
For excited states, this sophistication becomes even more important, because excited states nearly always exhibit degeneracies, for example (and most commonly) as a spin-triplet. 
They therefore inherit all the problems associated with a description of degenerate ground states.
The solution, as in the ground state, is to construct an ensemble where all degenerate excited states are weighted equally. 
For example, in the case that $\iket{\Psi_0}$ is unique, but $\iket{\Psi_{1,2,3}}$ are degenerate, we would write,
\begin{align}
    \Gammah^w_s=&(1-w)\iout{\Phi^{(w)}_0}
    + \frac{w}{3}\sum_{k=1}^3\iout{\Phi^{(w)}_k}\;,
\end{align}
which after minimization would yield $\E^w=(1-w)E_0 + wE_1$ because $E_1=E_2=E_3$.

Symmetries play a prominent role in
characterizing excitations, even if the ground state is not degenerate.
To address this, it is useful to work with \emph{equi-}ensembles,
\begin{align}
    \Gammah^{\wv}_s=&\sum_{\level}W_{\level}
    \Gammah_{\level}\;,
    &
    \E^{\wv}=&\sum_{\level}W_{\level}E_{\level}
    \label{eqn:Gammaw_level}
\end{align}
where $W_{\level}$ is the weight for the entire degenerate energy level and $\Gammah_{\level}=\tfrac{1}{N_{\level}}
\sum_{E_{\FE}=E_{\level}}\iout{\Phi_{\FE}}$ is an equal weighting of all $N_{\text{level}}$ degenerate states corresponding to the energy level.
Doing so provides a one-to-one mapping between the density and (KS) potential, while one may still (equivalently) unitarily transform the degenerate states in the degenerate subspace.\cite{GOK-1,Gould2020-SP}

Similarly, it is most natural to consider auxiliary ensembles which are formed by states that, individually, are 
non-interacting but also {\em symmetry-adapted} states.\cite{Ziegler1977,vonBarth1979,Daul1997,Frank1998,FILATOV1998,Filatov1999,Theophilou1998,Nagy1998,Pribram-Jones2014,Gould2017-Limits}
In this way, for example, we may be certain about the identity of the state we are addressing (e.g., assessing the energy of the lowest excited singlet of H$_2$ rather than the energy of its lowest excited unpolarized triplet). Symmetry-adapted {\em non-interacting}  many-body  states can be obtained by linearly combining a {\em finite} number of 
Slater determinants, which as mentioned above are known as configuration state functions (CSFs).~\footnote{
CSFs are best constructed starting  with symmetry-adapted {\em single-particle} states.
In EDFT, the latter can be obtained by using ensembles which, overall, are
totally symmetric (i.e., which remain unchanged under  symmetry operations of the
 system). The ensemble particle density is then totally symmetric and, thus,
the corresponding effective potential will have the symmetry of the actual external potential.
Symmetry adaptation also requires that    
spurious symmetry breaking in self-consistent calculations, due to strong correlation, is avoided.}
Practical examples of CSFs are given in the next section.
Naturally, there are situations in which one may find a  way  (via a set of properly selected calculations) to reduce the calculation to one involving single Slater determinants. In such cases, reuse of functional developments from regular DFT  becomes simpler. The next section considers the general case, including situations in which CSFs may be more effective or are the only consistent solutions.

Finally, we note that the `tracking' of KS states is particularly important and may be more complicated in excited state ensembles.
For the remainder of this article we therefore stick to excited state ensemble theory involving only a small number of easy-to-define excited states (because of symmetry and/or aufbau considerations) that can be trivially mapped to their ground state counterparts; and may thereby be understood using  conventional molecular orbital theory.
It should be kept in mind that higher-lying excited states can pose additional challenges.

\section{The road to ensemblization}
\label{sec:Ensemblisation}

As explained in previous sections, EDFT is more complex, both mathematically and in the physics it captures, than conventional DFT for non-degenerate ground states. It stands to reason, then, that developing approximations for EDFT would also be more complex.
In this section, we describe the process of ``ensemblization'' -- a term we use to describe the process of adapting exact functionals (e.g. $E_{\Hx}$, $E_{\xrm}$, $E_{\crm}$) from ground state to ensemble forms ($\EHx$, $\Ex$, $\Ec$) and the closely related process of adapting existing DFAs to ensemble DFAs (EDFAs).
The practical goal of ensemblization is to replicate the success of existing DFAs (in ground-state calculations and/or in adiabatic time-dependent DFT) with EDFAs that can be applied to novel problems (e.g., treatment of systems with fractional electron numbers or prediction of excited states from GOK-EDFT). Ensemblization is essential to EDFT because, as shown below, failure to carefully ensemblize can lead to major errors.

Key concepts in ensemblization arise from the fact that the ensembles considered above all follow  variational principles for the total energies, via appropriate extension of ground-state DFT. However, ensembles contain more states than ground states and therefore it is expected that the corresponding functional forms will contain extra components and that additional exact conditions will need to be phrased and satisfied by suitable approximations.
The flexibility associated with the additional degrees of freedom also suggests that more elaborate building blocks for approximations will need to be designed.

We now proceed to describe a series of steps, motivated by first principles arguments, and analyze the key exact features of the extended functionals that let us ensemblize DFAs in a systematic manner, ultimately concluding the section with a general protocol for addressing ensemblization.
We note that while the considerations below cover all ensembles considered in this article, below we focus mostly (but certainly not solely) on excited state ensembles, reflecting the fact that typical applications involving excited states bring about more complications than typical applications involving ground states or fractional electrons -- a situation also encountered in thermal DFT,\Response{}{\cite{Pittalis2011,Dufty2011,Hilleke2025,Ellaboudy2025}} where the free energy functional imposes \Response{}{peculiar} scaling considerations on key density functionals.
Moreover, \Response{}{formal} progress achieved in ensemblization pertinent to excited states is \Response{}{for the most part} more recent and not as well known.

\subsection{Step negative one: Naive adaptation}
\label{sec:naive}

In pure-state KS theory, a formal definition for the Hartree and exchange (Hx) energy is $E_{\Hx}=\ibraketop{\Phi_s}{\Wh}{\Phi_s}$.\cite{Perdew2003} Using the KS ensemble operator of Eq.\ \eqref{Gamma_ensemble_KS}, this definition is easily generalized to ensemble DFT as 
\begin{align}
\EHx[n] = \sum_{\FE}  w_\FE \ibraketop{\Phi_{s,\FE}}{\Wh}{\Phi_{s,\FE}}\;.
  \label{eqn:StatHx}
\end{align}
Historically, however, it has been customary to split H from xc (rather than Hx from c); and to define the ensemble Hartree energy as
\begin{align}
\EH^{\rm GI}[n] :=  E_{\Hrm}[n] = E_{\Hrm}[\sum_{\FE}w_{\FE}n_{s,\FE}]\;,
  \label{eqn:HGI}
\end{align}
i.e., to use the conventional form of the Hartree integral, $E_{\Hrm}[n]= \int \frac{n(\vr)n(\vrp) }{2|\vr-\vrp|} d\vr d\vrp$, applied to the {\em overall} ensemble density. 
The meaning of the superscript ``GI'' is clarified below.

\begin{figure}
    \includegraphics[width=\linewidth]{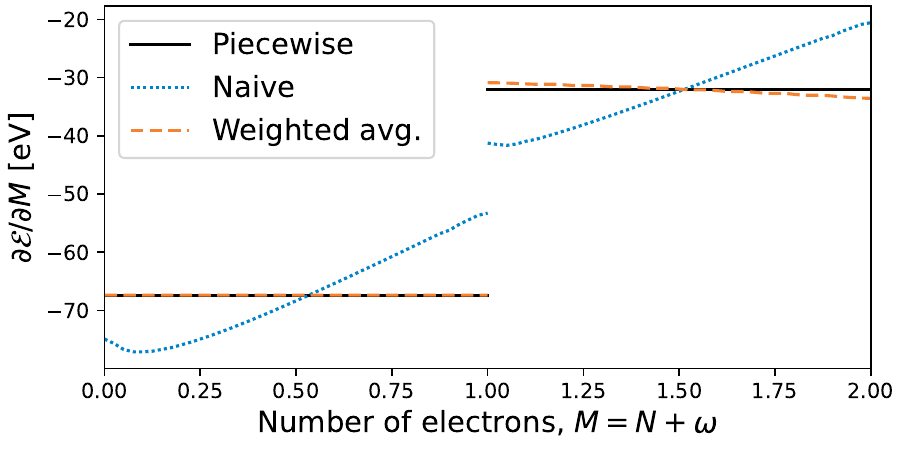}
    \caption{Energy derivative with respect to electron number, $M=N+\omega$,  for the H$_2$ molecule, computed using the LDA as adapted via naive [blue solid line, Eq.~\eqref{eqn:EDFA_trivial}] and via weighted average [orange dashed line, Eq.~\eqref{eqn:EDFA_wavg}] approaches.
    The ideal piecewise behaviour is shown as a black line.
    Data taken from Kraisler and Kronik, Ref.\ \onlinecite{Kraisler2013}, used with permission.
    \label{fig:H2Frac}}
\end{figure}

One can stick with the idea of employing the ground-state functional form with the overall density also for the xc energy (ignoring the  explicit dependence on the weights) to obtain a ``naive'' approximate ensemble-state total energy functional. For example, in this approach, the LDA-based total energy for a fractional-electron ensemble with $N+\omega$ electrons would be given by\cite{Kraisler2013}
\begin{align}
    \E_{\LDA}^{\text{naive}}[n^{N+\omega}]
    =&\Ts[n^{N+\omega}] + \int n^{N+\omega} v d\vr
    \nonumber\\&
    + E_{\Hrm}[n^{N+\omega}] + E_{\xc}^{\LDA}[n^{N+\omega}].
    \label{eqn:EDFA_trivial}
\end{align}
Unfortunately, this naive approach is insufficiently accurate.
Figure~\ref{fig:H2Frac} shows the derivative of the total energy with respect to electron number, $M=N+\omega$, as a function of the electron number, obtained for the hydrogen molecule using Eq.~\eqref{eqn:EDFA_trivial}.
As explained in Section \ref{sec:Fractional}, the exact energy curve is piecewise linear and its derivative is a stair-step function.
Figure~\ref{fig:H2Frac} shows that there is in fact a major problem with Eq. \eqref{eqn:EDFA_trivial}, as the result is quite far from being a stair-step.
Specifically, it is dominated by a linear term, which one can trace back to the quadratic dependence on the density in the Hartree energy.~\cite{Kraisler2013}

One could surmise that the above failure simply reflects shortcomings of the LDA. However, Eq.\ \eqref{eqn:StatHx} provides a first hint that much of the problem occurs already at the level of the Hartree energy. The Hartree energy as defined in  Eq.\ \eqref{eqn:HGI} is the {\em exact} semi-classical Coulomb energy for a pure state, but for ensemble states the Hartree energy expression should be modified to avoid spurious errors. Specifically, 
by expanding the electron density terms of the individual orbitals in the KS ensemble, one can readily see that the density of an individual orbital may interact with itself in the density-quadratic Hartree expression of Eq.\ \eqref{eqn:HGI} in two ways: (i) when the orbital
belongs to the same KS pure state, which is the well-known self-interaction error;\cite{Perdew1981}
(ii) when the same orbital is occupied in different KS pure states of the ensemble  -- this is an {\em additional} form of self-interaction, never encountered in conventional DFT.\cite{Gidopoulos2002-GI} We know that these latter interactions must be spurious because, by definition, the ensemble total energy is directly  expressed as a sum of the individual pure states. Hence, there cannot be any true cross-states. Pictorially, we can think of replicas of the same electron orbital in different states of the ensemble as ``ghosts'' to each other, in the sense that they do not interact. The spurious cross-terms are therefore referred to as ``ghost interactions''\cite{Gidopoulos2002-GI} and hence the superscript GI in Eq.\ \eqref{eqn:HGI}.

It is well-known that in standard DFT, self-interaction is canceled in full by using exact (Fock) exchange.\cite{Kummel08} Is this the case also for ensembles? To answer that, we recall that the exchange term corresponding to the Hartree expression of Eq.\ \eqref{eqn:HGI} may be defined as
\begin{align} 
\Ex^{\rm GI}[n] \ansatzeq
E^{\rm EXX}_{\xx}[ \sum_{\FE} w_{\FE}\rho_{s,\FE}(\vr,\vrp) ]\;
\label{eqn:ExGI}
\end{align}
where
$\rho_{s,\FE}(\vr,\vrp)$ are KS one-electron reduced density matrices.
$\Ex^{\rm GI}$ has the same  structure of $\EH^{\rm GI}$. As expected, pure-state self-interactions are indeed canceled in $\EHx^{\rm GI} = \EH^{\rm GI} + E^{\rm EXX}_{\xrm}$. However, expanding $\rho_{s,\FE}(\vr,\vrp)\rho_{s,\FE}(\vrp,\vr)$ in terms of single-particle orbitals, we find that terms involving different orbitals in different KS states do {\it not} reduce to simple products of particle densities. Therefore, ghost interactions remain! This is unfortunate, as it means that it is generally up to the more complicated and harder to approximate ensemble correlation functional to act as the ghostbuster. 

Further information on ghost interactions and various strategies to mitigate them can be found in \rcites{Gidopoulos2002-GI,Pastorczak2014-GI,Marut2020,Loos2020-EDFA,Tasnadi2003,Tasnadi2003a}.
In particular we point out the work of 
Loos and Fromager,\cite{Loos2020-EDFA} who proposed a weight-dependent LDA 
as a means of addressing GI directly while using a conventional density functional.
Here, having learned from the naive approach that ghost interactions can be a serious issue, the next sections discuss an approach that avoids these errors by letting the auxiliary states and functionals go beyond the typical DFT reference states.

Finally, while thermal DFT is not within the scope of this article, we still stress that it provides one case where the ghosts bust themselves -- the Hartree-Fock (i.e. Hartree-exchange) energy of Mermin's~\cite{Mermin1965} extension of DFT to grand canonical ensembles.
In such ensembles the Hartree-exchange energy defined by Eq.~\eqref{eqn:StatHx} is equal to the sum of Eqs~\eqref{eqn:HGI} and \eqref{eqn:ExGI}.
We refer the interested reader to Ref.~\onlinecite{Greiner2010}.

\subsection{Step one: explicit state averaging}

Now that we have gathered evidence against using an overly traditional-looking ansatz for the Hartree and exchange functionals,
let us return to their {\em joint} expression, Eq.~\eqref{eqn:StatHx}.
Invoking KS states in the form of single Slater determinants, H splits naturally  from x as follows\cite{Nagy2001}
\begin{align}
  \EH^{{\rm wavg}}[n] \ansatzeq &
  \sum_\FE w_\FE E_{\HH}[n_{s,\FE}]\;,
  \label{eqn:EH_wavg}
  \\
  \Ex^{{\rm wavg}}[n] \ansatzeq &
  \sum_\FE w_\FE E_{\xx}[\rho_{s,\FE}]\;.
  \label{eqn:Ex_wavg}
\end{align}

Let us reconsider the H$_2$ example we started with for fractional electron numbers. It can be shown that {\em ensemble} Hartree and  {\em ensemble} exchange functionals are piecewise-linear functions of $\omega$,\cite{Gould2013-LEXX,Kraisler2013,Kraisler2014,Goerling2015} up to (typically small) higher-order orbital effects. This feature is completely missed in the naive model of Eq.~\eqref{eqn:EDFA_trivial}.
Thus, a {\em second attempt} at ensemblization must involve taking a proper weighted average (wavg) of the $N$ and ($N+1$)-electron DFAs like in Eq. \eqref{eqn:EH_wavg} and Eq. \eqref{eqn:Ex_wavg}. Assuming we may approximate the treatment of correlation in a similar manner, we can express the ensemble energy as
\begin{align}
    \E_{\LDA}^{\text{wavg}}[n_s^{N+\omega}]
    =&\Ts[n_s^{N+\omega}] + \int n_s^{N+\omega} v d\vr
    \nonumber\\&
    + (1-\omega)\{ E_{\Hrm}[n_s^N] + E_{\xc}^{\LDA}[n_s^N] \}
    \nonumber\\&
    + \omega\{ E_{\Hrm}[n_s^{N+1}] + E_{\xc}^{\LDA}[n_s^{N+1}] \}
    \;,
    \label{eqn:EDFA_wavg}
\end{align}
where $n_s^N$ and $n_s^{N+1}$ are evaluated using the self-consistent KS orbitals of the $N+\omega$ system.
As can be seen in Figure~\ref{fig:H2Frac} for the H$_2$ example, this relatively simple step already remedies most shortcomings of Eq.~\eqref{eqn:EDFA_trivial}, and vindicates the LDA. For a recent systematic evaluation of this step for ionization potentials and fundamental gaps of atoms across the periodic table, see \rcite{Lavie2023}. 

While this second attempt at ensemblization greatly improves the treatment of the above fractional electron problem, we show below that it still does not offer a satisfactory treatment, even qualitatively, of degenerate ensembles. This is of crucial importance as in particular excited state ensembles often exhibit degeneracies. Therefore, additional ensemblization steps are called for.

\subsection{Step two: A unified derivation of Hartree-exchange energies}
\label{sec:Hx}

\subsubsection{The non-uniqueness disaster}
\label{subsec:disaster}

To understand why further improvements are needed, consider a GOK KS ensemble of Be, comprising the ground state, the first excited singlet state, and the first excited triplet state. 
Aiming for a description of these states in terms of an ensemble, we invoke a KS potential that, just like the actual external potential of Be, is spin-unpolarized and spherically symmetric.
The aforementioned pure states can then be constructed from the following Slater determinants (SDs):
$\iket{0_{\SD}}=\iket{1s^22s^2}$, $\iket{1_{\SD}}=\iket{1s^22s^{\up}2p_z^{\up}}$, $\iket{2_{\SD}}=\iket{1s^22s^{\down}2p_z^{\down}}$,
$\iket{3_{\SD}}=\iket{1s^22s^{\down}2p_z^{\up}}$, $\iket{4_{\SD}}=\iket{1s^22s^{\up}2p_z^{\down}}$
(for simplicity we ignore the $p_x$ and $p_y$ counterparts in our discussion).
An ensemble accounting for these states is therefore
\begin{align}
    \Gammah^{w_4}_{s}=&
    w_0\iout{0_{\SD}} + w_1\iout{1_{\SD}} + w_2\iout{2_{\SD}}
    \nonumber\\&
    + w_3\iout{3_{\SD}} + w_4\iout{4_{\SD}}
\end{align}
where $w_4=1-w_0-w_1-w_2-w_3$. 
Due to the different spatial symmetry of the ground state, we can set $w_0=0$ without issue
to obtain an ensemble comprised entirely of excited states.
Motivated by an anticipated ``triplet''-like symmetry, one can further choose fractional equi-occupation of the first three Slater determinants, namely, choose $w_4\equiv \omega$ and $\tfrac{1-\omega}{3} \equiv w_1=w_2=w_3$. Then, we obtain the SD ensemble,
\begin{align}
    \Gammah^{\omega}_{s,\text{SD}}=&\frac{1-\omega}{3} \sum_{\FE=1}^3 \iout{\FE_{\SD}}
    + \omega \iout{4_{\SD}},
    \;.\label{eq:GSD}
\end{align}
Note that all pure states comprising the above ensemble possess the same density, $n=2n_{1s} + n_{2s} + n_{2p_z}$.

The above SDs, however, are not the only solutions of the Kohn-Sham equations that possess the same density as the states in Eq.~\eqref{eq:GSD}.
The ``configuration state functions'' (CSFs), $\iket{1_{\CSF}}=\iket{1_{\SD}}$, $\iket{2_{\CSF}}=\iket{2_{\SD}}$, $\iket{3_{\CSF}}=\tfrac{1}{\sqrt{2}}[\iket{3_{\SD}}+\iket{4_{\SD}}]$, and $\iket{4_{\CSF}}=\tfrac{1}{\sqrt{2}}[\iket{3_{\SD}}-\iket{4_{\SD}}]$ are equally valid. Assigning the same weights as before to the CSFs leads to a \emph{different}, CSF ensemble:
\begin{align}
  \Gammah^{\omega}_{s,\text{CSF}}=&\frac{1-\omega}{3}\sum_{\FE=1}^3\iout{\FE_{\text{CSF}}}
    + \omega\iout{4_{\text{CSF}}}\; \label{eq:GCSF}
\end{align}
The SD and CSF ensembles of Eqs.\ (\ref{eq:GSD}) and (\ref{eq:GCSF}) are composed from the same set of weights and same elementary determinants, and both yield the same density.
It is also straightforward to show that both yield the same kinetic energy, $T_s=\tr[\Gammah\Th]=2t_{1s}+t_{2s}+t_{2p_z}$, where $t_k\equiv\ibraketop{\phi_k}{\th}{\phi_k}$ is the kinetic energy of orbital $k$.

\begin{figure}
\includegraphics[width=\linewidth]{{{FigSD_CSF}}}
\caption{
GOK-ensemble excitation energy (defined as $\tr[\Gammah_{s}^{\omega}\Hh]-E_{0,s}$) of Be computed using interacting ($\iket{\FE_{\text{exact}}}$, solid black), and non-interacting SD ($\iket{\FE_{s,\SD}}$, red dash-dot) and CSF $(\iket{\FE_{s,\CSF}}$, navy dashes) wavefunctions, as a function of $\omega$.
The SD ensemble [Eq.\ \eqref{eq:GSD}] mixes non-interacting states which are not necessarily spin eigenstates like the interacting states are. In contrast, the CSF ensemble [Eq.\ \eqref{eq:GCSF}] uses 
valid non-interacting spin eigenstates.
Energy values are taken directly from Yang~\emph{et al}~\cite{Yang2017-EDFT} or employed indirectly by using $\ibraketop{3_{\SD}}{\Wh}{3_{\SD}}=\ibraketop{4_{\SD}}{\Wh}{4_{\SD}}=\half(\ibraketop{3_{\CSF}}{\Wh}{3_{\CSF}}+\ibraketop{4_{\CSF}}{\Wh}{4_{\CSF}})$.
\label{fig:BeLevels}
}
\end{figure}

The situation becomes more complicated, however, for the electron interaction energies, because $W_{s,\SD}=\tr[\Gammah_{s,\SD}^{\omega}\Wh]$ and $W_{s,\CSF}=\tr[\Gammah_{s,\CSF}^{\omega}\Wh]$ are \emph{not} the same in general.
To demonstrate this, Figure~\ref{fig:BeLevels} shows the excitation energy, $\tr[\Gammah_{s}^{\omega}\Hh]-E_{0}$ (or $\tr[\Gammah_{s}^{\omega}\Hh]-E_{0,s}$ for KS states), obtained using the non-interacting ensemble energy, $E_s=\tr[\Gammah_s(\Th+\vh+\Wh)]=T_s + \int n_sv d\vr + W_s$ for the two options considered above, as well as for the interacting ensemble,
\begin{align}
    \Gammah^{\omega}=&\frac{1-\omega}{3}\sum_{\FE=1}^3 \iout{\FE_{\text{exact}}}
    + \omega\iout{4_{\text{exact}}}\;,
    \label{eq:GInt}
\end{align}
formed using the exact lowest excited singlet and triplet interacting eigenstates, $\iket{\FE_{\text{exact}}}$, of Be.
Here $E_{0}\equiv\tr[\Gammah\Hh]=\ibraketop{0_{\text{exact}}}{\Hh}{0_{\text{exact}}}$
($E_{0,s}=\ibraketop{0_s}{\Hh}{0_s}$ for KS).
Both KS ensembles, $\Gammah^{\omega}_{s,\text{SD}}$ and $\Gammah^{\omega}_{s,\text{CSF}}$, yield the same energy when $\omega=\tfrac14$ so that $w_1=w_2=w_3=w_4$.
But they are different for other values of $\omega$, an issue known as the ``non-uniqueness disaster''.\cite{Gould2017-Limits}
In view of the fact that energies are extracted by taking variations w.r.t. weights [see Eq. \eqref{edw} above], Figure~\ref{fig:BeLevels} provides a serious warning as to the fact that spin symmetries cannot be ignored in forming the KS ensembles.

In fact, non-uniqueness is even more complicated. In Eq.~\eqref{eq:GSD} the four SD states were assigned an essentially arbitrary order.
We could just as easily have set (e.g.) $\iket{1_{\SD}'}\equiv\iket{1s^22s^{\up}2p_z^{\down}}$ and $\iket{4_{\SD}'}\equiv\iket{1s^22s^{\up}2p_z^{\up}}$ and this alternative choice would lead to yet another energy curve.
By contrast, the CSFs were assigned their order by mirroring the symmetry properties of the exact eigenstates -- which also helps explain why their energy more closely follows the $\omega$-dependence of the exact curve.

The non-uniqueness disaster is a major problem for ensemble density functional theory because a functional should assign a unique output for a given density, but the SD and CSF options give different answers!
CSFs retain the appropriate spin-physics, however, which motivated their effective use in the symmetry eigenstate Hartree-exchange approach of Yang et al.\cite{Pribram-Jones2014,Yang2014,Yang2017-EDFT}

\subsubsection{Resolution and implications}
\label{sec:HxResolution}

Is there a way to {\em derive} a problem-free Hx energy functional from first principles? If so, the same procedure should also specify the KS states which are required in its calculation. 
Let us first consider the usual ground state case, where the KS wavefunction, $\iket{\Phi_s}$, is a SD.
In that case $E_{\Hx}=\ibraketop{\Phi_s}{\Wh}{\Phi_s}$.\cite{Perdew2003}
Next, consider $F^{\lambda}$ [eq.~\eqref{eqn:AC_KS}] in the limit of small $\lambda$.
From the Hellmann-Feynman theorem we obtain, $F^{\lambda}\approx \ibraketop{\Phi_s}{\Th+\lambda\Wh}{\Phi_s}$, where changes to the wavefunction can be ignored due to orthogonality.
Thus, $F^{\lambda\to 0}=\ibraketop{\Phi_s}{\Th}{\Phi_s} + \lambda\ibraketop{\Phi_s}{\Wh}{\Phi_s}=T_s + \lambda E_{\Hx}$,
it follows that $E_{\Hx}=\Dp{\lambda}F^{\lambda}|_{\lambda=0}$, an expression that does not resort to the KS state explicitly and therefore does not rely on its uniqueness.
One may therefore consider the following definition~\cite{Gould2017-Limits}:
\begin{align}
\EHx^{\wv}[n]:=\lim_{\lambda\to 0^+}\frac{\F^{\wv,\lambda}[n]-\F^{\wv,0}[n]}{\lambda}\;,
\label{eqn:HxLim}
\end{align}
as one that would be an appropriate generalized definition for ensembles.
This definition is appealing because it only involves $\F^{\wv,\lambda}[n]$, the {\em fundamental} functional of EDFT, which according to  Eq.~\eqref{eqn:Gammalambda}, is free from any spurious interaction and assigns one and the same value to proper ensemble densities.

The crucial element of Eq.~\eqref{eqn:HxLim} is that the case $\lambda=0^+$ allows fewer degeneracies than $\lambda=0$, because it has both one- (from $\Th$) and two-body (from $\Wh$) interactions.
The case of $\lambda=0$ allows the canonical Slater determinant solutions.
But, for $\lambda = 0^+$, degenerate perturbation theory guides us to choose a ``good'' basis: i.e., a basis in which the perturbation ($\Wh$) is block diagonal too.
A full mathematical proof and a more detailed discussion are provided in \rcite{Gould2017-Limits}.

Eq.~\eqref{eqn:HxLim}, in analogy to the case for  non-degenerate ground states, defines the ensemble Hx as a leading order approximation for $\EHxc=\F^1-\F^0$ in the coupling strength.
This expression only involves well-defined functionals -- $\F^{\wv,\lambda}[n]$ is uniquely defined for each $n$, $\lambda$ and $\wv$.
Therefore, the functional resulting from  Eq.~\eqref{eqn:HxLim} must be uniquely defined.
The variational principle and degenerate perturbation theory eventually yield,
\begin{align}
\EHx^{\wv}[n]=&\sum_{\FE}w_{\FE}\ibraketop{\FE_s}{\Wh}{\FE_s}\;,
\label{eqn:EHx}
\end{align}
where $\iket{\FE_s}$ may (and in all cases known to the authors do) acquire the form of CSFs.
This lets us obtain states, $\iket{\FE_s}$, that can be single KS Slater determinants, $\iket{\FE_s}\equiv\iket{\Phi_{\FE}}$, or unitary combinations thereof, $\iket{\FE_s}\equiv \sum_{\FE'}U_{\FE\FE'}\iket{\Phi_{\FE'}}$, i.e., the CSFs mentioned above, directly from a fundamental definition. 
In practice, these may be obtained by applying degenerate perturbation theory to $\Th$ using $\Wh$ as the perturbation.\cite{Gould2017-Limits}

Eq.~\eqref{eqn:EHx} avoids spurious interactions {\em and} uniquely assigns energies to densities.~\cite{Gould2017-Limits,Gould2020-SP}
By construction, it reduces to the usual KS DFT description of Hx for non-degenerate ground states.
Less trivially, it reproduces all successful ensemble forms proposed earlier for special cases.\cite{Theophilou1998,Nagy1998,Gidopoulos2002-GI,Yang2017-EDFT}
Eq.~\eqref{eqn:EHx} is also consistent with expressions for the joint Hartree-exchange energy of excited states which are derivable from state-specific (i.e., non-ensemble) symmetry-adapted DFT-like approaches -- see for example \rcite{Goerling2000}.
We also note that the analogue of Eq.~\eqref{eqn:HxLim}, for the case of fractional ensembles, yields the weighted average assumed in 
Eq.~\eqref{eqn:EDFA_wavg} (see also \rcites{Gould2013-LEXX,Kraisler2013,Kraisler2014}). More precisely, the form of Eq.~\eqref{eqn:EDFA_wavg} is exact for exchange -- the only approximation is replacing $E_{\xrm}$ by a DFA.

Despite the above considerations, given that single-Slater-determinant based DFT has worked remarkably well even with spurious symmetry breaking in ground states, one may be tempted to think that perhaps CSFs can be avoided in general.
That this is {\em not} the case is readily shown by considering the elementary CSFs [i.e., ingredients for Eq.~\eqref{eqn:EHx}] for the triplet states 
\begin{subequations}
\begin{align}
    &\iket{1s^{\up}2s^{\up}}\;,
    &
    &\tfrac{1}{\sqrt{2}}[\iket{1s^{\up}2s^{\down}}+\iket{1s^{\down}2s^{\up}}]\;,
    &
    &\iket{1s^{\down}2s^{\down}}\;,
\intertext{and their singlet counterpart}
&&
    &\tfrac{1}{\sqrt{2}}
    [\iket{1s^{\up}2s^{\down}}-\iket{1s^{\down}2s^{\up}}]\;
    \label{eqn:CSFsx}
\end{align}
\label{eqn:CSF}%
\end{subequations}%
for a two electron system, which, for consistency with the previous example, can be considered to be Be$^{2+}$.
Two of the states comprising the triplet ($\up\up$ and $\down\down$) are pure SDs. They also belong to the triplet which, among all the possible triplets, is lowest in energy.
So these SDs can be dealt with using the usual (ground-state-like) treatment.
This means that in principle we don't need to consider the third triplet as either of the SDs is sufficient to estimate the energy of the triplet.
However, the singlet does not have a SD counterpart.
Even breaking spin and spatial symmetries within a single SD (to yield a state $\iket{A^{\up}B^{\down}}$ with $\phi_B$ different to $\phi_A$)
cannot fix the problem, because the contribution of the Hxc potential, $v_{\Hxc\sigma}$, that can break symmetry is small compared to the spherically symmetric and spin-independent nuclear potential.
Thus, $\phi_{A^{\up}}\approx\phi_{1s^{\up}}\approx \phi_{1s^{\down}}$ and $\phi_{B^{\down}}\approx \phi_{2s^{\down}}\approx \phi_{2s^{\up}}$.
 and it follows that we cannot mimic the superposition physics of the singlet CSF using a single SD.
 In particular, a variational collapse to the singlet-SD ground state would be unavoidable without an extra constraint to exclude it.
The resulting singlet-triplet description may thus hardly be expected to be balanced  at all.

One way to set up a balanced description based only on SDs is to average the first excited triplet and the first excited singlet using equal weights for each of the four states - see, e.g., Refs.\ \onlinecite{GOK-3,Nagy2001,Gould2020-SP}.
It is readily shown that this is equivalent to averaging  the four SDs spanning the aforementioned four excited states equally.
Therefore, the corresponding Hx-description  yields a {\em vanishing} singlet-triplet splitting. This is acceptable as long as the
correlation functional can retrieve the {\em full} spin splitting, if needed.
Implicitly, however, this task requires to follow adiabatically the different states from the weakly interacting limit up to the full interacting limit.~\footnote{ Of course,  this point applies to more general spin states than singlet and triplet.}
But an ``incoherent'' mixing of SDs does not facilitate this directly.
\Response{}{We therefore see that one should embrace CSFs and weight them equally when they should be degenerate (to reflect the physics of the interacting system).
Still, it is important to allow for flexibility in the weights of non-degenerate states, to ensure that the structure of the excitation spectrum is captured by the non-interacting ensemble.}

We briefly also consider a more complicated (yet common) example of CSFs: the two-fold degenerate singlet found after double excitation to a degenerate orbital.~\cite{Gould2021-DoubleX}
The excited KS states required by Eq.~\eqref{eqn:EHx} are,
\begin{align}
    &\tfrac{1}{\sqrt{2}}[\iket{\cdots l_1^{\up}l_2^{\down}}
    - \iket{\cdots l_1^{\down}l_2^{\up}}]\;,    
    &
    &\tfrac{1}{\sqrt{2}}[\iket{\cdots l_1^2}
    - \iket{\cdots l_2^{2}}]\;,
    \label{eqn:CSFDouble}
\end{align}
which highlights that CSFs are also the natural KS states for the degenerate equi-ensembles described in Section~\ref{sec:Symmetries}.
Here $\cdots \equiv 1^2\cdots (h-1)^2$ indicates double occupation up to (but not including) the highest occupied molecular orbital (HOMO, $h$);
$l_1$ and $l_2$ label the degenerate lowest unoccupied molecular orbitals (LUMO), where we use symmetry to distinguish the HOMO from the LUMO(s).

Concluding, we stress that the CSFs of the KS system in EDFT do not requires us to step outside the realm of a single-particle approaches.
Rather, despite staying within a single-particle approach, they allow us to comply  with the fundamental symmetry of the problem at hand.
Functional approximations based on CSFs solve the problem of non-interacting excited states by avoiding spurious self- or ghost-interactions and non-uniqueness disasters.
They also allow us to gain an equal-footing  description of different spin states, without averaging to zero their (important) energy splittings at the level of the Hx energies. These CSFs come in handy also for computing 
a single excited state directly via EDFT (some examples are included in Table~\ref{tab:Hxc}). Thus, Eqs~\eqref{eqn:HxLim} and \eqref{eqn:EHx} are not only mathematically rigorous definitions of Hartree-exchange, but also reveal that in general the natural individual auxiliary states for ensembles are non-interacting yet symmetry-adapted pure state: i.e., CSFs.

\subsection{Step three: Separating Hartree and exchange components}
\label{sec:Handx}

\subsubsection{Novel H and x terms}

\begin{figure}
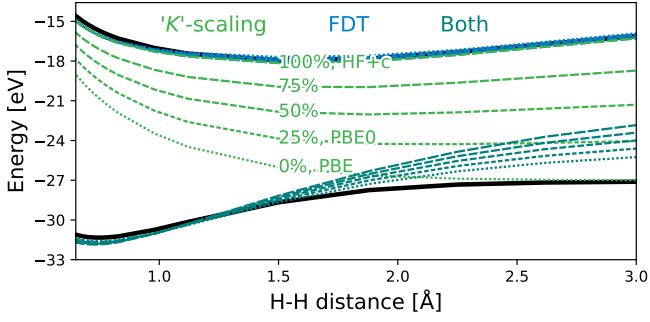

\includegraphics[width=\linewidth]{{{FigFDT}}}
\caption{Ground state and first excited singlet energy of H$_2$ at different bond lengths, for different ratios of PBE and Hartree-Fock exchange (indicated by dash lengths).
`$K$'-scaling (blue lines) means that we scale all `exchange-like' integrals by the fraction of exact exchange, $\alpha$.
FDT (red lines) means that we scale only the terms in Eq.~\eqref{eqn:EExK}. Ground state energies (green lines) are the same in both approaches.
Reference energy (black lines) results are from
full configuration-interaction (FCI). All results were obtained using a minimal basis set (def2-msvp).
\label{fig:FDT}}
\end{figure}

We motivate the need for yet further advancements for carrying out correct ensemblization steps by starting from a seemingly very simple example: 
H$_2$ treated via \emph{ensemble} Hartree-Fock (EHF) theory.
Regular HF theory may be thought of as calculating the energy of the Slater-determinant wavefunction instead of the interacting one, i.e. $E_{\text{HF}}:=\ibraketop{\Phi_s}{\Hh}{\Phi_s}$.
This may be generalized to ensembles as $\E_{\text{EHF}}:= \tr[\Hh\Gammah_s]$, where $\Gammah_s\equiv \Gammah^{0^+}[n]$ is the ensemble of KS wave functions.
The results of the previous sections reveal that $\E_{\text{EHF}}\equiv \Ts+\int n vd\vr + \EHx$.
In practical terms, at least for the cases considered here, the resulting approach boils down to (an ensemble of) restricted-open-shell HF theory.

It is well known that the ground state of H$_2$ involves a doubly occupied `gerade' ($g$) molecular orbital, i.e., $\iket{S_0} =  \iket{g^{\up}g^{\down}}$.
The lowest (triplet and singlet) excited states of H$_2$ have the same general form as Eq.~\eqref{eqn:CSF}, but with $1s\to g$ and $2s\to u$ (where $u$ indicates the ungerade molecular orbital), namely, $\iket{T_1}$ is spanned by \{$\iket{g^{\up}u^{\up}}$, $\tfrac{1}{\sqrt{2}}[\iket{g^{\up}u^{\down}}+\iket{g^{\down}u^{\up}}]$, $\iket{g^{\down}u^{\down}}$\};
and $\iket{S_1} =  \tfrac{1}{\sqrt{2}}[\iket{g^{\up}u^{\down}}-\iket{g^{\down}u^{\up}}]$.
We work with restricted orbitals, in which the spatial part is separable from the spin part -- for the triplet this requires taking the unpolarized state $\iket{T_1^0}$ only or, mathematically equivalently, an average over all three states. We shall continue to denote this by $\iket{T_1}$ for simplicity.

Using the above-defined expression, $\E_{\text{EHF}}=\Tr[\Hh\Gammah_s]$, lets us define EHF energies for the individual states, by defining $\Gammah^{S_0}:=\iout{S_0}$, $\Gammah^{T_1}:=\iout{T_1}$ and $\Gammah^{S_1}:=\iout{S_1}$.
Eq.\ \eqref{eqn:EHx} then leads to,
\begin{align}
    E_{\text{EHF}}^{S_0}:=&2 t_g + \int 2n_g v d\vr + 4J_{gg} - 2K_{gg}\;,
\label{eqn:H2_1S0}
    \\
    E_{\text{EHF}}^{T_1}:=&t_g + t_u + \int (n_g+n_u) v d\vr
    + J_{gg} + 2J_{gu} + J_{uu}
    \nonumber\\&
    - K_{gg} - 2K_{gu} - K_{uu}\;,
\label{eqn:H2_3T1}
    \\
    E_{\text{EHF}}^{S_1}:=&t_g + t_u + \int (n_g+n_u) v d\vr
    + J_{gg} + 2J_{gu} + J_{uu}
    \nonumber\\&
    - K_{gg} + 2K_{gu} - K_{uu}\;,
\label{eqn:H2_1S1}
\end{align}
as the ensemble Hartree-Fock expressions for the three states in question.
In the above, $t_k$ indicates the kinetic energy associated with orbital $\phi_k$;
the Hartree-exchange energy expressions involves Hartree-like integrals, 
$J_{ij}=\half\int n_i(\vr) n_j(\vr) \tfrac{d\vr d\vrp}{|\vr-\vrp|}$,
and exchange-like integrals,
$K_{ij}=\half\int \rho_i(\vr,\vrp)\rho_j(\vrp,\vr) \tfrac{d\vr d\vrp}{|\vr-\vrp|}$,
where $\rho_i(\vr,\vrp)=\phi_i^*(\vr)\phi_i(\vrp)$ and $n_i(\vr)=\rho_i(\vr,\vr)$. 
We note that in the above expressions the `self-Hartree' terms, $J_{gg}$ and $J_{uu}$, are exactly equal to the `self-exchange' terms, $K_{gg}$ and $K_{uu}$, respectively. We nonetheless retain them explicitly, in order to follow the typical division between Hartree and exchange terms in standard DFT calculations.

Let us now go beyond Hartree-Fock theory and mix in a fraction $\alpha$ of exact exchange with a complementary fraction $1-\alpha$ of a density functional approximation (DFA) for exchange, to form what is known as a hybrid density functional.\cite{Kummel08} For $S_0$ and $T_1$ this proceeds readily -- we simply scale all Fock exchange (`$K$') energy terms in Eqs.\ \eqref{eqn:H2_1S0} and \eqref{eqn:H2_3T1} by $\alpha$ and then add $(1-\alpha)E_{\xrm}^{\DFA}$ from a DFA of our choice. The case of $S_1$ is less obvious because unlike $S_0$ and $T_1$, it cannot be represented by any single SD with a well-defined Fock exchange term. Nonetheless, the most straightforward extension of the ground- and triplet-state results is to use  the same `$K$'-scaling argument for $S_1$, to obtain
\begin{align}
    &E_{\text{$K$-scaling}}^{S_1}:=t_g + t_u + \int (n_g+n_u) v d\vr
    \nonumber\\&
    + J_{gg} + 2J_{gu} + J_{uu}
    + \alpha (-K_{gg} + 2  K_{gu} - K_{uu})
    \nonumber\\&
    + (1-\alpha)E_{\xrm}^{\text{DFA}}
    + E_{\crm}^{\text{DFA}}\;.
    \label{eqn:H2_1S1_hyb}
\end{align}
as a hybrid version of Eq.~\eqref{eqn:H2_1S1}. Here, the DFA terms for x and c are computed using the triplet 1RDM as it: a) has the same density as the excited singlet and; b) has well-developed approximations for it.

The results obtained from the above scheme [i.e., the usual ground state functional for $S_0$ and Eq.~\eqref{eqn:H2_1S1_hyb} for $S_1$], using the Perdew-Burke-Ernzerhof (PBE) functional\cite{DFA:pbepbe} as the DFA,  are shown in Figure~\ref{fig:FDT} as green ($S_0$) and blue ($S_1$) lines, for various values of $\alpha$. They are additionally compared to reference full configuration-interaction (FCI) calculations. 
First, it is clear that these DFAs do not do a good job of describing the H$_2$ ground state, $S_0$, in the dissociation limit; and that the quality of the job is sensitive to the fraction of exact exchange.
This failure is well known and is a manifestation of the static correlation error, considered to be a major shortcoming of conventional DFAs, including hybrid functionals, with dissociating H$_2$ being a paradigmatic example.~\cite{Cohen2012}
More interestingly, it is equally clear that the variation with $\alpha$ of the singlet excited state, $S_1$, energy is \emph{much greater} -- by a factor of about five -- than the static correlation error. Specifically, the predicted excited state energy ranges from being surprisingly accurate for Hartree-Fock + PBE correlation (HF+PBEc, $\alpha=1$) to downright terrible for the parent PBE functional ($\alpha=0$).

Why is the predicted singlet energy so variable? Naively, one could simply interpret this behavior as yet another quirk of approximate density functionals. A significant hint that this is not so, however, is that the triplet, $T_1$, energy of H$_2$ (not shown) does not suffer from static correlation errors and is found to be consistently well-described, largely independently of our choice of $\alpha$.
$S_1$ does not suffer from severe static correlation errors either, otherwise HF+PBEc would actually have produced the worst result, as it does for $S_0$. Why is $S_1$ so poorly described, then? The answer can be found by considering the ``singlet-triplet energy gap'', $E_{ST}:=E_{S_1} - E_{T_1}$, which in Hartree-Fock theory is easily obtained from Eqs.\ \eqref{eqn:H2_3T1} and \eqref{eqn:H2_1S1} as $E_{ST}^{\text{HF}}=4K_{gu}$.
Thus, based on the strong performance of DFT on the triplet state and the success of HF+PBEc for the excited singlet state, the excited singlet energy should be approximately given as $E_{S_1}\approx E_{T_1} + 4K_{gu}$. However, in Eq.\ \eqref{eqn:H2_1S1_hyb}  we scaled all the $K$ terms such that $E_{S_1}^{\text{hybrid}}= E_{T_1}^{\text{hybrid}} + \alpha (4K_{gu})$. Thus, the singlet energy is expected to be excellent when $\alpha=1$, but identical to the triplet energy when $\alpha=0$, which indicates a qualitative failure.

The above example clearly illustrates that even when we follow all the rules of the previous sub-section on how to define the ensemble Hartree-exchange energy from first principles, it is still important to carefully \emph{define} separate ensemble Hartree and exchange energies from first principles -- otherwise one can find enormous and unphysical variations in energies with respect to the hybrid functional parameter. In practical terms, our goal is to find a general expression, $\Ex^{\text{ens}}$, for which the total Hxc energy,
\begin{align}
    \EHxc^{\wv,\alpha}:=\EH^{\wv,\text{ens}} + \alpha\Ex^{\wv,\text{ens}} + (1-\alpha)\Ex^{\wv,\DFA} + \Ec^{\wv,\DFA}
    \label{ensemble_hybrid}
\end{align}
varies only reasonably (i.e., as a result of limitations of the underlying DFA) as we change $\alpha$.

In the H$_2$ singlet-triplet case, our above identification of the problem holds its solution therein: the best partitioning between Hartree and exchange would be achieved by moving  $4K_{gu}$, despite its naive '$K$' (exchange) designation, from the exchange energy to the Hartree energy, such that   $\EH^{S_1,\text{ens}}= J_{gg}+2J_{gu}+J_{uu} + 4K_{gu}$ and $\Ex^{S_1,\text{ens}}=-K_{gg}-2K_{gu}-K_{uu}$, with the overall Hartree-exchange energy unaltered. The $4K_{gu}$ term would then {\em not} scale with $\alpha$ and good results can be expected.  
The red curves in Figure~\ref{fig:FDT} represent the results obtained through such a partition [i.e., through Eq.~\eqref{eqn:H2_1S1_hyb} but with $2\alpha K_{gu}\to 2(2-\alpha)K_{gu}$] and indeed they provide excellent agreement with the black FCI curves while removing nearly all of the variation with respect to $\alpha$.

\subsubsection{Systematic Derivation}

In the example above, we used a combination of known success stories (energy of the first triplet state with PBE-based hybrids, energy of the first excited singlet state with HF+PBEc) with theoretical insights (understanding of singlet-triplet gaps in molecular orbital theory) to intuit the best division of Hx into H and x.
Intuition may also guide us in other useful special cases. However, generally we do not wish to work out (and test for different $\alpha$) a new approximation every time we encounter a new CSF. Nor do we want intuition to lead us astray by nudging us toward SD-based expressions when CSFs are required, as it did with the `$K$'-scaling approach! Last but not least, a per-case solution defeats the general philosophy of first principles calculations as the suppliers, rather than the consumers, of physical and chemical intuition. 
We therefore \emph{need a first principles definition of Hartree or exchange energies} -- such that one defines the other via $\EHx=\EH+\Ex$ -- to complement the first principles Hartree-exchange functional of Eq.~\eqref{eqn:EHx}.
We can then rigorously generalize all ground state functionals to their ensemble counterparts.

The first principles solution, as it turns out, comes from a possibly surprising place.
For ground-state DFT, it has been long known that the exchange energy of a pure state is related to its KS (retarded) density-density response function, $\chi_s$, via the fluctuation dissipation theorem (FDT),\cite{Harris1974,Langreth1977} which states that
\begin{align}
    E_{\xrm}[n]=&
- \int \frac{d\vr d\vrp}{2|\vr-\vrp|}
\Bigl\{ n(\vr) \delta(\vr - \vrp) 
\nonumber\\&
~~~~+ \int_{\zm}^{\infty} \frac{d \omega}{\pi} ~
\Im{ \chi_s[n](\vr,\vr';\omega) } \Bigr\},
\label{FDT_GS}
\end{align}
where $\chi_s$ is the frequency dependent (yet a ground-state functional) response function that describes the infinitesimal change in density induced by an infinitesimal change in the KS potential.
We note that the FDT is often discussed together with correlation -- an issue we return to in Eqs~\eqref{eqn:EEcACFpure}--\eqref{eqn:EEcACFLast} in Section~\ref{sec:c} below.

For a non-degenerate ground state, the use of Eq.~\eqref{FDT_GS} or the usual HF expression are completely equivalent -- thus choosing one or the other makes no difference in practice.
However, the KS response function has the advantage of being a well-defined quantity, whether it comes from a pure state or an ensemble, which makes it a safer starting point for ensembles.\cite{Farid1998}
Based on this, the FDT has been extended to ensemble states,\cite{Gould2020-FDT, Gould2021-DoubleX}
yielding the formal expression,
\begin{align}
 \Ex^{\wv,\FDT}[n] := & \int n_{2,\xrm}^{\wv,\FDT}(\vr,\vrp)\frac{d\vr d\vrp}{2|\vr-\vrp|}\;,
 \label{eqn:EEx}
 \intertext{where} 
 n_{2,\xrm}^{\wv,\FDT}(\vr,\vrp)=&
 - n(\vr) \delta(\vr - \vrp) 
 \nonumber\\&
 -\int_{\zm}^{\infty} \frac{d \omega}{\pi} ~
 \Im{ \chi_s^{\wv}[n] (\vr,\vr';\omega) }\;.
 \label{eqn:n2x}
\end{align}
Here, $\chi_s^{\wv}\equiv \chi^{\wv,\lambda=\zp}$ is the KS (retarded density-density) response function for the considered ensemble;
and $n_{2,\xrm}^{\wv,\FDT}$ is an effective `exchange' pair-density obtained from the FDT, the physical meaning of which is elaborated below. The Coulomb integral of $n_{2,\xrm}^{\wv,\FDT}$ yields the exchange energy.

Using the response function has an additional advantage over other definitions.
Technically, $\chi_s^{\wv}$ involves an expansion in terms of CSFs because it is defined for $\lambda=\zp$, not $\lambda=0$.
However, because it is defined via changes to the KS density, a one-body property, it can readily be shown that it must depend only on orbital occupation factors and orbital solutions of the KS equation, i.e., the $f_i$ and $\phi_i$ in $n=\sum_i f_i |\phi_i|^2$, and $\epsilon_i$.
Thus, unlike in the ensemble Hx expression [Eq.~\eqref{eqn:EHx}], one does not need to worry about the specifics of the CSFs.

Eqs.~\eqref{eqn:EEx} and \eqref{eqn:n2x} are essentially the same as Eq.~\eqref{FDT_GS}, except that they now extend to ensemble states (note the $\wv$ designators). They involve only well-defined mathematical quantities, therefore intrinsically avoid any non-uniqueness issues, and can be safely used to \emph{define} the ensemble exchange energy.
Importantly, analytic integration over the frequency integral~\cite{Gould2020-FDT,Gould2021-DoubleX}  (with some caveats%
\footnote{The use of $\max(I,J)$ is not strictly correct for degenerate states. The more precise expression is to use $f_I$ if $\epsilon_I>\epsilon_J$, $f_J$ if $\epsilon_J>\epsilon_I$, or $\tfrac{f_I+f_J}{2}$ if $\epsilon_I=\epsilon_J$ (or, more generally, the average over the whole level).
In practical terms we usually pick equi-ensembles so that $f_I=f_J$ when $\epsilon_I=\epsilon_J$ and so bypass this issue.
For spin-dependent potentials and orbitals we must use a spin-resolved response function, $\chi_{s,\sigma}$, and replace $f^{\wv}_{\max(I,J)}K_{IJ}$ by $\sum_{\sigma} f^{\wv}_{\max(I,J)\sigma}K_{IJ\sigma}$ where ordering of $i=I\sigma$ and $j=J\sigma$ is defined separately for each spin channel, and exchange integrals are different in different spin-channels.})
yields the more practical expression,
\begin{align}
    \Ex^{\wv,\FDT}[n] = & -\sum_{IJ}f^{\wv}_{\max(I,J)} K_{IJ}
    \label{eqn:EExK}
\end{align}
where $I$ is the spatial index of orbital label, $i=I\sigma$; $f^{\wv}_I=f^{\wv}_{I\up}+f^{\wv}_{I\down}$ is the spin-summed occupancy of spatial orbital, $\varphi_I$, ordered by energy $\epsilon_I$; and $K_{IJ}$ is the above-defined exchange integral.
Here, the somewhat unusual `max' comes from terms proportional to  $f_I\sgn(\epsilon_I-\epsilon_J)$ in the frequency integrated KS response, that either cancel (for $\epsilon_I>\epsilon_J$) or enhance (for $\epsilon_J>\epsilon_I$) equivalent terms in $n(\vr)\delta(\vr-\vrp)$.

The FDT then also lets us readily define,
\begin{align}
  \EH^{\wv,\FDT}[n] \equiv & \EHx^{\wv}[n] - \Ex^{\wv,\FDT}[n]\;.
  \label{eqn:EEH0a}
\end{align}
A more detailed mathematical analysis~\cite{Gould2020-FDT,Gould2021-DoubleX} provides a closed-form expression [true for equi-ensembles, but with similar caveats to those given for Eq.\ \eqref{eqn:EExK}],
\begin{align}
    \EH^{\wv,\FDT}[n]:=\sum_{\FE\FE'}w_{\max(\FE,\FE')}\HInt[n_{s,\FE\FE'}]\;,
  \label{eqn:EEH0direct}
\end{align}
that invokes KS transition densities, $n_{s,\FE\FE'}=\ibraketop{\FE_s}{\nh}{\FE_s'}$, between the KS-CSFs. 
Here, we have introduced
\begin{align}
\HInt[q]=\HInt[q^*]=\int q(\vr)q^*(\vrp)\frac{d\vr d\vrp}{2|\vr-\vrp|},
\label{eqn:Hint}
\end{align}
which is a real-valued functional accepting a complex-valued function argument, $q$, to accommodate complex-valued quantities, especially transition densities. This definition reduces to the standard Hartree expression, $E_{\Hrm}[n]=\HInt[n]$, for real-valued densities.
However, in practice it is almost always better to use Eqs.~\eqref{eqn:EHx} and \eqref{eqn:EExK} to evaluate \eqref{eqn:EEH0a} as their difference.
This is because the apparent mathematical simplicity of the closed form expression can be deceptive when applied to real systems.

Does the FDT-based definition solve the above-discussed quandary of the $S_1$ state of H$_2$? In this case we have $f_g=f_u=1$ (since both gerade and ungerade orbitals are occupied by one electron) and $f_{i>u}=0$ (all other orbitals being empty). Therefore, from Eq.~\eqref{eqn:EExK} $\Ex^{S_1,\FDT}=-K_{gg}-K_{gu}-K_{ug}-K_{uu}=-K_{gg}-2K_{gu}-K_{uu}$, which is precisely the useful result we derived previously, but it is now based on rigorous definitions rather than on intuition. Importantly, this result is manifestly, and correctly, different than the one obtained by naively counting all `$K$'-containing terms in Eq.~\eqref{eqn:H2_1S1} as exchange [note the term $-2K_{gu}$ here, rather than $+2K_{gu}$ in Eq.~\eqref{eqn:H2_1S1}].

Having now defined Hartree and exchange components from first principles, and having earlier defined Hx exactly, we are now ready to define hybrid functionals.
Recognizing that Hx and x are easier to define and compute than H, it is advantageous to rewrite the ensemble hybrid expression of Eq.\ \eqref{ensemble_hybrid} in a mathematically equivalent but practically more useful form:
\begin{align}
\EHxc^{\wv,\DFA}=\EHx^{\wv} + (1-\alpha) (\Ex^{\wv,\DFA}-\Ex^{\wv}) + \Ec^{\wv,\DFA} \;.
\label{eqn:Hybrid}
\end{align} 
Here, quantities without a DFA label indicate exact (ensemble Hartree-Fock) results, and DFA indicates our choice of an approximate density functional, with the ``parent'' DFA  obtained for $\alpha = 0$.

Importantly, because they are derived from the FDT, Eqs.~\eqref{eqn:EExK}-\eqref{eqn:Hybrid} necessarily reduce to the expected energy expressions when applied to single-SD states.
In particular, Eq.~\eqref{eqn:EExK} reproduces, where appropriate, the standard expression for the exchange functional of ground state DFT -- which is good, as it allows re-use of existing DFT results.
Here, the expressions build ensemble quantities by summing over the contributions of {\em each} state in the ensemble; i.e., $\Ex^{\wv,\FDT}$ is an ensemble state-driven (i.e., state resolved) quantity. 

\begin{figure}
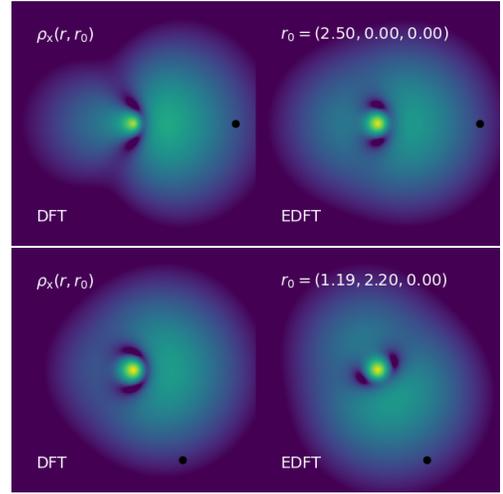

    \includegraphics[width=0.75\linewidth]{{{Frame038_par}}}
    \includegraphics[width=0.75\linewidth]{{{Frame079_par}}}
    \caption{Exchange pair-hole densities, $\rho_{\xrm}(\vr,\vr_0)$, for the B atom, obtained from regular DFT [left, based on a conventional single-SD model] and from symmetry-preserving EDFT [right, based on Eq.~\eqref{eqn:n2x}].
    The red dots denote the position $\vr_0$ and the colours indicate the value of $\rho_{\xrm}(\vr,\vr_0)$ at $\vr$.
    The top and bottom panels illustrate the effect of rotation of $\vr_0$.
    Colors range from navy (no depletion) through blue to light green (maximal depletion).
    \label{fig:Symmetry}}
\end{figure}

Before concluding, we mention two important issues.
Firstly, recall that hybrid DFAs in ground-state DFT are almost always applied within GKS theory,\cite{Seidl1996} i.e., with a non-multiplicative Fock operator.
In EDFT, this requires a suitable generalization of ensemble KS theory.
The recently-derived\cite{Gould2021-EGKS} ensemble GKS equations are somewhat more complicated than their ground state counterparts, but still amenable to practical solutions.

Secondly, an interesting ``side-effect'' of using first principles to define $\Ex^{\FDT}$ and $\EH^{\FDT}$ for degenerate states (see Sec.~\ref{sec:Symmetries}), that shows how first principles consideration help ensure good properties.
The effective pair-density, $n_{2,\xrm}^{\wv,\FDT}$ used in Eq.~\eqref{eqn:EEx}, is the probability that a ``hole'' in the electron density appears at $\vrp$, given an electron at $\vr$.
For equi-ensembles over rotational symmetries, like those discussed in Sec.~\ref{sec:Symmetries}, it can be shown that the x pair-density (and its H counterpart) are invariant to rotations (see the Supplementary Material of \rcite{Gould2020-FDT} for details).
This invariance is demanded by physical laws, but is \emph{not always reproduced} in standard HF theory, even when symmetries are preserved at the orbital level and energies are predicted correctly.

Figure~\ref{fig:Symmetry} illustrates the above result for the B atom, by considering exchange pair-hole densities, $\rho_{\xrm}(\vr,\vr_0):=n^{\wv,\FDT}_{2,\xrm}(\vr,\vr_0)/n(\vr_0)$. The exchange pair-hole density represents the depletion in the probability of finding an electron at $\vr$ given an electron at $\vr_0$.
Therefore, rotation of $\vr_0$ (per the top and bottom panels) ideally should be equivalent to rotation of the system.
Figure~\ref{fig:Symmetry} compares $\rho_{\xrm}(\vr,\vr_0)$  obtained from a conventional Slater determinant-based ground state (i.e., from standard DFT exchange definitions) with those obtained from applying Eq.~\eqref{eqn:n2x} to an equal mix over all degenerate ground states (i.e., from the above EDFT considerations). 
The EDFT hole faithfully rotates to follow the black dot and thus reproduces the correct physical behaviour.
The conventional  DFT hole, however, does not. 
Because all calculations use the same set of orbitals, obtained from a centro-symmetric potential, the difference must come from the symmetry adaptation in $\rho_{\xrm}(\vr,\vr_0)$ by EDFT.

In summary, using the FDT to derive exchange leads to Eq.~\eqref{eqn:EExK}, an expression that is well-suited to being combined with exchange DFAs in hybrids [Eq.~\eqref{eqn:Hybrid}], as illustrated in Figure~\ref{fig:FDT}.
The same expression also has advantages for degenerate states treated via equi-ensembles, as illustrated in Figure~\ref{fig:Symmetry}.

\subsubsection{Approximating exchange terms}
\label{sec:XApprox}

So far, we have focused on exact properties of exchange, with the ensemble hybrid of Eq.~\eqref{eqn:Hybrid} motivating why a division into `H' and `x' terms is necessary.
But practical application of Eq.~\eqref{eqn:Hybrid} additionally requires useful approximations for $\Ex^{\wv,\DFA}$.
This can be accomplished by adapting  existing state-of-art ground state exchange DFAs for use in ensemble problems.

Because ground-state exchange approximations have been developed for  pure states described by a Slater determinant, 
it is sensible to first resolve the ensemble exchange energy, $\Ex^{\wv}$, in terms of  contributions from individual pure states.
Each ensemble-member state, $\iket{\FE}$ or $\iket{\FE_s}$, can be associated with its weight, $w_{\FE}$, in Eq.~\eqref{eqn:EExK},
by recognising that the occupation factors therein, $f_I^{\wv}$,  are themselves weighted averages, i.e., $f_I^{\wv}=\sum_{\FE}w_{\FE}\theta_I^{\FE}$.
We therefore obtain $\Ex^{\wv,\FDT}=\sum_{\FE}w_{\FE}E_{\xrm,\FE}^{\FDT}$, where
\begin{align}
    E_{\xrm,\FE}^{\FDT} 
    =-\sum_{IJ}\theta_{\max(I,J)}^{\FE}K_{IJ}\;
    \label{eqn:ExState}
\end{align}
is the exchange energy associated with $\iket{\FE_s}$ and $\theta_I^{\FE}$ stands for the spin-summed occupation factor of the orbital $I$ in state $\FE$.

It can be verified that Eq.~\eqref{eqn:ExState} yields expected results when dealing with (spin-polarized) ground states, which may therefore be approximated by standard (spin-polarized) DFAs.
Intuition is harder to use, however, when approximations must be devised for more general states.
But, having made a precise identification of the exact individual components, we are guided towards appropriate approximations.

The key insight is to recognise that Eq.~\eqref{eqn:ExState} depends only on the set of \emph{total} (i.e. spin-summed) occupation factors, $\theta_I^{\FE}$, of orbitals in state $\FE$.
From this we immediately recognise that --- given a fixed set of orbitals --- any two states with the same spin-summed occupation factors (e.g. $\up$- and $\down$-majority triplet states) should have the same exchange energy, as well as the same KS kinetic energy.
Note that in the above it is important to equally weight all degenerate spin-states so that the ensemble is described by restricted KS theory (i.e., spatial orbitals are the same for $\up$ and $\down$ spin).

As an example, consider again the case of singly-promoted triplet, $\iket{T_1^{-1,0,1}}$, and a singlet, $\iket{S_1}$.
All four states in this singlet-triplet set have the same total occupation factors.
Eq.~\eqref{eqn:ExState} dictates that they have the same exchange energy -- for any given set of orbitals.
Therefore, consistency requires that we demand that $\iket{S_1}$, $\iket{T_1^0}$ and $\iket{T_1^{\pm 1}}$ all have the same exchange energy within any give approximate expression.
This exact condition can be very useful, as we already know how to deal with DFAs for spin-polarized triplets and, consequently, can reuse without modification any known triplet DFA for the spin-unpolarized triplet and singlet, namely, $E_{\xrm,T_1^0}=E_{\xrm,S_1}=E_{\xrm,T_1^{\pm 1}}\approx E_{\xrm}^{\DFA}[\rho_{T_1}]$, where $\rho_{T_1}$ is the 1RDM for the SD describing a polarized triplet. 
We remind that this does not preclude singlet-triplet splitting, because in this point of view the splitting energy is contained in the Hartree term,  $E_{ST}=E_{\Hrm,S_1}-E_{\Hrm,T_1}$.

For a fractional ensemble problem the situation is simpler.
One can show that $\Ex^{N-q}=(1-q)E_{\xrm}^N + qE_{\xrm}^{N-1}$, where the superscript indicate a ground state with the given number of electrons.\cite{Kraisler2013,Goerling2015}
Therefore a similar relation, $\Ex^{\EDFA,N-q}\approx (1-q)E_{\xrm}^{\DFA,N} + qE_{\xrm}^{\DFA,N-1}$, which is based on SD ground states at integer electron number, is immediately amenable to use of existing DFAs.

For a general excited state, $\iket{\FE_s}$, we first seek an exact relationship (combination law) for the total occupation factor $\theta^{\FE}_I=C_1^{\FE} \theta^{\SD_1}_I + C_2^{\FE} \theta^{\SD_2}_I + \ldots$, which expresses $\theta_I^{\FE}$ as a linear combination of total occupation factors of pertinent SD states corresponding to the lowest energy state within a given spin multiplicity.
By construction, the same combination law then applies to the exact exchange and therefore one can express an approximate exchange for state $\iket{\FE_s}$ as 
\begin{align}
    E_{\xrm,\FE}^{\EDFA}
    \approx C_1^{\FE} E_{\xrm}^{\DFA}[\rho_{\SD_1}] + C_2^{\FE} E_{\xrm}^{\DFA}[\rho_{\SD_2}] + \ldots,
    \label{eqn:EDFA_x}
\end{align}
i.e., in terms of nothing but SD states that we already know how to handle with conventional DFAs.

While for some low-lying states deducing the coefficients, $C_{1,2,\ldots}$, in the above equation can be done by inspection, in general this is a non-trivial task.
Combinations for typical important excitations are given in Table~\ref{tab:Hxc} (which also includes correlations for reasons discussed in Section~\ref{sec:CApprox}).
In particular, the table reveals that coefficients can generally be either positive or negative. For example, even for the simplest double excitation,~\cite{Gould2021-DoubleX} the combination law is $E_{\xrm,h^2\to l^2}^{\EDFA}\approx 2E_{\xrm}[\rho_{T_1}]-E_{\xrm}[\rho_{S_0}]$; i.e.,  $E_{\xrm,h^2\to l^2}^{\EDFA} \neq E_{\xrm}[\rho_{S_2}]$, where $S_2$  is the doubly excited determinant. 
This shows that the doubly-excited state requires a different and non-trivial ensemble treatment, reflecting its excited state nature, despite being described by an SD Kohn-Sham state.

\subsection{Step four: Deriving ensemble correlations}
\label{sec:c}

\subsubsection{A hidden type of correlation: State-driven and density-driven correlations}
\label{sec:c_ACF}

The final energy contribution we must deal with is the correlation. 
Generally, the correlation term is the hardest many-electron term to address. This is because the Hartree and exchange energies, while also reflecting many-electron interactions, can ultimately be computed directly in terms of pairs of occupied KS states, i.e. single-electron orbitals. Correlation, however, reflects ``true many-electron'' interactions.

Let us first recall some aspects of pure-state correlation.
The correlation energy in KS-DFT is defined as $E_{\crm}=\ibraketop{\Psi}{\Hh}{\Psi}-\ibraketop{\Phi_s}{\Hh}{\Phi_s}$, where $\iket{\Psi}$ is the interacting wave function and $\iket{\Phi_s}$ is the non-interacting KS wave function.
For any many-electron system, one can express the correlation as $E_{\crm}=T_{\crm}+W_{\crm}$, i.e., as the sum of two many-body interactions: a positive kinetic energy term, $T_{\crm}=\ibraketop{\Psi}{\Th}{\Psi}-\ibraketop{\Phi_s}{\Th}{\Phi_s}>0$, and a negative Coulomb term, $W_{\crm}=\ibraketop{\Psi}{\Wh}{\Psi}-\ibraketop{\Phi_s}{\Wh}{\Phi_s}< -T_{\crm}$, such that the overall correlation energy is negative.\cite{Perdew2003} 

The mixture of positive kinetic and negative electrostatic terms makes deeper analysis difficult. However, the problem can be simplified by invoking an adiabatic connection of the KS and interacting electron systems, already mentioned briefly in Section \ref{sec:KS}. The adiabatic connection formula (ACF) rids us of the explicit kinetic energy contribution by expressing the correlation energy as\cite{Harris1974,Langreth1975}
\begin{equation}
E_{\crm}[n]=\int_0^1 \Big(\ibraketop{\Psi_\lambda}{\Wh}{\Psi_\lambda}-\ibraketop{\Phi_s}{\Wh}{\Phi_s}\Big)d\lambda,
\label{eqn:EEcACFpure}
\end{equation}
where $\iket{\Phi_s}\equiv \iket{\Psi_{\lambda\to 0^+}}$ and the integral is over the interaction energies of wave functions, $\iket{\Psi_{\lambda}}$, that all yield the ground-state density, $n$, from an \emph{adiabatically connected} Hamiltonian, $\Hh^{\lambda}=\Th+\vh^{\lambda}+\lambda\Wh$.
In this approach, the Coulomb interaction is scaled by the parameter $0 \leq \lambda \leq 1$, connecting smoothly between the KS system ($\lambda=0$) and the original interacting-electron system ($\lambda=1$), while $\vh^{\lambda}$ changes with $\lambda$ so as to retain the same density throughout.

Using the ACF in conjunction with the fluctuation-dissipation theorem (FDT), following steps similar to those outlined in the previous sub-section, yields the ACF-FDT expression for pure-state correlation:
\begin{align}
    E_{\crm}^{\text{pure}}[n]=&\int n_{2,\crm}(\vr,\vrp) \frac{d\vr d\vrp}{2|\vr-\vrp|}\;,
    \label{eqn:EcACFDTPure}
    \\
    n_{2,\crm}^{\text{pure}}(\vr,\vrp)=&-\int_0^1\int_{0^-}^{\infty}\frac{d\omega}{\pi}
    \Im\Delta\chi^{\lambda}[n](\vr,\vrp;\omega) d\lambda ,
    \label{eqn:EEcACFLast}
\end{align}
in terms of the difference, $\Delta\chi^{\lambda}=\chi^{\lambda}-\chi_s$, between the response function at $\lambda$ and its non-interacting KS counterpart, $\chi_s\equiv \chi^0$.

Naively, all we need to do in order to extend the above equation for \emph{ensemble} correlation is to use it with the ensemble density rather than the pure state density. 
This yields
\begin{align}
 \Ec^{\wv,\sd}[n] :=& \int n_{2,\crm}^{\wv,\sd}(\vr,\vrp)\frac{d\vr d\vrp}{2|\vr-\vrp|}
 \label{eqn:EEcSD}
 \\
 n_{2,\crm}^{\wv,\sd}(\vr,\vrp) := &  - 
  \int_{\zp}^{1} \int_{\zm}^{\infty} \frac{d \omega}{\pi}
  \Im{ \Delta \chi^{\wv,\lambda}[n] (\vr,\vr'; \omega) } d \lambda \;,
  \label{eqn:n2cSD}
\end{align}
with the meaning of the superscript `sd' to be explained below.
Eq.\ (\ref{eqn:n2cSD}) is now defined in terms of the \emph{ensemble} density-density response function,
\begin{align}
\Delta\chi^{\wv,\lambda}=\chi^{\wv,\lambda}-\chi_s^{\wv}
=&\sum_{\FE}w_{\FE}[\chi_{\FE}^{\wv,\lambda}-\chi_{s,\FE}^{\wv}]\;,
\end{align}
of interacting (no subscript) and non-interacting (KS CSF, '$s$' subscript) systems. In this way, Eq.~\eqref{eqn:EcACFDTPure} is simply a special case of Eq.~\eqref{eqn:EEcSD}.

\begin{figure}
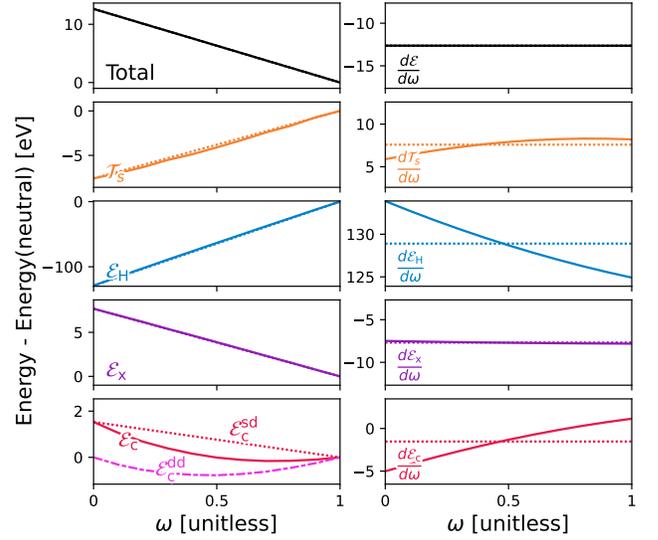

    \includegraphics[width=\linewidth]{{{FigDDC}}}
    \caption{Left panels: Total energy and its DFT components for a fractional cation of H$_2$O with $9+\omega$ electrons, as a function of $\omega$, shown as solid lines.
    In the bottom left panel, the dotted line represents the contribution of the sd correlation of Eq.\ \eqref{eqn:EEcSD} and the magenta dash-dotted line represents the dd correlation energy of Eq.~\eqref{eqn:EEcDD}.
    Right panels: derivative of energies in the left panel with respect to electron number.
    \label{fig:DDC}
    }
\end{figure}

At this point, the discerning reader should already be wary of naive ensemblization. And indeed, we show that it is also problematic here.
In fact, we'll soon find out that there are extra correlations to be taken care of! This has originally been derived in the context of GOK-EDFT,~\cite{Gould2019-DD} but we take the opportunity to illustrate that similar types of correlations also occur for ensembles with a non-integer number of electrons.

To demonstrate this point, consider exact ensemble-DFT results for the fractionally charged water molecule, with the number of electrons ranging from 9 (H$_2$O$^+$) to 10 (H$_2$O).\footnote{Specifically, we consider three-member ensembles composed of $\omega$ times the neutral atom, and $\tfrac{1-\omega}{2}$ times each of the two degenerate doublet states ($\up$- and $\down$-majority) of the cation.} 
These are obtained by considering accurate reference densities (in this case coming from coupled cluster calculations) as  `exact' values and inverting the ensemble KS equation to obtain the KS orbitals and thus the `exact'  energies.\cite{Gould2023-JCP,Gould2019-Hx}

Figure~\ref{fig:DDC} shows the obtained total energy, as well as various ensemble energy components, as a function of the fractional charge, $0\leq \omega\leq 1$, added to the water molecule cation such that it has $9+\omega$ electrons. As expected from the piecewise linearity of the exact functional, the total energy is a linear function of $\omega$. The external energy (not shown) is also piecewise linear in $\omega$, by construction, as it is a linear functional of the density, which is linear in $\omega$.
We see that the exchange energy, $\Ex^{\omega}$, is very close to linear, but not exactly linear. This is  
because the KS response function in Eq.~\eqref{eqn:n2x} need not be perfectly linear.
All other energies, namely the non-interacting kinetic energy ($\Ts^{\omega}$), the Hartree energy ($\EH^{\omega}$), and the correlation energy ($\Ec^{\omega}$), exhibit substantial non-linearities.
All non-linearities cancel once they are combined to form the total energy.

The non-linearity of the correlation energy in Fig.\ \ref{fig:DDC} reveals that Eq.~\eqref{eqn:EEcSD} is incomplete, owing to the following reasoning. 
First,  the KS response ($\lambda=0^+$) is nearly linear, as revealed by the near-linearity of $\Ex^{\omega}$.
Second, the interacting ($\lambda=1$) response is exactly linear, which follows from its definition as $\delta n^{\omega} / \delta v$, where $n^{\omega}$ is linear in $\omega$, and $v$ is the $\omega$-independent external potential. 
Because the interacting response is exactly linear, the response $\chi^{\omega,\lambda}$ for $0<\lambda<1$ is likely to be even closer to linear than the KS response, from which it follows that the correlation energy [Eq.~\eqref{eqn:EEcSD}] should be more nearly linear than its exchange counterpart [Eq.~\eqref{eqn:EEx}].
However, Fig.\ \ref{fig:DDC}  clearly reveals that $\Ec$ is significantly non-linear. 
It follows that there must be an additional term, which we define as $\Ec^{\dd}=\Ec-\Ec^{\sd}$, that captures the deviation from linearity.

To arrive at a formal ensemblization of the correlation energy, which would resolve the above quandary, we need to carefully reconsider both the ACF\cite{Perdew1985,Nagy1995,Borgoo2015} and the FDT\cite{Gould2020-FDT} for ensembles. 
Working with Eq.~\eqref{eqn:Gammalambda}, the \emph{ensemble} adiabatic connection is  carried out at a fixed ensemble density, i.e., we choose a potential, $v^{\wv,\lambda}[n]$, such that the Hamiltonian $\Hh^{\wv,\lambda}[n]=\Th+\vh^{\wv,\lambda}[n]+\lambda\Wh$ ensures that the ensemble density
\begin{align} \Tr[\Gammah^{\wv,\lambda}\nh]=\sum_{\FE}w_{\FE}\ibraketop{\FE^{\wv,\lambda}[n]}{\nh}{\FE^{\wv,\lambda}[n]}=n,
    \label{eqn:nwConstraint}
\end{align}
is the same for all $0 \leq \lambda \leq 1$, where $\iket{\FE^{\wv,\lambda}[n]}$ are eigenstates of $\Hh^{\wv,\lambda}[n]$ and
\begin{align}
    \Gammah^{\wv,\lambda}[n]\equiv \sum_{\FE}w_{\FE}\iout{\FE^{\wv,\lambda}[n]}\;
\end{align}
is thereby well-defined for every $\lambda$, under the usual assumption that $v^{\wv,\lambda}[n]$ exists.
Then, using the Hellmann-Feynman theorem on Eq.\ (\ref{eqn:Gammalambda}) yields $\tfrac{\partial\F^{\wv,\lambda}}{\partial \lambda}=\tr[\Gammah^{\wv,\lambda}[n]\Wh]$. 
One finally obtains $\EHxc^{\wv}=\F^{\wv,1}-\F^{\wv,0}=\int_0^1 \frac{\partial\F^{\wv,\lambda}}{\partial \lambda}d\lambda$, which leads to
\begin{align}
    \Ec^{\wv}[n]=&\EHxc^{\wv}[n]-\EHx^{\wv}[n]=\int_{\zp}^1 \tr[(\Gammah^{\wv,\lambda}-\Gammah_s^{\wv})\Wh] d\lambda
    \nonumber \\
    =&\int_{\zp}^1 \sum_{\FE}w_{\FE}
    \big\{ \ibraketop{\FE^{\lambda}}{\Wh}{\FE^{\lambda}}
    - \ibraketop{\FE_s}{\Wh}{\FE_s}
    \big\} d\lambda
    \;.
    \label{eqn:EEcACF}
\end{align}
Note that we connect from $\zp$ to avoid the non-uniqueness issues mentioned in Section~\ref{sec:Hx}.
Also note that Eq.\ (\ref{eqn:EEcACF}) can be thought of as a weighted ensemble sum over each of the pure states comprising the example. When there is only one ``ensemble member'', i.e., the state is pure, Eq.\ (\ref{eqn:EEcACF}) properly reduces to Eq.\ (\ref{eqn:EEcACFpure}).  
Here and henceforth we use $\iket{\FE^{\lambda}}$ as shorthand for $\iket{\FE^{\wv,\lambda}[n]}$ and use the subscript $s$ as shorthand for the KS CSF states with $\lambda=\zp$. Thus, for $\lambda\neq 1$, all wave functions and quantities derived therefrom carry an implicit weight-dependence.

Consider now the (extended) FDT.
\rcite{Gould2020-FDT} has shown that the adiabatically-connected pair-density, $n_2^{\lambda}(\vr,\vrp)$, of a GOK-ensemble may be separated into response ($\RES$) and  density- and transition density ($\ENS$) parts, in the form:
\begin{align}
    n_2^{\lambda,\wv}(\vr,\vrp)=&
    \sum_{\FE}w_{\FE}\big\{
    n_{2,\FE}^{\RES,\lambda}(\vr,\vrp) + n_{2,\FE}^{\ENS,\lambda}(\vr,\vrp)
    \big\}\;,
    \label{eqn:n2lambda}
\end{align}
where,
\begin{align}
    n_{2,\FE}^{\RES,\lambda}(\vr,\vrp)=&-n_{\FE}^{\lambda}(\vr)\delta(\vr-\vrp)
    \nonumber\\&
    -\int_{0^-}^{\infty}\frac{d\omega}{\pi}\Im\chi_{\FE}^{\lambda}[n](\vr,\vrp;\omega)
    \\
    n_{2,\FE}^{\ENS,\lambda}(\vr,\vrp)=&
    n_{\FE}^{\lambda}(\vr)n_{\FE}^{\lambda}(\vrp)
    \nonumber\\&
    + 2\Re\sum_{\FE'<\FE}n_{\FE\FE'}^{\lambda}(\vr)n_{\FE'\FE}^{\lambda}(\vrp)
    \label{eq:ANSterm}
    \;.
\end{align}
Here, $n_{\FE\FE'}^{\lambda}=\ibraketop{\FE^{\lambda}}{\nh}{\FE^{\prime\lambda}}$ is a transition density between states $\FE$ and $\FE'$ and $n_{\FE}^{\lambda}\equiv n_{\FE\FE}^{\lambda}$ is the density of state $\FE$.
$\FE'<\FE$ indicates that $E_{\FE'}\leq E_{\FE}$ and we used $n_{\FE\FE'}=n_{\FE'\FE}^*$ to combine $\FE'>\FE$ terms with their $\FE'<\FE$ counterparts. 
Degenerate ensembles are also covered by \eqref{eqn:n2lambda}, by assigning an arbitrary (but fixed once set) order to the degenerate states
and using same weights for states with same energy.
Eq.\ \eqref{eqn:n2lambda} can also be used for a fractional ground state ensemble with $N+\omega$ electrons, by choosing $\kappa$ to be $N$ and $N+1$, with ensemble weights of $1-\omega$ and $\omega$, respectively, and noting that the transition density terms in Eq.\ (\ref{eq:ANSterm}) vanish as they involve different particle numbers.

The first term in Eq.\ \eqref{eqn:n2lambda}, i.e., the $\RES$ term, precisely gives rise to $\Ec^{\wv,\sd}$ of Eq.~\eqref{eqn:EEcSD}, where, for reasons elaborated in the next sub-section, `sd' stands for state-driven correlation.
But we notice that the second term, i.e., the $\ENS$ term, was neglected entirely in writing the naive Eq.\ \eqref{eqn:EEcSD}!
This means that certain contributions to the correlation energy, due to the ensemble treatment of densities or the excited state character of the states included, were missed.
Eq. \eqref{eq:ANSterm} implies that the $\ENS$ terms contribute via the electrostatic energy of (transition) densities, in the same fashion as the corresponding terms for $\lambda\to 0^+$ contribute to Eq.~\eqref{eqn:EEH0direct}.
Hence, we refer to them as density-driven (dd) correlations.

Taking the ACF [Eq.~\eqref{eqn:EEcACF}] finally reveals that the missing dd correlation energy is given by:
\begin{align}
    \Ec^{\wv,\dd}[n]
    :=& \int_{0^+}^1\sum_{\FE} w_{\FE}\bigg\{ 
    (E_{\Hrm}[n_{\FE}^{\lambda}] - E_{\Hrm}[n_{s,\FE}])
    \nonumber\\
    & + 
    2\sum_{\FE'<\FE}
    (\HInt[n_{\FE\FE'}^{\lambda}] - \HInt[n_{s,\FE\FE'}])
    \bigg\} d\lambda\;,
    \label{eqn:EEcDD}
\end{align}
for GOK/degenerate ensembles with $\HInt$ defined above in Eq.~\eqref{eqn:Hint} to address (possibly complex-valued) transition densities, $n_{\FE\FE'}$, and using $E_{\Hrm}[n]=\HInt[n]$ for (guaranteed real-valued) densities $n_{\FE}$; or,%
~\footnote{This result is easily obtained by following the same derivation as the FDT for excited states, but recognising that $\ibraketop{\Phi^N}{\nh}{\Phi^{N\pm 1}}=0$ due to the difference in electron numbers.
As a result, the transition density terms disappear, leaving only the diagonal terms.}
\begin{align}
    \Ec^{N+\omega,\dd}[n]
    :=& \int_{0^+}^1
    (1-\omega)\{E_{\Hrm}[n_{N}^{\lambda}]-E_{\Hrm}[n_{s,N}]\}
    \nonumber\\&
    +\omega\{E_{\Hrm}[n_{N+1}^{\lambda}]-E_{\Hrm}[n_{s,N+1}]\}
    d\lambda\;,
    \label{eqn:EEcDD_Frac}
\end{align}
for fractional ensembles, by evaluating the electrostatic energy of the $\ENS$ terms. 

\subsubsection{Aspects and implications of state-driven and density-driven correlations}
\label{sec:CAspects}

Let us now stress some important aspects of sd and dd correlations.
First, notice that $\Ec^{\wv,\dd}$ does not appear in any discussion of pure-state correlation because it is \emph{exactly zero for conventional DFT ground states}.
All transition density terms in Eq.\ (\ref{eq:ANSterm}) trivially disappear for the pure ground state as there are no lower energy states to transition to.
But correlation due to the first term in Eq.\ (\ref{eq:ANSterm}) is also zero for a pure state, because $n^{\lambda}=n$ is obeyed for all $\lambda$ by construction.
This means that the first term in Eq.\ (\ref{eq:ANSterm}) produces the same value for the first and second bracket in Eq. (\ref{eqn:EEcDD}) and the integral vanishes.

For a pure excited state, the first term of Eq.~\eqref{eqn:EEcDD} vanishes, by construction, for the same reason as its counterpart in the ground state.
Furthermore, if the lower energy terms have a different fundamental spin structure (e.g. triplet vs singlet) the second term also vanishes because the difference in spin means that $n_{\FE\FE'}^{\lambda}=0$.
Thus, the lowest energy state of each given spin also has no dd correlations.
But, in general, the second term can be non-zero even for a pure excited state.

In ensemble DFT, \emph{both} of the terms of Eq.\ (\ref{eq:ANSterm}) can result in the non-zero correlation expressions of Eqs~\eqref{eqn:EEcDD} and \eqref{eqn:EEcDD_Frac}, which represent correlation beyond that of the naive expression of Eq.\ (\ref{eqn:EEcSD}).
Eq.~\eqref{eqn:EEcSD} represents an expected weighted contribution, $\chi^{\wv}=\sum_{\FE}w_{\FE}\chi_{\FE}$, of the response, $\chi_{\FE}$, of each state in the ensemble.
Hence it was designated `sd', for ``state-driven'' (sd) correlation energy.
Due to its dependence on response functions, $\Ec^{\wv,\sd}$ is the ``natural'' correlation-companion for $\Ex^{\wv,\RES}$ of Eq.~\eqref{eqn:EEx}.
We therefore expect typical pure-state DFAs, which benefit from cancellation of errors between exchange and correlation, to be effective for this term.
Eq.~\eqref{eqn:EEcDD} entails two \emph{new} correlation terms arising solely from ensemble density terms.
Hence it was designated `dd', for density-driven correlation.%
\footnote{We stress that the concept of density-driven \emph{correlations} presented here should not be confused with the concept of density-driven \emph{errors} [M.\ C.\ Kim, E. Sim, and K. Burke, ``Understanding and reducing errors in density functional calculations'', \emph{Phys.\ Rev.\ Lett.} {\bf 111}, 073003 (2013); S. Song, S. Vuckovic, E. Sim, and K. Burke, ``Density-corrected DFT explained: questions and answers'', \emph{J.\ Chem.\ Theo.\ Comp} {\bf 18}, 817 (2022)].
Density-driven correlation are an \emph{exact} energy term that results from difference in densities between interacting and partly- or non-interacting ensembles constrained to have the same total ensemble density.
They may or may not be accounted for in suitably constructed ensemble DFAs. Density-driven errors may refer to both exchange and correlation and may appear in DFAs only (i.e., cannot appear in exact DFT).
These errors arise from failure of a given approximation to produce highly-accurate densities.} 
The first term is a typical weighted average of Hartree-like differences between pure state densities ($n_{\FE}=\ibraketop{\FE}{\nh}{\FE}$).
The second is a Hartree-like term that brings in fluctuations, i.e., {\em off-diagonal} matrix elements ($n_{\FE\neq\FE'}=\ibraketop{\FE}{\nh}{\FE'}$).
The fluctuation terms are zero, e.g., when they involve  states with different spins (as in an ensemble of a ground-state singlet and an excited-state triplet) or electron numbers (as in a fractional density ensemble), but they are not generally zero.

We emphasize that one may be rather easily tempted to think that $\Ec^{\dd}=0$, because the adiabatic connection is obtained at a fixed density, $n$.
However, as discussed in Section \ref{sec:Fractional} and demonstrated in Fig.\ \ref{fig:Water} therein, it is only the {\em total} ensemble density that is kept fixed.
Individual pure-state-resolved densities and transition densities, $n_{\FE}^{\lambda}$ and $n_{\FE\FE'}^{\lambda}$, typically vary with $\lambda$ and differ from the corresponding fully-interacting or non-interacting densities.
Specifically, the surprising density-driven correlation revealed in Fig.~\ref{fig:DDC} for the water molecule is a direct consequence of the difference between true and KS densities for individual pure-state ensemble members, shown explicitly in Figure~\ref{fig:Water} for the same molecule.

A closer inspection of Fig.~\ref{fig:DDC} also shows that, as expected, $\Ec^{\dd}$ is trivially zero for $\omega$=1 as the neutral water molecule is in a pure state.
More interestingly, it is also zero for $\omega=0$, despite the water molecule cation being treated here using an ensemble average over $\up$- and $\down$-majority doublet states (just like the lithium example of Figure~\ref{fig:LiLevels} in Section~\ref{sec:Symmetries}).
This is because any ensemble that involves an average only over the degenerate levels of a lowest doublet or triplet (etc) yields $\Ec^{\dd}=0$, because the density of all states in the ensemble is the same and thus does not vary with $\lambda$, and (as explained above) there are no non-zero transition densities.
In contrast, spatial degeneracies can lead to density-driven correlations. 

The existence of density-driven correlations peculiar to ensemble DFT was recognized somewhat earlier than the FDT argument through which it was presented here. 
These correlations were \Response{}{reported first} in \rcite{Gould2019-DD} and shortly thereafter refined in \rcite{Fromager2020-DD}, \Response{}{with different definitions from each other} and from the FDT one of Eq.~\eqref{eqn:EEcDD}. 
Both studies were in the context of neutral excitations.
The ACF-FDT definition provided here was given later in \rcite{Gould2020-FDT} and is expected to play an important role in deriving approximations. 
Above, we showed that similar quantities emerge also in fractional ensembles of ground states.

The existence of multiple definitions for the division of correlation into state-driven and density-driven contributions is not a problem in itself because in all definitions the total correlation remains the same and is well defined.
Here, we preferred the ACF-FDT definition because its framework affords direct connections with the exchange (for sd) and Hartree (for dd) terms.
In other words, the approach provides us with a unified framework to develop approximations for all relevant energy components of density functionals for ensembles.

In summary, ensemble correlation consists of two contributions. $\Ec^{\wv,\sd}$ extends our conventional understanding of correlation to an ensemble of states.
In all cases considered in this Perspective it is approximated by adapting existing DFAs to ensemble cases via an appropriate weighted average.
By contrast, $\Ec^{\wv,\dd}$ is a form of correlation which is specific to ensembles.\cite{Gould2019-DD}
It arises from the difference of each pure-state individual KS density from its true one, rather than from the energy of each individual state.\cite{Gould2019-DD,Fromager2020-DD}

\newcommand{\Et}{\tilde{E}}
\newcommand{\gap}{\underline{~~}}

\begin{table*}[t!]
    \caption{EDFA combination rules for the ground and various excited states.
    Expressions for $E_{(\Hrm)\xc,\FE} := E_{\Hxc,\FE}-E_{\Hrm}[n_{\FE}]$ [Eq.~\eqref{eqn:(H)xc}] allow the reuse of conventional (HF) exchange and the pairing of it with state-driven correlation contributions. Contributions from the  unconventional Hartree and density-driven correlation terms are also reported.
    $E_{\xc}^{\DFA}$ is a chosen DFA evaluated using density matrices  $\rho_{S_0,D_0,T_1,Q_1}$, for spin-polarized Slater determinant states.
    $\gap$ indicates a hole, namely, an unoccupied KS state.
    $\Et_{hl}=(1-\xi)(4K_{hl})$ is the modified singlet-triplet splitting term that accounts for density-driven correlations. 
    $T_2$ (with orbital $\varphi_{l_2}$) indicates a spatially-different degenerate counterpart to $T_1$ (with orbital $\varphi_{l_1}$); and $T_f^t$ indicates the triplet formed by promoting an electron from $\varphi_f$ to $\varphi_t$.
    `+perms'  indicates the addition of spin-based permutations. Results for quadruplets from \rcite{BerrellThesis}.
    \label{tab:Hxc}}
    \begin{ruledtabular}%
    \begin{tabular}{ccc}
    State, $\iket{\FE_s}$ & Description & $E^{\EDFA}_{(\Hrm)\xc,\FE}$ \\\hline
    Filled $1^2\cdots (h-1)^2$ \\\hline
    $\iket{S_0}\equiv\iket{h^2}$ & Ground state & $E^{\DFA}_{\xc}[\rho_{S_0}]$  \\
    $\iket{T_1^1}\equiv\iket{h^{\up}l^{\up}}$ & Triplet & $E^{\DFA}_{\xc}[\rho_{T_1}]$ \\
    $\iket{T_1^0}\equiv\tfrac{1}{\sqrt{2}}[\iket{h^{\up}l^{\down}} + \down\leftrightarrow \up]$ & Triplet & $E^{\DFA}_{\xc}[\rho_{T_1}]$\\
    $\iket{S_1}\equiv \tfrac{1}{\sqrt{2}}[\iket{h^{\up}l^{\down}} - \down\leftrightarrow \up]$ & Singlet excitation & $E^{\DFA}_{\xc}[\rho_{T_1}] + \Et_{hl}$\\
    $\iket{S_2}\equiv \iket{\gap l^2}$ & Double excitation & $2E^{\DFA}_{\xc}[\rho_{T_1}] - E^{\DFA}_{\xc}[\rho_{S_0}] + \Et_{hl}$ \\
    $\tfrac{1}{\sqrt{2}}[\iket{\gap l_1^{\up}l_2^{\down}} - \up\leftrightarrow\down]$ & Double excitation & $E^{\DFA}_{\xc}[\rho_{T_1}] 
    + E^{\DFA}_{\xc}[\rho_{T_2}] - E^{\DFA}_{\xc}[\rho_{S_0}]
    + \half(\Et_{hl_1} + \Et_{hl_2}) + \Et_{l_1l_2}$ \\
    $\tfrac{1}{\sqrt{2}}[\iket{\gap l_1^2} - \iket{\gap l_2^2}]$ & Double excitation & $E_{\xc}^{\DFA}[\rho_{T_1}] + E^{\DFA}_{\xc}[\rho_{T_2}] - E^{\DFA}_{\xc}[\rho_{S_0}]
    + \half(\Et_{hl_1} + \Et_{hl_2}) + \Et_{l_1l_2}$ \\
    \hline
    Filled $1^2\cdots (h-2)^2$ \\\hline
    $\iket{D_0}\equiv\iket{(h-1)^2h^{\up}}$ & Doublet & $E_{\xc}^{\DFA}[\rho_{D_0}]$ \\
    $\iket{Q_1}\equiv\iket{(h-1)^{\up}h^{\up}l^{\up}}$ & Quadruplet & $E_{\xc}^{\DFA}[\rho_{Q_1}]$ \\
    $\tfrac{1}{\sqrt{3}}[\iket{(h-1)^{\up}h^{\up}l^{\down}} + \text{perms} ]$ & Quadruplet & $E_{\xc}^{\DFA}[\rho_{Q_1}]$ \\
    $\tfrac{1}{\sqrt{2}}[\iket{(h-1)^{\up}h^{\down} l^2} - \up\leftrightarrow\down]$ & Double excitation & $E^{\EDFA}_{\xc}[\rho_{T_{h-1}^l}] 
    + E^{\DFA}_{\xc}[\rho_{T_h^l}] - E^{\DFA}_{\xc}[\rho_{S_0}]
    + \half(\Et_{(h-1)l} + \Et_{hl}) + \Et_{(h-1)h}$ \\
    \end{tabular}
    \end{ruledtabular}
\end{table*}

\subsubsection{From weakly to strongly correlated ensembles}
\label{sec:HLDensity}

\begin{figure}
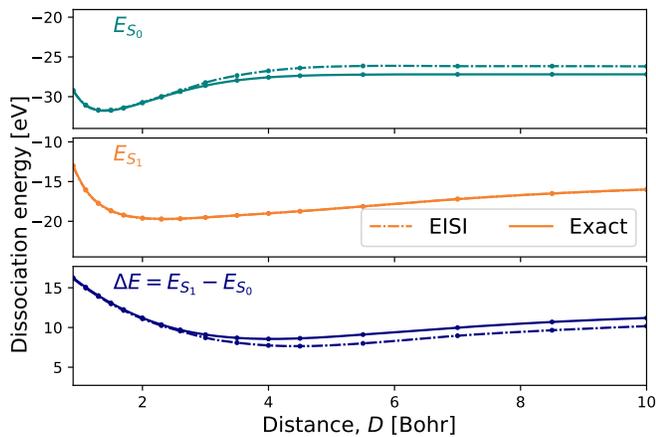

    \includegraphics[width=\linewidth]{{{FigH2-ISI}}}
    \caption{Dissociation curves of H$_2$ for the lowest two singlet states
    and their energy difference, computed using ensemble interaction strength interpolation between the high- and low-density limits (EISI), compared to exact theory.
    Data taken from \rcite{Gould2023-PRL}.
    }
    \label{fig:H2_ISI}
\end{figure}

The next obvious question is how to approximate the sd [Eq.~\eqref{eqn:EEcSD}] and dd [Eq.~\eqref{eqn:EEcDD}] terms usefully. Before we address this, we first highlight a different aspect of the correlation energy in ensembles of many-electron systems, as it uncovers another surprise and turns out to be useful for considering approximations.

Development of ground-state DFAs has benefited enormously from understanding and using uniform density scaling relationships and bounds, i.e., from considering the $\gamma$-dependence of exact density functionals for $n(\vr)\to \gamma^3 n(\gamma \vr)=:n_{\gamma}(\vr)$.\cite{Perdew2003, Levy_1991} 
Large values of $\gamma$ lead to high densities localized in space, while small values of $\gamma$ lead to low densities spread over wide spatial regions -- the two limits therefore exhibit different electronic quantum regimes.
Scaling is closely related to the ACF, with the high-density limit ($\gamma\to\infty$) being related to $\lambda\to 0$ and the low-density limit ($\gamma\to 0^+$) being related to $\lambda\to\infty$. Analysis of these limits can help inform the design of approximations.
One may expect ensemble DFAs to benefit from similar considerations, as demonstrated for thermal ensembles.\cite{Pittalis2011,Dufty2011,Dufty2015,Burke2016:195132}

Scaling was originally introduced to GOK-EDFT by Nagy.~\cite{Nagy1995}
Gould \emph{et al.}~\cite{Gould2023-PRL} recently obtained exact results for $\EHxc$ (and thus $\Ec=\EHxc-\EHx$) in the high- and low-density limits.
Additional scaling relations for components of the correlation energy have recently been derived by Scott et al.~\cite{Scott2024}
The high-density limit was found to be,
\begin{equation}
\lim_{\gamma\to +\infty} \EHxc^{\wv}[n_{\gamma}]  =  \gamma \EHx^{\wv}[n] +\Ec^{\text{GL2},\wv}[n]+\ldots,
\end{equation}
where $\EHx$ is the Hx energy discussed in Sec~\ref{sec:Hx} and $\Ec^{\text{GL2},\wv}$ is the ensemble extension of second-order G\"orling-Levy perturbation theory,\cite{Yang2021} (GL2) with both involving CSFs naturally as explained above.

The low-density limit was found to be,
\begin{equation}
\lim_{\gamma\to 0^+} \EHxc^{\wv}[n_{\gamma}] = \gamma \ESCE[n] + \gamma^{3/2}F^{\ZPE}[n] +\ldots, 
\end{equation} 
wherein the first {\em two} leading order terms are described by \emph{existing} strictly correlated electrons (SCE) and zero point energy (ZPE) terms for {\em ground} states.\cite{Seidl2000,SeiGorSav-PRA-07,GorVigSei-JCTC-09,Gori-Giorgi2010}  
Most interestingly, the low-density limit of matter implies that -- to two leading orders -- \emph{every} GOK ensemble has the same Hxc behaviour as \emph{every} pure state.
This means that the aggregate of ensemble effects in kinetic and electrostatic terms must cancel completely.

Is this surprising result useful?
To illustrate its potential, Figure~\ref{fig:H2_ISI} shows dissociation energy curves for the ground- and first-excited singlet states of H$_2$, computed using the ensemble interaction strength interpolation (EISI) approximation~\cite{Seidl2000,SDGF22} and exact theory.
ISI, extended to ensembles in \rcite{Gould2023-PRL}, uses both \emph{low- and high-density limits} in its construction and thereby is able to reproduce the entire dissociation curve for both the strongly correlated ground- and weakly correlated excited-state, and thus their gap.
It should be noted, however, that H$_2$ is a relatively simple case and ISI ground state energies of more complex systems are less accurate.~\cite{Vuckovic2017,Vuckovic2019}
Still, ISI considerations can be used to improve more generally useful approximations, as of the types shown in the next sub-section.

\subsubsection{Approximating state-driven and density-driven correlation terms} \label{sec:CApprox}

With the results of Sections \ref{sec:CAspects} and \ref{sec:HLDensity} in hand, we are now ready to discuss approximate correlation terms.
We first remind that the exchange and sd correlation terms are defined via response functions, $\chi_{s,\FE}^{\wv}$ and $\chi_{\FE}^{\wv,\lambda}$, for non-interacting and interacting electrons, respectively.
A natural ansatz (inexact but often quite effective in practice~\cite{Gould2020-Molecules,Gould2022-HL,Gould2024-GX24}) is to assume that key relations obeyed by the exact non-interacting response function should also be obeyed by other (i.e. approximate and/or interacting) response functions.
It follows~\cite{Gould2024-GX24} from this response ansatz that state-driven correlation energies may be approximated by the same combinations rules for approximate exchange, adapted from laws for exact exchange, discussed in Section~\ref{sec:XApprox} above.
That is, rather than applying Eq.~\eqref{eqn:EDFA_x} to exchange only, we set
\begin{align}
    E_{\xc,\FE}^{\sd-\EDFA}
    \approx C_1^{\FE} E_{\xc}^{\DFA}[\rho_{\SD_1}] + C_2^{\FE} E_{\xc}^{\DFA}[\rho_{\SD_2}] + \ldots,
    \label{eqn:ExcSD_S}
\end{align}
with the same SD reference states and weighting coefficients as given in 
Table~\ref{tab:Hxc} and discussed in  Section~\ref{sec:XApprox} above.

Approximating dd correlations is more challenging, as $E^{\dd}_{\crm,\FE}[n]$ cannot resemble anything available from conventional DFT.
This is because, as discussed in the opening paragraphs of Section~\ref{sec:CAspects}, Eq.~\eqref{eqn:EEcDD} yields exactly zero for the regular non-degenerate ground state that is at the center of all popular DFA constructions.

A gateway to understanding dd correlations is to exploit the adiabatic connection by considering it from $0<\lambda<\infty$ instead of from $0<\lambda\leq 1$ [as in Eq.~\eqref{eqn:EEcDD}].
As discussed in Section~\ref{sec:HLDensity} above, the limit $\lambda\to\infty$ corresponds to the well-understood low-density limit, which can be used to motivate an approximation.
A detailed analysis is provided in \rcite{Gould2024-GX24}.
Here, we summarise its key elements.

The first step towards an approximation is to define the Hxc energy via an extended ACF.
We write,
\begin{align}
    \EHxc^{\wv}[n]
    :=&\int_{0^+}^1 \W^{\wv,\lambda}[n] d\lambda
    \nonumber\\
    :=&\int_{0^+}^1 \W^{\wv,\RES,\lambda}[n] + \W^{\wv,\ENS,\lambda}[n] d\lambda\;,
\end{align}
where $\W^{\wv,\lambda}$ was split into response (\RES) and density (\ENS) parts by using results from Section~\ref{sec:c_ACF}.
By inspection, the response (\RES) terms capture exchange and sd-correlations, as per Eqs.~\eqref{eqn:EEx} and \eqref{eqn:EEcSD}, and it follows from the above arguments (see also Section~\ref{sec:XApprox}) that the corresponding energy can be handled by combination rules.

The density (\ENS) term, $\EH^{\wv}+\Ec^{\wv,\dd}:=\int_{0^+}^1  \W^{\wv,\ENS,\lambda} d\lambda$, includes Hartree and dd-correlations, and is therefore our target for approximations.
To this end we recognise that the $\W^{\wv,\ENS,\lambda}$ term is known in two limits:\cite{Gould2023-PRL}
1) for $\lambda\to 0$, it is the ensemble Hartree energy, $\EH^{\wv}$;
2) for $\lambda\to\infty$, it is consistent with a typical ground state division ($E_{\Hrm}[n]+E_{\xc}^{\text{SCE}}[n]=\HInt[n]+E_{\xc}^{\text{SCE}}[n]$) for the low-density limit of matter.~\cite{Gould2024-GX24}
Interpolating between these limits by defining $\W^{\wv,\ENS,\lambda}=[1-f(\lambda)]\W^{\wv,\ENS,0} + f(\lambda)\W^{\wv,\ENS,\infty}$, where $f(\lambda)$ obeys $f(0)=0$ and $f(\infty)=1$.
It follows that
\begin{align}
    \Ec^{\wv,\dd}[n]\approx \xi \left\{ \W^{\wv,\ENS,\infty}[n] - \W^{\wv,\ENS,0^+}[n]  \right\} \;.
\end{align}
where $\xi=\int_0^1 f(\lambda)d\lambda$. Note that this result is exact for some $\xi^{\wv}[n]$ -- the only approximation is to set $\xi$ as a $\wv$- and $[n]$-independent constant.

We now focus on the low-density limit for our final step.
Here, the key is to recognise that the low-density limit of matter is completely independent of ensemble effects.\cite{Gould2023-PRL}
Therefore, $\W^{\wv,\ENS,\infty}[n]=W^{\ENS,\infty}[n]$ must be the same as its pure state counterpart and, consequently, $W^{\ENS,\infty}[n]=E_{\Hrm}[n]$ must be the traditional Hartree energy. 
We therefore find that
\begin{align}
\Ec^{\wv,\dd}[n]\approx \xi \left\{ E_{\Hrm}[n] - \EH^{\wv}[n]  \right\} \;.
\label{eqn:EEcDD_Approx}
\end{align}
An equivalent expression for the H + ddc terms is $\EH^{\wv}[n]+\Ec^{\wv,\dd}[n]=(1-\xi)\EH^{\wv}[n] + \xi E_{\Hrm}[n]$.
It relates the combined approximate {\ENS} terms to a weighted average of the FDT-derived Hartree energy and its traditional counterpart.

Eq.~\eqref{eqn:EEcDD_Approx} can also be applied directly to individual excited states by applying the same reasoning directly to an ensemble containing only the target state.
Eq.~\eqref{eqn:EEH0direct} then yields a well-defined ``ensemblized'' Hartree contribution,
\begin{align}
E_{\Hrm.\FE}:=\HInt[n_{s,\FE}]+2\sum_{\FE'<\FE}\HInt[n_{s,\FE\FE'}]\;,
\label{eqn:EEH_State}
\end{align}
that can be used for a specific state, $\iket{\FE}$.
For example, $E_{\Hrm,^1S_2}=\HInt[n_{s,^1S_2}] + 2\HInt[n_{s,^1S_2~^1S_1}]+2\HInt[n_{s,^1S_2~^1S_0}]=E_{\Hrm}[n_{s,^1S_2}]+4K_{hl}+0$.
Note that $E_{\Hrm,\gs}=\HInt[n_{s,\gs}]+0=E_{\Hrm}[n_{s,\gs}]$ reduces to its usual form for ground states.
In the absence of spatial symmetries (but sometimes even in their presence), the first term of Eq.~\eqref{eqn:EEH_State} cancels its low-density limit equivalent in Eq.~\eqref{eqn:EEcDD_Approx} to yield~\cite{Gould2024-GX24}
\begin{align}
    E_{\crm,\FE}^{\dd-\EDFA}
    =&-2\xi\sum_{\FE'<\FE}\HInt[n_{s,\FE\FE'}]\;,
\label{eqn:EEcDD_State}
\end{align}
as the approximate dd correlation energy of the excited state, involving only the downward transition densities, $n_{s,\FE\FE'}$.
Note that this is zero (as expected) for the lowest energy state of any given spin symmetry, i.e., the states that are amenable to conventional DFA treatment.

\subsection{Step five: Solving ensemblized DFT problems}
\label{sec:OO}

As mentioned above, Steps 1--4 provide a clear path towards extending existing ground state DFA understanding to the physical complications of ensemble problems. However, they also introduce a new challenge -- the resulting energy expressions are not amenable to standard self-consistent field (SCF) algorithms because even the most simple functionals become orbital dependent upon ensemblization. While this is not an issue for the formal theory, which is well-defined in both regular KS~\cite{GOK-1,GOK-2} and GKS~\cite{Gould2021-EGKS} approaches, it has a major impact on how variational solutions of ensemblized problems are found numerically using electronic structure software.

To illustrate the problem, consider the energy of an ensemble consisting of the ground and first excited singlet with weights $1-w$ and $w$, respectively.
Even if density-driven correlations are ignored, Steps 1--4 (see Table~\ref{tab:Hxc} for further details) yield,
\begin{align}
    \E^w[n^w]\approx&
    T_s[\rho^w] + \int n^w(\vr)v(\vr)d\vr
    \nonumber\\&
    + (1-w) E^{\DFA}_{\Hxc}[\rho_{\gs}] + w E^{\DFA}_{\Hxc}[\rho_{\ts}]
    \nonumber\\&
    + 4wK_{hl}\;,
    \label{eqn:EwNoSCF}
\end{align}
as the ensemblized energy.
Clearly, Eq.~\eqref{eqn:EwNoSCF} is an \emph{implicit} functional of the ensemble 1RDM, $\rho^w=(1-w)\rho_{\gs} + w\rho_{\ts}$, because: i) $E_{\Hxc}[\rho]$ depends non-linearly on $\rho$; and ii) $K_{hl}$ depends on the HOMO and LUMO only.
SCF algorithms require the explicit evaluation of $\delta\E^w/\delta\rho^w$ which is not available. Implicit evaluation of the derivative is in principle possible via the optimized effective potential (OEP) equation,\cite{Kummel08} but in practice the OEP approach is expensive and difficult. Thus, Eq.~\eqref{eqn:EwNoSCF} is incompatible with standard SCF techniques.

A relatively straightforward solution is to employ orbital optimization (OO) algorithms instead.\cite{Hait2021,Levi2020}
That is, rather than seek $\E_0^{\wv}:=\min_{\rho^{\wv}}\E^{\wv}[\rho]$, which is the goal of a typical DFT code, we instead seek,
\begin{align}
    \bar{\E}_0^{\wv}:=&\min_{\{\phi\}}\E^{\wv}[\{\phi\}]\;,
    &
    \int\phi_i^*(\vr)\phi_j(\vr)d\vr=&\delta_{ij}\;.
\end{align}
\Rcite{Gould2024-Stationary} discusses further details of this approach, as well as the formal  connections between (G)KS and OO theory for ensembles.
In practice, the solution of minimizing orbitals can be found iteratively in some important cases by adapting restricted open-shell Hartree-Fock ideas, as done by Filatov.\cite{Filatov2015-Review} Alternatively, one can repeatedly make unitary transformations of already orthogonal orbitals (e.g., from the ground state). This involves an iterative transformation of the orbitals via $\phi_i\to \sum_j U_{ij}\phi_j$ using $\mat{U}=\exp(\mat{A})$, which is chosen to accelerate the iterations.
OO algorithms inherit the formal scaling of SCF, as well as any exploitation of basis set properties (e.g., fast Fourier transforms for planewaves) or compactness (e.g., density fitting for local basis functions), but generally have a larger prefactor.

We refer the interested reader to  \rcites{Hait2021,Levi2020} and references therein for overviews of recent advances in OO algorithms and their application to both molecules and solids.
The adaptation of OO strategies for ensemble DFAs is reasonably well developed (see, e.g. Supplementary Material of \rcites{Gould2024-ELDA,Gould2024-GX24} for detailed examples) and is especially simple and effective when the `promoted from' and `promoted to' orbitals have different symmetries, as is often the case in low-energy excitations.
Specifically, the \texttt{Broadway} code\cite{Broadway2024} is designed to explicitly solve ensemblized functionals.

\section{From approximations to applications}
\label{sec:Results}

\begin{figure*}[t!]
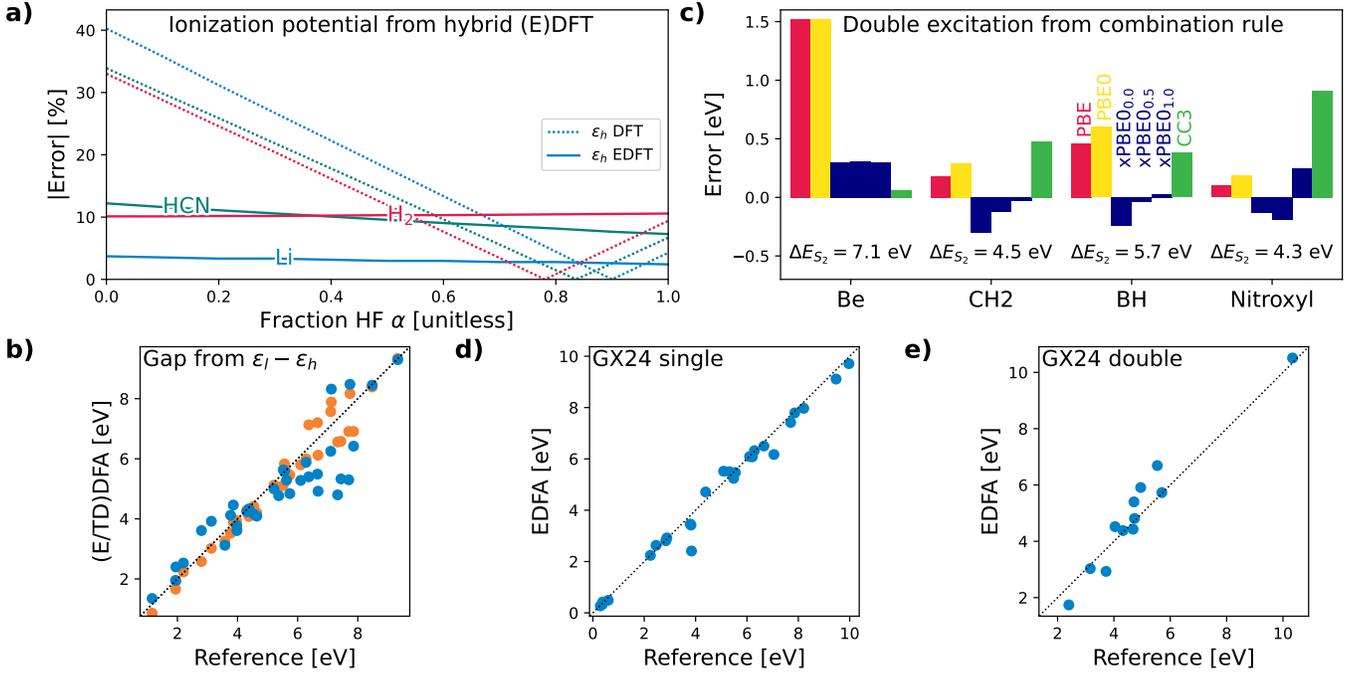

\includegraphics[width=\linewidth]{{{FigKeyResults}}}
\caption{
Absolute deviation (in \%), compared to the experimental value, of the highest-occupied eigenvalue obtained from PBE-based hybrid functionals without (DFT) and with (EDFT) ensemble corrections, as a function of the fraction of Fock exchange, for HCN, Li, and H$_2$.
{\bf b)} singlet-triplet and singlet-singlet excitation energies obtained directly from ensemblized $\epsilon_l-\epsilon_h$ (blue), and TDA-TDDFT (orange), both based on the B3LYP density functional, shown as a function of reference values.
{\bf c)} Errors in double excitation energies obtained from PBE-based hybrids for small molecules using EDFAs (red, yellow), exchange-only EDFAs (navy) and CC3 (green).
{\bf d)} single (singlet-singlet and singlet-triplet)  and {(\bf e)} double (singlet-singlet only) excitation energies obtained using the GX24 ensemble density functional. 
Data and technical details from: a) new results, based on \rcite{Kraisler2015b}; b) \rcite{Gould2022-HL}; c) \rcite{Gould2021-DoubleX};  d), e) \rcite{Gould2024-GX24}.
\label{fig:KeyResults}}
\end{figure*}

\subsection{``Ensemblization'' in action}
\label{sec:AppOurs}

We are now ready to illustrate the value of the concepts and approximations presented above, via some key examples.
We avoid technical details, which can be found in the original articles, and instead focus on pertinent elements of ensemblization and their effect.

We follow the chronological order of development and begin with the fractional electron number ensemble case.
Various studies~\cite{Kraisler2013,Kraisler2014,Kraisler2015,Lavie2023} have shown that using the (exact for H and x but inexact for c) combination rule of Eq.~\eqref{eqn:EDFA_wavg} can already lead to novel insights and practical improvements in adherence of approximate density functionals to the ionization potential (IP) theorem.\cite{Perdew1982,Levy1984,Almbladh1984,Perdew1987} The latter states that for the exact functional the highest occupied eigenvalue is equal and opposite to the ionization potential, i.e.,  $E(N-1)-E(N)=-\epsilon_h$.
Specifically, Eq.\ \eqref{eqn:EDFA_wavg} has been used in combination with left and right derivative relations at the integer ($N$) electron point to derive an additive correction to $\epsilon_h$ that properly accounts for the derivative discontinuity (see Section \ref{sec:Fractional}).

Figure~\ref{fig:KeyResults}a, motivated by results reported in \rcite{Kraisler2015b}, reveals that the derivative discontinuity ensemble correction can impart major improvements to IP calculations with hybrid functional calculations. This is demonstrated in the Figure by considering PBE-based hybrids of the form $E_{\xc}^{\alpha}=\alpha E_{\xrm}^{\text{HF}} + (1-\alpha)E_{\xrm}^{\text{PBE}} + E_{\crm}^{\text{PBE}}$.
In the absence of ensemble corrections, the DFT-computed $\epsilon_h$ (dotted lines) can disobey the IP theorem by as much as tens of percent, with a substantial dependence on the hybrid mixing parameter, $\alpha$, arising from the different binding behaviour of PBE and HF exchange. Using EDFT (solid lines) to correct $\epsilon_h$ not only generally improves attainment of the IP theorem, but also reduces the sensitivity to $\alpha$ dramatically. This specific case hints at a more general principle, which is that the EDFT approach of Eq.~\eqref{eqn:EDFA_wavg} can strongly mitigate the infamous ``parameter dilemma'',\cite{Kraisler2015b} i.e., that an accurate description of energy-related quantities requires a certain value for a free parameter in a functional, whereas the accurate description of potential-related quantities requires a different value.

We now turn to the practical role of ensemblization in excited state (GOK) ensembles. Figure~\ref{fig:KeyResults}b shows excitation energies predicted using ``perturbative EDFT'' (pEDFT),\cite{Gould2022-HL} compared to accurate reference energies.
pEDFT is based on similar ensemblization principles to those behind Figure~\ref{fig:KeyResults}a, in that it seeks to capture ensemblization effects by applying the most minimal ensemble extension of any given DFA, using the above-mentioned extension of EDFT to GKS theory,~\cite{Gould2021-EGKS} which facilitates the use of hybrid functional approximations.
Rather than targeting the IP theorem, pEDFT seeks to make the ``HOMO-LUMO gap'' exactly equal to the optical gap by introducing an ensemble with an infinitesimally weighted excited state contribution, so that the orbitals that are occupied in the ground state are left unchanged but the unoccupied orbitals reflect properties of the excitation.
pEDFT guarantees that $\epsilon_l^{T_1}-\epsilon_h^{S_0}=E_{T_1}-E_{S_0}$ or $\epsilon_l^{S_1}-\epsilon_h^{S_0}=E_{S_1}-E_{S_0}$, depending on the target excitation, where $\epsilon_h^{S_0}$ is obtained from a conventional ground state calculation but $\epsilon_l^{T_1}$ or $\epsilon_l^{S_1}$ are obtained based on $\phi_l^{T_1}$ or $\phi_l^{S_1}$ in the ensemble, by invoking a special case of the extended Koopmans theory for excited states.\cite{LevyNagy1991}
As can be seen in the Figure, predictions made using the pEDFT HOMO-LUMO gap (in this case based on B3LYP\cite{Stephens1994}) are nearly as good as those made using time-dependent DFT (within the Tamm-Damcoff approximation), but can be obtained at a fraction of the computational cost.
pEDFT has also been used to gain insights into the structure and limitations of ensemble DFAs in the charge transfer limit.\cite{Amoyal2024}

Next, we focus on double excitations, as they are of special interest in GOK-EDFT, due to the fact that TDDFT with standard DFAs cannot capture them at all.~\cite{Neepa2004-Double}
Figure~\ref{fig:KeyResults}c shows double excitation energies of Be and small molecules, computed using ensemblized EDFAs based on the results of Section~\ref{sec:Ensemblisation}.
Note that Be and BH also require consideration of spatial symmetries, per Section~\ref{sec:Symmetries}.
The results reveal that EDFT can yield effective predictions for double excitation into both non-degenerate and degenerate unoccupied orbitals, and that properly ensemblized EDFAs can even out-perform sophisticated wave function theory approximations (here CC3~\cite{CC3}) in predicting difficult excitation energies.
The results also validate using the combination laws of  Table~\ref{tab:Hxc} in Eq.~\eqref{eqn:ExcSD_S} as an effective strategy for exchange, because the errors are not only small, but also vary minimally as $\alpha$ is varied, just as with the ionisation potentials studied in Figure~\ref{fig:KeyResults}a.

The main remaining theoretical deficiency in the approximations used in Figures~\ref{fig:KeyResults}a,b is the failure to include density-driven correlations. Before proceeding to demonstrate the consequences of their inclusion, we briefly summarize how they are to be used in practice.

For fractional ensembles, the full ensemblization procedure yields,
\begin{align}
    E^{\EDFA,N-q}\approx& (1-q)E^{\DFA,N} + qE^{\DFA,N-1} 
    \nonumber\\& -\xi q(1-q) J_{hh}\;,
    \label{eqn:GX24Frac}
\end{align}
where the first line is the sd-EDFA and the second line is the dd-EDFA.
That is, the sd-EDFA is the intuitive weighted sum over (e.g.) the energies of $N$- and $N-1$-electron systems.
The dd-EDFA, $-\xi q(1-q)J_{hh}$, can be obtained by using $E_{\Hrm}[n^{N-q}]=E_{\Hrm}[n^{N-1}+(1-q)|\phi_h^2|]$ and $\EH^{N-q}=(1-q)E_{\Hrm}[n^{N-1}+|\phi_h|^2]+qE_{\Hrm}[n^{N-1}]$ in Eq.~\eqref{eqn:EEcDD_Approx}.
Neglecting DD-correlations, as above, is equivalent to setting $\xi=0$.

For excited states, ensemblization yields \emph{a unique EDFA for each energy level of a given system}, because each level interacts differently with lower-lying states.
It is convenient to express the energy of each state as,
\begin{align}
    E_{\FE}^{\EDFA}:=T_{s,\FE} + \tint n_{s,\FE}(\vr)v(\vr) d\vr + E_{\Hrm}[n_{s,\FE}] + E_{(\Hrm)\xc,\FE}^{\EDFA}\;,
    \label{eqn:EDFA_S}
\end{align}
where $T_{s,\FE} = \ibraketop{\Phi_{s,\FE}}{\hat{T}}{\Phi_{s,\FE}}$ and $n_{s,\FE}= \ibraketop{\Phi_{s,\FE}}{\hat{n}}{\Phi_{s,\FE}}$ can (as in ground states) be expanded into orbital contributions, $v(\vr)$ is the external potential,  $E_{\Hrm}[n_{s,\FE}]$ is the Hartree energy of $n_{s,\FE}$, and the last term in Eq.~\eqref{eqn:EDFA_S} is,
\begin{subequations}\begin{align}
    E_{(\Hrm)\xc,\FE}^{\EDFA}:=&E_{\Hxc,\FE}^{\EDFA} - E_{\Hrm}[n_{s,\FE}]\;, \label{eqn:(H)xc_td_1}
    \\
    =& E_{\Hrm,\FE} + E_{\xc,\FE}^{\sd-\EDFA}  + E_{\crm,\FE}^{\dd-\EDFA} 
    \nonumber\\& - E_{\Hrm}[n_{s,\FE}]\;, \label{eqn:(H)xc_td_2}
    \\
    =&E_{\xc,\FE}^{\sd-\EDFA} + 2(1-\xi)\sum_{\FE'<\FE}\HInt[n_{s,\FE\FE'}]\;.
    \label{eqn:(H)xc_td}
\end{align}\label{eqn:(H)xc}\end{subequations}
In detail, Eq. \eqref{eqn:(H)xc_td}
captures the regular (sd) xc energy from Eq.~\eqref{eqn:ExcSD_S} and any \emph{extra} H-like terms (involving transition densities) from the FDT Hartree energy of Eq.~\eqref{eqn:EEH_State} and the dd correlation approximation of Eq.~\eqref{eqn:EEcDD_State}.
The (H) in $E_{(\Hrm)\xc,\FE}^{\EDFA}$ reminds us that the conventional Hartree energy, $E_{\Hrm}[n_{s,\FE}]$, has been subtracted as in Eq.~\eqref{eqn:(H)xc_td_1}.
Eq.~\eqref{eqn:EDFA_S} reduces to a typical DFA for ground states, because $E_{\xc,\gs}^{\sd-\EDFA}=E^{\DFA}_{\xc}[n_{s,\gs}]$, $E_{\Hrm,\gs}=\HInt[n_{s,\gs}]=E_{\Hrm}[n_{s,\gs}]$ and $E_{\crm,\gs}^{\dd-\EDFA}=0$ yielding $E^{\EDFA}_{(\Hrm)\xc,\gs}=E^{\DFA}_{\xc}[n_{\gs}]$.
Importantly, Eqs.~\eqref{eqn:EDFA_S} and \eqref{eqn:(H)xc} can also be immediately used to evaluate the overall ensemble energy via $\E^{\EDFA,\wv}:=\sum_{\FE}w_{\FE}E_{\FE}^{\EDFA}$.

As already mentioned above, the combinations rules of Table~\ref{tab:Hxc} [per Eq.~\eqref{eqn:ExcSD_S}] allow us to (re)use ground state DFAs to approximate exchange and sd correlation. The dd correlation terms (when non-zero) are approximated using Eq.~\eqref{eqn:EEcDD_State}, which lets us combine the transition density [i.e. (H)] terms of $E_{\Hrm,\FE}$ with their counterparts in the dd-correlation approximation, via a $(1-\xi)$ prefactor [see Eq.~\eqref{eqn:(H)xc_td}].
Table~\ref{tab:Hxc} is not comprehensive and other excited state are amenable to EDFAs of similar form.
Extending the table involves: i) obtaining combination laws from total occupation factors and applying them as combination rules for x and sdc terms; and ii) obtaining (H) and ddc terms from the non-zero downward transition densities in Eq.~\eqref{eqn:EEcDD_State}.
Degenerate energy levels may sometimes require evaluation of Eq.~\eqref{eqn:EEcDD_Approx} and lead to extra $J$-like ddc terms similar to that of Eq.~\eqref{eqn:GX24Frac}.

Figures~\ref{fig:KeyResults}d,e show results for single and double excitations energies obtained from GX24. The latter is a very new EDFA which, to the best of our knowledge for the first time, is based on \emph{all} the ensemblization results reviewed here, including contributions from dd-correlations.~\cite{Gould2024-GX24}
GX24 adopts the full strategy behind the computations shown in Figures~\ref{fig:KeyResults}b,c, but also introduces an explicit density-driven correlation approximation based on the scaling properties (low- and high-density limits) of excited state ensembles.
It thus represents the most theoretically complete ensemble approximation to date, and incorporates almost all key theory results from Section~\ref{sec:Ensemblisation}.
Specifically, GX24 combines the ensemblization strategy described here with a range-separated hybrid (RSH) for the sd-EDFA in Eq.~\eqref{eqn:(H)xc_td} and a dd correlation parameter ($\xi=0.32$) found by minimizing errors of 22 difficult low-lying excitations.
The RSH involves a 3:5 ratio of HF and PBE exchange at short-range and full HF at long-range, connected through a range separation parameter $\mu=0.2$~Bohr$^{-1}$, along with PBE correlation.
The resulting energy functionals are then solved for each state individually -- i.e., using Eq.~\eqref{eqn:EDFA_S} for each state of interest -- through orbital optimisation (see Section~\ref{sec:OO}).
Importantly, GX24 uses only standard RSH ingredients that are available in any DFT code.
Thus its excited state energy functionals, including for double excitations that cannot be predicted by usual linear-response time-dependent DFT, are easily implemented in any DFT code with an OO algorithm.
Full details, including basic guidance on implementation, may be found in the supplementary material of \rcite{Gould2024-GX24}.

Figure~\ref{fig:KeyResults}d shows that GX24 has excellent performance for single excitations, with small errors in almost all cases, including cases with very small energy differences $E_{S_1}-E_{T_1}$.
TDDFT based on standard approximate functionals can predict many of these excitations, but yields much less accurate excitation energies.
Results for double excitations, shown in Figure~\ref{fig:KeyResults}e, are not quite as accurate as the single excitations.
They are significantly more impressive, however, given that time-dependent DFT can at all predict only three of the excitations in the figure, and performs extremely poorly (errors of several eV) even for these three.~\cite{Gould2024-GX24}

Before concluding this demonstration of the power of ensemblization, it is worth noting that its benefits can sometimes be enjoyed even if ensembles are bypassed completely, as in the case of GX24.
Specifically, excitation energies can often be computed directly in a ``$\Delta$SCF'' formalism by independently evaluating the energy of each state via Eq.~\eqref{eqn:EDFA_S} (i.e. obtaining a self-consistent `field', SCF), and then evaluating the excitation energy as the difference ($\Delta$) of minima.
As mentioned in Section~\ref{sec:OO} this may necessitate the use of OO strategies.
Importantly, if one is only interested in a few low-lying excited states, ensemblized $\Delta$SCF has better computational scaling than LR-TDDFT or TDDFT.~\cite{Gould2019-DD,Gould2022-HL}
It is thus particularly appealing for very large systems (e.g. bio-molecules) that are inaccessible to existing excited state techniques.
The ensemblized $\Delta$SCF approach has indeed been used for the calculations in Figure~\ref{fig:KeyResults}d,e.

\subsection{A broader perspective of other DFT ensembles and applications}
\label{sec:Others}

Before concluding, it is important to emphasize that not all EDFT innovations must follow the ensemblization path discussed throughout this review. In this section, we highlight \Response{}{paths different to those} discussed above.
We do not aim at a comprehensive report of all alternative approaches, but we do point to other ideas and areas where ensembles have been used to attain advantages and insights not accessible within conventional DFT.

Beyond its obvious application to Coulomb systems, DFT is often analyzed using Hubbard model systems.\cite{Capelle2013}
The Hubbard model falls within the realm of lattice DFT and thus can exhibit different physics to electronic models.
Nevertheless, it can be a useful source of analytic exact results for interacting electronic systems within appropriate limits.~\cite{Sobrino2023}
In this spirit, Burke, Pribram-Jones, Ullrich, and co-workers have addressed approximations and exact conditions in excited state EDFT, using either the Hubbard model~\cite{Sagredo2018,Scott2024} or exact inputs treated via CSFs~\cite{Yang2014,Pribram-Jones2014,Yang2017-EDFT} (for example, the case of Be from their work has already been discussed in Section~\ref{sec:Hx}).

A particularly notable recent finding by Scott et al.,\cite{Scott2024} obtained by considering Hubbard model systems, is that ensemble GKS and ensemble KS theories can exhibit markedly different behaviours in certain low-density-like regimes.
This behavior differs from that of the ground state electronic problem, where differences between GKS and KS tend to be insignificant compared to errors in approximations.~\cite{Kummel2003,Garrick2020,Garrick2022}
We note, however, that Hubbard models were also found to  have a different low-density limit~\cite{Deur2018,Scott2024} to the {\em ab initio} Coulomb case discussed in Section~\ref{sec:HLDensity}, meaning that KS and GKS solutions may be more similar in Coulomb systems.

In a different line of research, and in order to address fundamental gap prediction, Fromager and co-workers have developed an ``$N$-centered'' ensemble framework.\cite{Senjean2018,Senjean2020,Cernatic2022} 
This formalism preserves a net integer electron number by mixing an $N$-electron system with both its $(N+1)$- and $(N-1)$-electron counterparts.
The simplest $N$-centered ensemble is described by the operator,
\begin{align}    \Gammah^{\Xi}:=&\Xi\big(\iout{\Psi^{N+1}} + \iout{\Psi^{N-1}}\big)
    \nonumber\\&
    + (1-2\Xi)\iout{\Psi^{N}}\;,
    \label{eqn:NCentered}
\end{align}
as well as its KS equivalent (i.e., with interacting wave functions, $\iket{\Psi^N}$, replaced by non-interacting counterparts, $\iket{\Phi_s^N}$), which is constructed such that the derivative of its energy, $E^{\Xi}=\tr[\Gammah^{\Xi}\Hh]$, with respect to $\Xi$,  yields the fundamental gap (ionisation potential minus electron affinity), i.e., 
\begin{align}
\text{IP}-\text{EA}=\Dp{\Xi}E^{\Xi}\;.
\label{eqn:FundamentalGap}
\end{align}
Recently, extensions that combine features of $N$-centered (for fundamental gaps, associated with charged excitations) and excited state (for optical gaps, associated with neutral excitations) ensembles have also been derived.~\cite{Cernatic2024,Cernatic2024b}.

\begin{figure}
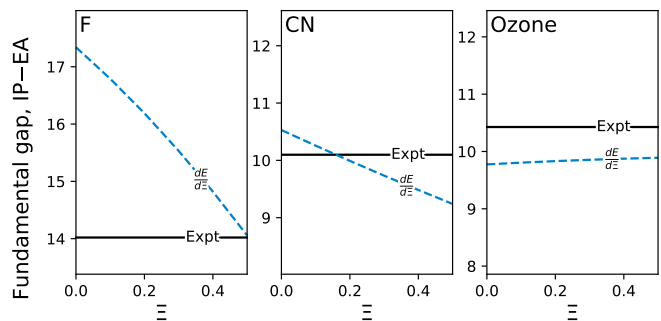

    \includegraphics[width=\linewidth]{{{FigNCentered}}}
    \caption{Fundamental gap (where appropriate as a function of $\Xi$) from experiment (black) and from $\Dp{\Xi}E^{\Xi}$ (blue dashes) using GX24.
    Note that the exact theoretical result should be independent of $\Xi$ and the dependence is introduced by the approximation.
    Results computed for this work using the approach described in \rcite{Gould2024-GX24}.\label{fig:NCentered}}
\end{figure}

Work on $N$-centered ensembles has so far focused on Hubbard models, where analytic or numerically exact results are obtainable.
Here, we take advantage of its mathematical similiarity to fractional ensembles to: i) apply the ensemblisation process described in Section~\ref{sec:Ensemblisation} to Eq.~\eqref{eqn:NCentered}; ii) thereby obtain the appropriate analogue%
\footnote{Specifically, $E^{\EDFA,\Xi}:=\Xi( E^{\DFA,N+1} + E^{\DFA,N-1} ) + (1-2\Xi) E^{\DFA,N} - \xi \Xi (2K_{hh})$ for F and CN and $E^{\EDFA,\Xi}:=\Xi( E^{\DFA,N+1} + E^{\DFA,N-1} ) + (1-2\Xi) E^{\DFA,N}E^{\EDFA,\Xi}:=\Xi( E^{\DFA,N+1} + E^{\DFA,N-1} ) + (1-2\Xi) E^{\DFA,N}-\xi \Xi [K_{hh} + K_{ll} + 2\sum_{i\leq h} (K_{ih}-K_{il})]$ for ozone. We set $\xi=0.32$.}
of Eq.~\eqref{eqn:GX24Frac}; and iii) solve it for atomic and molecular systems.
Figure~\ref{fig:NCentered} shows $N$-centered ensemble based calculations of F, CN, and ozone -- all of which have small electron affinities which are difficult to evaluate using DFAs.
The Figure reports fundamental gap calculations (blue dashed lines) obtained from Eq.\ (\ref{eqn:FundamentalGap}) using 
$\Dp{\Xi}E^{\Xi,\text{GX24}}$, where $E^{\Xi,\text{GX24}}$ is an appropriate  generalization of Eq.~\eqref{eqn:GX24Frac}; compared to experimental values (black lines).
In all cases, setting $\Xi=\tfrac12$ yields results within 1~eV of experiment.

In order to address both ground and excited states that exhibit strong correlations,
Filatov and co-workers (see \rcite{Filatov2015-Review,Filatov2016} for an overview) have been developing the restricted ensemble-referenced Kohn–Sham (REKS) approach. 
This is achieved by combining key elements of EDFT with key elements of wave function theories (WFT).
The underlying ideas are general, but practical aspects are currently tailored mostly towards photochemistry problems.\cite{Filatov1999-REKS}

In the framework of the present work, REKS may be viewed as beginning from a singlet-only KS ensemble of ground and low-lying excited states consistent with the results of Section~\ref{sec:HxResolution}, i.e., singlet eigenstates of the ensemble KS system.
Then, quasi-degenerate perturbation theory is applied to selected low-lying KS CSFs, $\iket{\kappa_s}$, to deduce interacting energy expressions that improve on the usual mean-field treatment of (E)DFAs.
A model energy is then constructed as a functional of the ensemble density ($n=2\sum_i |\phi_i|^2 + f_h |\phi_h|^2 + f_l |\phi_l|^2$ in the simplest REKS approach) by including an explicit dependence on both the orbitals ($\phi_i$, $\phi_h$ and $\phi_l$) and fractional occupation factors ($f_h$ and $f_l$).
Finally, the ground state energy or ensemble energy (depending on approach) is minimized with respect to the orbitals \emph{and} occupation factors.

\begin{figure}
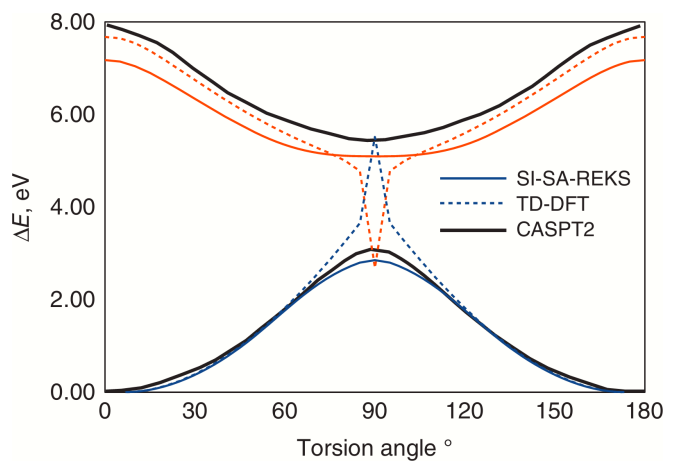

    \includegraphics[width=\linewidth]{{{Figure9-Filatov2015}}}
    \caption{
    Profile of the potential energy surfaces of the ground ($^1A_1$) and excited ($^1B_1$) states (under the D$_2$ symmetry) along the double
bond torsion mode of C$_2$H$_4$. Black lines: CASPT2 results. Colored lines: REKS results. Dashed colored lines—TD-DFT results. DFT calculations employed the CAM-B3LYP density
functional.
    Reproduced from \cite{Filatov2015-Review}, used with permission.
    }
    \label{fig:REKS}
\end{figure}

As one example, Fig.~\ref{fig:REKS} (taken from \rcite{Filatov2015-Review}) illustrates the power of REKS to address difficult, strongly-correlated electronic structure problems of realistic systems. Specifically,
the Figure demonstrates how State-Interaction State-Average REKS (SI-SA-REKS) can capture the 
profile of the potential energy surfaces of the ground $^1A_1$ and excited $^1B_1$ states (under the D$_2$ symmetry) along the double
bond torsion mode of C$_2$H$_4$, whereas TDDFT fails to do so even with the CAM-B3LYP\cite{DFA:CAM-B3LYP} range-separated hybrid functional. 

Finally, we briefly mention some very recent work that is at the interface of degenerate ensembles and fractional ensembles; and degenerate ensembles and excited state ensembles.
The groups of O'Regan and Kraisler independently revealed~\cite{Burgess2024,Goshen2024} complicated dependence of ground state energies on the electron number, $N=N_{\up}+N_{\down}$, and total magnetization, $M=N_{\up}-N_{\down}$, within ensemble spin-DFT.
In a different direction, Theophilou~\cite{Theophilou2018} derived a centro-symmetric variant of DFT that avoids any issues with spatial symmetries, that was recently extended by Nagy~\cite{Nagy2025} to excited state EDFT.
The implications from these works on general (E)DFT have yet to be explored in detail.

\section{Summary and Outlook}
\label{sec:Future}

\subsection{Summary}

In this overview, we have attempted to convey the  power and flexibility of EDFT -- a multifaceted extension of DFT that can be applied effectively (i.e., provide a useful balance of first-principles methodology with efficienct computation) to many situations in which traditional DFT is either inapplicable in principle or struggles in practice. 
Yet, we have stressed throughout that DFT is still at the core of EDFT.
The latter reduces to the conventional theory in appropriate limits and greatly benefits from the adaptation of the results of decades of DFT research.

After a brief introduction to the topic, in Section \ref{sec:EDFT} we discussed basic concepts of EDFT. Specifically, we introduced the ensemble Kohn-Sham theory and presented three different types of ensembles, corresponding to systems in a degenerate ground state, systems with a non-integer number of electrons, and systems in an excited state. 

In Section~\ref{sec:Ensemblisation}, we presented the important concept of ensemblization, namely, that ground-state DFT cannot be used naively for EDFT, but rather \Response{}{needs} to be extended appropriately. This idea is based on the philosophy that exact density functional components should be defined as rigorously and generally as possible, with the exact results then used to inspire approximations (EDFAs) that reuse the most appropriate existing DFA infrastructure.

We have presented numerous examples of failures associated with naive use of ground-state DFT and, perhaps more importantly, discussed in detail steps needed for systematic ensemblization. In particular, we explained the need for explicit state averaging, presented a unified derivation of Hartree-exchange energies, followed by their separation into individual components, and then discussed ensemble correlations. In the latter case, we showed that a ``hidden'' type of correlation, completely absent in ground-state DFT, emerges, explained its physical origins, and discussed how it may be approximated.

In Section \ref{sec:Results}, we explained how the concept of ensemblization can be used to turn the potential of EDFT into a practical reality. First, we presented some applications of ensemblization to Coulomb systems, using the types of ensembles discussed in Section \ref{sec:EDFT}. 
Next, we surveyed a broader perspective of additional systems, ensembles, and DFT-based mapping.

We believe that the ideas and concepts summarized above  offer a fresh path to unleashing the potential of EDFT for practical applications to modern problems in chemistry, physics, and materials science. With the notable exception of REKS~\cite{Filatov1999-REKS}, until relatively recently this potential went mostly unrealised.
The examples in Section~\ref{sec:Results} illustrate that ensemblization can already yield useful results by overcoming the limits of intuition.
Extension to other kinds of ensembles, notably  thermal ensembles~\cite{Mermin1965,Pittalis2011,Dufty2011,Dufty2015}, or the extended $N$-centered ensembles of Fromager et al.,~\cite{Cernatic2024,Cernatic2024b} will hopefully yield similar benefits.

Importantly, we emphasize that EDFAs have plenty of scope for improvement.
Ground-state DFT is now a mature topic with a rich history of some 60 years of research, and conventional DFAs have benefited from at least 40 years of steady improvement targeted at ground states, \Response{}{leading occasionally to new functionals embraced by a large number of users}.
In contrast, the majority of the EDFT work discussed in this article has been published in the last decade and especially in the last five years.
There is thus plenty of room for improvement and multiple research opportunities await.

\subsection{Outlook}

We conclude by listing what we view as some major outstanding problems in ensemble (and ensemblization of) DFT, and some possible ideas towards their solution.

An obvious focus for improvements is the state-driven correlation energy model.
Combination rules (see Section~\ref{sec:CApprox}) are reasonably effective, yet are based on rules that are exact for non-interacting response functions, yet are inexact for their interacting counterparts.
An improved treatment of state-driven correlation physics is therefore required to attain higher accuracy.
Potential sources of improvement include using the recently derived ensemble LDA~\cite{Gould2024-ELDA} for excited states as a starting point for better EDFAs; incorporating ensemble G\"orling-Levy perturbation theory terms~\cite{Yang2021} in ensemble double hybrids; drawing insights from the ensemble random-phase approximation (see, e.g. \rcite{Gould2021-DoubleX}); or applying data-driven approaches.

Density-driven correlations also offer scope for improvement.
A recently derived approximation [Eqs~\eqref{eqn:EEcDD_Approx} and \eqref{eqn:EEcDD_State}] for density-driven correlations seems to be surprisingly effective,\cite{Gould2024-GX24} but involves empiricism in the choice of $\xi$.
Both theory and approximation can certainly be refined, with the challenge of retaining practically useful and computationally tractable expressions.

Applications of ensemblization work have so far focused on molecules based on light elements.
Some key results of EDFT theory rely on the properties of molecules and therefore need to be reevaluated in order to underlie calculations in the solid state.
Application of EDFT to a particular class of non-thermal states of the uniform gas is a first step towards this goal.\cite{Gould2024-ELDA,Loos2025-XLDA}
Ensemblization of of methods based on spin-current-DFT \cite{desmarais2025torques,Desmarais-mSCAN,Desmarais-JSCAN,Desmarais-ELF,Desmarais-GKS-SCDFT} will be required to study magnets and spin-orbit-coupled systems.
These advances would facilitate the estimation of energy splittings which are of great importance in various optoelectronics and spintronics applications.

Another important line of future development would  be to merge the strategies  of Section~\ref{sec:Ensemblisation} with those of the SI-SA-REKS scheme~\cite{Filatov2015-Review}.
This may necessitate extension into ensemble DFT\cite{Pernal2016,Senjean2015,Alam2016,Alam2017} of long-explored connecting schemes between KS-DFT and wave function theory,\cite{Savin1995-BookChapter} which have led to improved understanding and treatment of ground states.
Success in this endeavor could yield a rather general, practical solution that not only deals consistently with weakly and strongly correlated molecular excited states, but also everything in between.

Additional approaches and problems that gain inspiration from and potentially extend the reach of GOK-EDFT are time-dependent EDFT,\cite{Li1985:0556,Li1985:3970,Daas2025}
the quest of overcoming the limits of the Born–Oppenheimer approximation, \cite{Filatov2018-BBO,Fromager2024-BBO}
ensemble generalization of reduced-density-matrix-functional theory,\cite{Valone1980b,Schilling2021,Liebert2022} and a  unitary coupled-cluster approach for purified mixed states.\cite{Benavides-Riveros2022}

As exemplified by Eq.~\eqref{eqn:EDFA_S} and Table~\ref{tab:Hxc}, it is worth stressing  that ensemblization \emph{does not} always require an ensemble calculation, but may instead be used to develop extended functionals for state-specific approaches that can be used in ``$\Delta$SCF'' calculations for excitations.
Recent analyses of this particular use of EDFT identify complications that arise and call for practical solutions.\cite{Gould2024-Stationary,Fromager2024-Stationary} 
Emerging optimization approaches\cite{Hait2021,Levi2020} and theory developments in state-specific DFT\cite{Su2024,LuGao2022,Giarrusso2023,Loos2025-ESP,Filatov2025,Gould2024-GX24} appear to hold promise in that regard.

Finally, we stress that ground-state DFT also continues to progress, in reducing its computational cost (e.g. \rcite{Grimme2021}), in improving its scaling (e.g. \rcite{Gavini2023}), and in improving its accuracy (e.g. \rcite{Teale22} provides a useful survey of the state-of-art in DFT).
Ensemblization means that much of this progress can be extended to excited states.
Here, we have shared our view on how this may be achieved.
While the above account is by no means comprehensive, we do hope that this Review will stimulate new ideas, developments, and applications.

\acknowledgments
TG and LK were supported by an Australian Research Council (ARC) Discovery Project (DP200100033).
TG was supported by an ARC Future Fellowship (FT210100663). LK was supported by the Aryeh and Mintzi Katzman Professorial Chair and the Helen and Martin Kimmel Award for Innovative Investigation.
TG would like to thank E. Fromager for many interesting discussions regarding ensemble DFT.
We would like to thank all the participants of the EDFT2024 conference in Durham for the wonderful brainstorming on the past and future of EDFT.

%

\end{document}